\documentclass[12pt]{article}
\pdfoutput=1

\usepackage{draft,hyperref,cancel,subfig,float,ifthen,xcolor,tikz-3dplot}
\usepackage[nosort]{cite}

\begin{document}

\begin{titlepage}

\preprint{CALT-TH-2020-054}

\begin{center}

\title{
Lorentzian Dynamics and 
\\
Factorization Beyond Rationality
}

\author{Chi-Ming Chang$^{a,b}$ and Ying-Hsuan Lin$^{c,d}$}

\address{\small${}^a$Yau Mathematical Sciences Center (YMSC), Tsinghua University, Beijing, 100084, China}
\vspace{-.1in}
\address{\small${}^b$Beijing Institute of Mathematical Sciences and Applications (BIMSA), Beijing, 101408, China}
\vspace{-.1in}
\address{\small${}^c$Jefferson Physical Laboratory, Harvard University, Cambridge, MA 02138, USA}
\vspace{-.1in}
\address{\small${}^d$Walter Burke Institute for Theoretical Physics,
\\ 
California Institute of Technology, Pasadena, CA 91125, USA}

\email{cmchang@tsinghua.edu.cn, yhlin@fas.harvard.edu}

\end{center}

\vfill

\begin{abstract}
We investigate the emergence of topological defect lines in the conformal Regge limit of two-dimensional conformal field theory. We explain how a local operator can be factorized into a holomorphic and an anti-holomorphic defect operator connected through a topological defect line, and discuss implications on analyticity and Lorentzian dynamics including aspects of chaos. We derive a formula relating the infinite boost limit, which holographically encodes the ``opacity'' of bulk scattering, to the action of topological defect lines on local operators. Leveraging the unitary bound on the opacity and the positivity of fusion coefficients, we show that the spectral radii of a large class of topological defect lines are given by their loop expectation values. Factorization also gives a formula relating the local and defect operator algebras and fusion categorical data. We then review factorization in rational conformal field theory from a defect perspective, and examine irrational theories. On the orbifold branch of the $c = 1$ free boson theory, we find a unified description for the topological defect lines through which the twist fields are factorized; at irrational points, the twist fields factorize through ``non-compact'' topological defect lines which exhibit continuous defect operator spectra. Along the way, we initiate the development of a formalism to characterize non-compact topological defect lines.
\end{abstract}

\vfill

\end{titlepage}

\tableofcontents

\section{Introduction}

Two-dimensional conformal field theory enjoys special kinematics that lead to holomorphically factorized continuous symmetries \cite{Belavin:1984vu}. However, except in very special cases, the full theory is not holomorphically factorized. The local operators transform as bi-modules of the left and right-moving chiral algebras, but a generic local operator cannot be regarded as the composite of holomorphic and anti-holomorphic {\it local} operators. In rational conformal field theory \cite{Friedan:1983xq} there is a weaker sense of holomorphic factorization. Loosely speaking, on an oriented manifold $\cM_2$, the holomorphic and anti-holomorphic degrees of freedom dwell on two separate copies of $\cM_2$ (more precisely, $\cM_2$ and its orientation reversal $\overline{\cM}_2$), connected through a bulk topological quantum field theory \cite{Witten:1988hf,Fuchs:2002cm,Fuchs:2003id,Fuchs:2004dz,Fuchs:2004xi,Fjelstad:2005ua}. The truly holomorphically factorized case is when the bulk theory is trivial. Extensive studies in the past have revealed that rational conformal field theory, three-dimensional topological quantum field theory, modular tensor category, and various other mathematical structures are different facets of the same underlying truth \cite{Moore:1988qv,Moore:1988ss,Moore:1988uz,Elitzur:1989nr,Moore:1989vd,Moore:1989yh,Behrend:1999bn,Petkova:2000dv,Petkova:2001ag,Petkova:2001zn,Fuchs:2002cm,Fuchs:2003id,Fuchs:2004dz,Fuchs:2004xi,Fjelstad:2005ua}. In particular, the nontrivial dynamics of the conformal field theory, encoded in the three-point structure constants, can be explicitly expressed in terms of invariant data of modular tensor category, or equivalently as link invariants of the topological quantum field theory; crossing symmetry is solved by solutions to the pentagon identity.\footnote{This formulation is ignorant of the explicit form of the chiral algebra blocks, and in particular, the normalization of the blocks is a gauge ambiguity. There is no purely categorical way to decide which gauge gives the canonically normalized blocks (corresponding to normalizing the two-point function of chiral algebra primaries). Other means such as solving the null state decoupling equation \cite{Knizhnik:1984nr} or the Wronskian method \cite{Mukhi:2017ugw} are necessary to determine this piece of the conformal field theory data. An explicit illustration of this point will be given in Section~\ref{Sec:Ising}.
}
This paper investigates whether some of this rich structure and insight survive when we venture beyond rationality. Since general conformal field theory has no relation to bulk topological quantum field theory, it is instructive to first reformulate holomorphic factorization in a purely two-dimensional framework. The role of line defects in the bulk topological quantum field theory is replaced by topological defect lines (TDLs) of the conformal field theory, and a local operator can be regarded as the composite of a holomorphic and an anti-holomorphic {\it defect} operator connected by a topological defect line.\footnote{Topological defect lines in two-dimensional quantum field theory have been investigated in \cite{Oshikawa:1996ww,Oshikawa:1996dj,Petkova:2000ip,Fuchs:2002cm,Fuchs:2003id,Fuchs:2004dz,Fuchs:2004xi,Frohlich:2004ef,Fjelstad:2005ua,Frohlich:2006ch,Quella:2006de,Fuchs:2007vk,Fuchs:2007tx,Bachas:2007td,Kong:2008ci,Frohlich:2009gb,Petkova:2009pe,Drukker:2010jp,2012CMaPh.313..351K,Carqueville:2012dk,Davydov:2013lma,Brunner:2013xna,Kong:2013gca,Petkova:2013yoa,Gaiotto:2014lma,Bischoff:2014xea,Bhardwaj:2017xup,Chang:2018iay,Ji:2019ugf,Lin:2019hks,Thorngren:2019iar,Lou:2020gfq,Komargodski:2020mxz}. The modernized view of (generalized) symmetries as topological defects was developed in \cite{Drukker:2010jp,Davydov:2010rm,Kapustin:2014gua,Gaiotto:2014kfa}.
\label{Foot}
} 
For rational theory, this reformulation is a superficial one, obtained essentially by ignoring the third dimension of the bulk, and giving a new name, Verlinde lines \cite{Verlinde:1988sn,Petkova:2000ip,Drukker:2010jp,Gaiotto:2014lma}, to the projected shadows of line operators in the bulk theory. Nonetheless, this new perspective permits the extrapolation of key ideas to theories that need not have a bulk correspondence. Mathematically, only the structure of fusion category \cite{Etingof:aa,etingof2016tensor}, and not modular tensor category, is required to describe the dynamics of topological defect lines. Less is more. 

Loosely speaking, a local operator $\cO$ on the Euclidean plane $\bar z = z^*$ is holomorphically-defect-factorized if
\ie
\begin{gathered}
\begin{tikzpicture}
\draw (-2,0) \dott{right}{$\cO(z, \bar z)$};
\draw (0,0) node {$\sim$} ;
\draw [line,->-=.6] (1.5,0) \dotsol{left}{$\cD(z)$} -- node [above] {$\cL$} node [below] {$\substack{~\\~}$} (2.5,0) \dotemp{right}{$\ocD(\bar z)$};
\end{tikzpicture}
\end{gathered} \, ,
\fe
where $\cL$ is a topological defect line, and $\cD$ and $\ocD$ are holomorphic and anti-holomorphic defect operators. These objects are introduced in Section~\ref{Sec:Factorization}, and a precise definition of factorization is given in Definition~\ref{Hypo}. To avoid confusion with the stronger sense of holomorphic factorization (of the full theory), the factorization described above will be referred to as ``holomorphic-defect-factorization'' throughout this paper. 

Holomorphic-defect-factorization obscures the meaning of spacetime signature. Starting from a Euclidean correlator, Lorentzian dynamics are obtained by continuing the complex coordinates $z, \, \bar z$ of local operators independently to real $z$ and $\bar z$\cite{Schwinger:1958qau,Schwinger:1959zz,wightman1960quantum,Osterwalder:1973dx,Osterwalder:1974tc}. However, for a holomorphically-defect-factorized local operator, a new interpretation is available: The correlator stays in the Euclidean regime, but becomes one involving defect operators and topological defect lines. This dual perspective suggests that aspects of Lorentzian dynamics are dictated by fundamental properties of topological defect lines. In particular, for a four-point function involving holomorphically-defect-factorized local operators, the conformal Regge limit \cite{Cornalba:2007fs,Costa:2012cb} at infinite boost is completely fixed by the action of the topological defect line on local operators. For rational theories, this connection was explored from a bulk perspective by \cite{Gu:2016hoy} in the context of out-of-time-ordered correlators and chaos. We reformulate this connection in a purely two-dimensional way, and generalize beyond rationality. In particular, the ``opacity" of a Lorentzian four-point function is related to the matrix elements of the factorizing topological defect line. By a unitarity bound on the opacity proven in \cite{Caron-Huot:2017vep}, we show that the spectral radii of factorizing topological defect lines are determined by the loop expectation values. Interestingly, in higher dimensional conformal field theory, light-ray operators \cite{Kravchuk:2018htv} dominate the Regge limit of four-point functions, and explain the analyticity in spin of the Lorentzian inversion formula \cite{Caron-Huot:2017vep}. The central role played by line operators in the conformal Regge limit appears to be a common theme.

The connection between topological defect lines and Lorentzian dynamics is bidirectional. The Regge limit of correlators allows the discovery of topological defect lines given the correlators of local operators. Traditionally, a topological defect line $\cL$ is characterized by a topological map $\widehat\cL$ on the Hilbert space $\cH$ of local operators, subject to stringent consistency conditions, including the condition that the modular S transform of the twisted partition function ${\rm Tr}_{\cH} \, \widehat\cL \, q^{L_0-\frac{c}{24}} \bar q^{\bar L_0-\frac{\bar c}{24}}$ gives a sensible partition function for the defect Hilbert space \cite{Petkova:2000ip}. A close analogy is the characterization of a consistent conformal boundary condition as a (closed string) state satisfying the Cardy condition \cite{Cardy,Cardy:1986gw,Cardy:1989vyr}. Due to this analogy, we also call this condition for topological defect lines the Cardy condition. At the level of principle, it would be desirable to have a direct formula for $\widehat\cL$ in terms of correlators of local operators. As we will explain, assuming that a local operator is holomorphically-defect-factorized through $\cL$, the conformal Regge limit provides such a formula. Conversely, the conformal Regge limit serves as a nontrivial test of whether a local operator is holomorphically-defect-factorized. We call this the strong holomorphic-defect-factorization criterion (Definition~\ref{Strong}). We also formulate the weak holomorphic-defect-factorization criterion (Definition~\ref{Weak}), for topological defect lines satisfying a weaker version of the Cardy condition.

The holomorphic-defect-factorization criteria are put to test in the $c=1$ free boson theory, on both the toroidal branch and the orbifold branch. On the toroidal branch, all local operators are holomorphically-defect-factorized through ${\rm U}(1)$ symmetry defect lines, regardless of rationality. On the orbifold \cite{Dixon:1985jw,Dixon:1986jc} branch, although the cosine operators are always factorized, for the twist fields we find a dichotomy between rational and irrational points. At rational points, the twist field correlator satisfies the strong holomorphic-defect-factorization criterion, and we obtain a uniform formula describing the map $\widehat\cL$ for the topological defect line $\cL$ through which the twist field factorizes; in particular, at $r^2 = u/v$ with $u, v$ coprime, the planar loop expectation value is $\vev\cL_{\bR^2} = \sqrt{uv}$. At special rational points, it can be explicitly checked that our formula agrees with the Verlinde formula \cite{Verlinde:1988sn}. At irrational points, only the weak holomorphic-defect-factorization criterion is satisfied. More precisely, the twist field factorizes through a ``non-compact" topological defect line with the defining property that its defect Hilbert space exhibits a continuous spectrum (Definition \ref{NCP}).\footnote{Topological defect lines exhibiting continuous spectra in compact theories were previously encountered in \cite{Fuchs:2007tx}.}
A non-compact topological defect line cannot be described by a semi-simple object in a fusion category. We initiate a preliminary development of a more general framework---{\it TDL category}---that includes non-compact topological defect lines and relaxes semi-simplicity. In many examples, the more general TDL categories (which contain non-compact topological defect lines) arise in the limit of sequences of fusion categories, in which sequences of simple topological defect lines converge to non-compact topological defect lines.

This paper is organized as follows. Section~\ref{Sec:Factorization} introduces topological defect lines, explains the meaning of holomorphic-defect-factorization, expresses the three-point function of local operators in terms of defect data, discusses the properties of factorizing topological defect lines, and introduces non-compact topological defect lines and TDL categories. Section~\ref{Sec:FRL} studies correlators of holomorphically-defect-factorized local operators, and connects the conformal Regge limit to fundamental properties of topological defect lines. In particular, it is explained how the conformal Regge limit provides a way to discover topological defect lines. Section~\ref{Sec:Lorentzian} explores further aspects of Lorentzian dynamics, including a unitarity bound on the opacity of the four-point function in the conformal Regge limit, its relation to a formula on the spectral radii of the topological defect lines, and the connection to chaos via out-of-time-order correlators. Section~\ref{Sec:Rational} examines holomorphic-defect-factorization in rational theories, first from a purely two-dimensional perspective, and then reviews the three-dimensional bulk perspective. Section~\ref{Sec:Free} tests holomorphic-defect-factorization beyond rationality, by studying the $c = 1$ free boson theory on both toroidal and orbifold branches. Section~\ref{Sec:Conclusion} ends with a summary and further comments. Appendix~\ref{Sec:Crossing} proves that the crossing symmetry of holomorphic defect operators implies the crossing symmetry of holomorphically-defect-factorized local operators. Appendix~\ref{Sec:Bounded} proves the spectral radius formula by utilizing the Perron-Frobenius theorem and its generalizations. Appendix~\ref{App:Free} collects formulae and computations relevant for the study of the free boson orbifold theory in Section~\ref{Sec:S1Z2}.

\section{Holomorphic-defect-factorization of local operators}
\label{Sec:Factorization}

\subsection{Topological defect lines}
\label{Sec:Defect}

Let us first review basic properties of topological defect lines (TDLs), which encompass and generalize symmetry defect lines. The exposition here largely follows \cite{Chang:2018iay}; for other relevant references see footnote~\ref{Foot}. TDLs can reverse orientation, act on local operators by circling and shrinking, end on defect operators, join in junctions, undergo isotopic transformations without changing the correlation functionals, and different configurations of TDLs are equivalent under the so-called $F$-moves. The direct sum of two TDLs gives another TDL, and correlation functionals are additive under direct sums.

A TDL $\cL$ has an orientation reversal $\ocL$, meaning the equivalence of 
\ie
\begin{gathered}
\begin{tikzpicture}[scale=.5]
\draw [line,->-=.55] (0,-1.5) -- (0,0) node [left] {$\cL$} -- (0,1.5);
\end{tikzpicture}
\end{gathered} 
\quad = \quad
\begin{gathered}
\begin{tikzpicture}[scale=.5]
\draw [line,-<-=.55] (0,-1.5) -- (0,0) node [right] {$\ocL$} -- (0,1.5);
\end{tikzpicture}
\end{gathered} 
\, .
\fe
It acts on a local operator by circling and shrinking,
\ie
\label{Act}
\begin{gathered}
\begin{tikzpicture}[scale=.75]
\draw [line,-<-=.26] (0,0) circle (1.5);
\draw \dott{below}{$\phi(z,\bar z)$};
\node at (-2,0) {$\cL$};
\end{tikzpicture}
\end{gathered} 
\quad = \quad
\begin{gathered}
\begin{tikzpicture}
\draw \dott{right}{$\widehat\cL(\phi)(z, \bar z)$};
\end{tikzpicture}
\end{gathered} 
\, .
\fe
In particular, the loop expectation value of a TDL $\cL$ on the plane is\footnote{The planar loop expectation value $\vev{\cL}_{\bR^2}$ is related to the quantum dimension $d_{\cL}$ in the categorical language by a factor of the Frobenius-Schur indicator $\chi_{\cL}$
\ie
\label{LEV}
\vev{\cL}_{\bR^2} = {d_{\cL} \over \chi_{\cL}} \, .
\fe
The quantum dimension $d_{\cL}$ is equal to the vacuum expectation value of $\cL$ wrapping the non-contractible cycle on a cylinder, {\it i.e.}
\ie
d_{\cL} = \vev{\cL}_{{\rm S}^1 \times \bR} \, .
\fe
The two loop expectation values are related by at most a phase arising from the extrinsic curvature improvement term \cite{Chang:2018iay}.
\label{VEV}
}
\ie
\vev{\cL}_{\bR^2} ~=~
\begin{gathered}
\begin{tikzpicture}[scale=1]
\draw [line,-<-=.26] (0,0) circle (.75);
\node at (0,0) {$\cL$};
\end{tikzpicture}
\end{gathered} \, .
\fe
A TDL is associated with a defect Hilbert space obtained by quantizing on the cylinder with twisted (by the TDL) periodic boundary conditions. The defect partition function is
\ie
Z_\cL(\tau, \bar\tau) \ = \ 
\begin{gathered}
\begin{tikzpicture}[scale=.75]
\draw [line,->-=.55] (-.5,-1.5) -- (-.5,1.5);
\draw [line] (-2,-1.5) -- (1,-1.5) -- (1,1.5) -- (-2,1.5) -- (-2,-1.5);
\draw (0,0) node {$\cL$} ;
\end{tikzpicture}
\end{gathered} \, .
\fe
Via the state-operator map, states in the defect Hilbert space $\cH_\cL$ correspond to defect operators on which the TDL can end. Since the defect Hilbert space has a norm, every defect operator $\cD \in \cH_\cL$ has a hermitian conjugate $\cD^\dag \in \cH_{\ocL}$ of the same weight, 
\ie
(h_\cD, \, \bar h_\cD) = (h_{\cD^\dag}, \, \bar h_{\cD^\dag}) \, ,
\fe
and the two are related by charge conjugation.

A TDL $\cL$ is called {\it simple} if the defect Hilbert space $\cH_{\cL\ocL}$ has a unique ground state with $(h,\bar h)=(0,0)$, and called {\it semi-simple} if can be uniquely expressed as a direct sum of finitely many simple TDLs. Any TDL $\cL'$ such that the defect Hilbert space $\cH_{\cL \ocL'}$ has a ground state with $(h, \bar h) = (0,0)$ is said to be {\it isomorphic} to $\cL$, in the sense that there is a Virasoro-equivariant isomorphism between $\cH_{\cL}$ and $\cH_{\cL'}$. A category of TDLs is called {\it semi-simple} if every TDL is semi-simple. We assume semi-simplicity for now, and comment on the more general situation later.

A trivalent junction of TDLs is depicted as
\ie
\begin{gathered}
\begin{tikzpicture}[scale=1]
\draw [line,->-=1] (-1,0) -- (-.8,0) node {$\times$} -- (0,0) node [right] {$\cL_3$};
\draw [line,->-=1] (-1,0) -- (-1.5,.87) node [above left=-3pt] {$\cL_1$};
\draw [line,->-=1] (-1,0) -- (-1.5,-.87) node [below left=-3pt] {$\cL_2$};
\end{tikzpicture}
\end{gathered}
\fe
The marking $\times$ labels the ordering of edges at trivalent junctions, and can be permuted around by the {\it cyclic permutation map} $V_{\cL_1, \cL_2, \cL_3} \to V_{\cL_2, \cL_3, \cL_1}$. The junction vector space $V_{\cL_1, \cL_2, \cL_3}$ associated to a trivalent junction is defined as the subspace of topological weight $(0, 0)$ states in the defect Hilbert space $\cH_{\cL_1, \cL_2, \cL_3}$. The space of possible trivalent junctions is encoded in the {\it fusion rule} of the simple TDLs; the fusion coefficients correspond to the dimensions of the junction vector spaces.

There is a trivial TDL $\cI$ that represents no TDL insertion. However, when it ends on another TDL $\cL$ forming a trivalent junction, it introduces a map from the junction vector space $V_{\cL,\ocL,\cI}$ (resp. other permuted orderings) to $\bC$. Such a trivalent junction could be removed by evaluating the map on the identity junction vector $1_{\cL,\ocL,\cI}$ (resp. other permuted orderings). 

A configuration of TDLs is a (linear) correlation functional of junction vectors, and different configurations are equivalent under $F$-moves
\ie
\label{F}
\begin{gathered}
\begin{tikzpicture}[scale=1]
\draw [line,-<-=.56] (-1,0) -- (-.8,0) node {$\times$} -- (-.5,0) node [above] {$\cL_5$} -- (0,0);
\draw [line,->-=1] (-1,0) -- (-1.5,.87) node [above left=-3pt] {$\cL_1$};
\draw [line,->-=1] (-1,0) -- (-1.5,-.87) node [below left=-3pt] {$\cL_2$};
\draw [line,->-=1] (0,0) -- (.5,-.87) node [below right=-3pt] {$\cL_3$};
\draw [line,->-=1] (0,0) -- (.1,.175) node {\rotatebox[origin=c]{60}{$\times$}} -- (.5,.87) node [above right=-3pt] {$\cL_4$};
\end{tikzpicture}
\end{gathered}
\quad = \quad
\sum_{\cL_6} 
\begin{gathered}
\begin{tikzpicture}[scale=1]
\draw [line,-<-=.56] (0,-1) -- (0,-.8) node {$\times$} -- (0,-.5) node [right] {$\cL_6$} -- (0,0);
\draw [line,->-=1] (0,-1) -- (.87,-1.5) node [below right=-3pt] {$\cL_3$};
\draw [line,->-=1] (0,-1) -- (-.87,-1.5) node [below left=-3pt] {$\cL_2$};
\draw [line,->-=1] (0,0) -- (-.87,.5) node [above left=-3pt] {$\cL_1$};
\draw [line,->-=1] (0,0) -- (.175,.1) node {\rotatebox[origin=c]{30}{$\times$}} -- (.87,.5) node [above right=-3pt] {$\cL_4$};
\end{tikzpicture}
\end{gathered}
~\circ~
(F^{\cL_1, \cL_2, \cL_3}_{\ocL_4})_{\cL_5, \cL_6} \, ,
\fe
where the $F$-symbols are bilinear maps
\ie
(F^{\cL_1, \cL_2, \cL_3}_{\ocL_4})_{\cL_5, \cL_6} \ : \ 
V_{\cL_1, \cL_2, \ocL_5} \otimes V_{\cL_5, \cL_3, \cL_4} \to 
V_{\cL_2 ,\cL_3, \ocL_6} \otimes V_{\cL_1, \cL_6, \cL_4} \, .
\fe
In particular, the planar loop expectation value \eqref{LEV} is related to an $F$-symbol by
\ie
\label{FDim}
(F^{\cL, \ocL, \cL}_{\cL})_{\cI, \cI} ~:~ 1_{\cL, \ocL, \cI} \otimes 1_{\cI, \cL, \ocL}
~\mapsto~ {1\over \vev{\cL}_{\bR^2}} \times ( 1_{\ocL, \cL, \cI} \otimes 1_{\cL, \cI, \ocL}) \, .
\fe
The aforementioned cyclic permutation map is related to an $F$-symbol the $F$-move
\ie
&
\begin{gathered}
\begin{tikzpicture}[scale=1]
\draw [line] (-1,0) -- (-.8,0) node {$\times$} -- (0,0);
\draw [line,->-=1] (-1,0) -- (-1.5,.87) node [above left=-3pt] {$\cL_1$};
\draw [line,->-=1] (-1,0) -- (-1.5,-.87) node [below left=-3pt] {$\cL_2$};
\draw [line,->-=1] (0,0) -- (.5,-.87) node [below right=-3pt] {$\cL_3$};
\draw [line,dashed,->-=1] (0,0) -- (.1,.175) node {\rotatebox[origin=c]{60}{$\times$}} -- (.5,.87) node [above right=-3pt] {$\cI$};
\end{tikzpicture}
\end{gathered}
~=~
\begin{gathered}
\begin{tikzpicture}[scale=1]
\draw [line] (0,-1) -- (0,-.8) node {$\times$} -- (0,0);
\draw [line,->-=1] (0,-1) -- (.87,-1.5) node [below right=-3pt] {$\cL_3$};
\draw [line,->-=1] (0,-1) -- (-.87,-1.5) node [below left=-3pt] {$\cL_2$};
\draw [line,->-=1] (0,0) -- (-.87,.5) node [above left=-3pt] {$\cL_1$};
\draw [line,dashed,->-=1] (0,0) -- (.175,.1) node {\rotatebox[origin=c]{30}{$\times$}} -- (.87,.5) node [above right=-3pt] {$\cI$};
\end{tikzpicture}
\end{gathered}
~\circ~
(F^{\cL_1, \cL_2, \cL_3}_{\cI})_{\ocL_3, \ocL_1} \, .
\fe
For simplicity, the marking $\times$ will be ignored subsequently, which means that our formulae will be correct up to cyclic permutation maps.

\subsection{Holomorphic-Defect-Factorization Hypothesis}

\begin{defn}[Holomorphic-Defect-Factorization]
\label{Hypo}
A local operator $\cO$ on the Euclidean plane $\bar z = z^*$ with definite conformal weight $(h, \bar h)$ is said to be {\it holomorphically-defect-factorized} if it can be obtained in the following coincidence limit:
\ie
\label{Coincidence}
\begin{gathered}
\begin{tikzpicture}
\draw \dott{right}{$\cO(z, \bar z)|_{\bar z = z^*}$};
\end{tikzpicture}
\end{gathered}
\ = \
{\sqrt{{\vev{\cL}_{\bR^2}}}} \times 
\lim_{\bar z' \to \bar z = z^*}
\begin{gathered}
\begin{tikzpicture}
\draw [line,->-=.6] (0,0) \dotsol{left}{$\cD(z)$} -- node [above] {$\cL$} node [below] {} (1,0) \dotemp{right}{$\ocD(\bar z')$};
\end{tikzpicture}
\end{gathered} \, ,
\fe
where $\cL$ is a simple topological defect line, $\cD$ is a holomorphic defect operator of weight $(h, 0)$ in the defect Hilbert space $\cH_\cL$, and $\ocD$ is an anti-holomorphic defect operator of weight $(0, \bar h)$ in the dual defect Hilbert space $\cH_{\ocL}$.
\end{defn}

\begin{defn}[Factorizing topological defect line]
\label{Factorizing}
A simple topological defect line $\cL$ is said to be factorizing if there exists is a holomorphic defect operator in the defect Hilbert space $\cH_\cL$, and an anti-holomorphic defect operator in the dual defect Hilbert space $\cH_{\ocL}$.
\end{defn}

Throughout this paper, we use solid dots to represent holomorphic defect operators, empty dots to represent anti-holomorphic defect operators, and solid-inside-empty dots to represent local operators. The limit in \eqref{Coincidence} is well-defined because there is no singularity. As we will see in Section~\ref{Sec:OPE} the overall factor is such that if $\cD$ and $\ocD$ are each properly normalized, 
\ie\label{eqn:D2ptfns}
\la\hspace{-4pt}
\begin{gathered}
\begin{tikzpicture}[scale=1]
\draw [line,->-=.6] (0,0) \dotsol{left}{$\cD(0)$} -- node [above] {$\cL$} node [below] {} (1,0) \dotsol{right}{$\cD^\dag(1)$};
\end{tikzpicture}
\end{gathered}
\hspace{-4pt}\ra
~=~
\la\hspace{-4pt}
\begin{gathered}
\begin{tikzpicture}[scale=1]
\draw [line,->-=.6] (0,0) \dotemp{left}{$\ocD^\dag(0)$} -- node [above] {$\cL$} node [below] {} (1,0) \dotemp{right}{$\ocD(1)$};
\end{tikzpicture}
\end{gathered}
\hspace{-4pt}\ra
~=~ 
1
\, ,
\fe
then $\cO$ is too,
\ie
\la \cO(0) \, \cO^\dag(1) \ra ~=~ 1 \, .
\fe
We write
\ie
\cO = \cD \stackrel{\cL}{\text{---}} \ocD
\fe
for brevity.

When studying local operators in a conformal field theory, it is often natural to choose a real basis, in which the two-point function of every basis operator with itself is nonzero. However, holomorphically-defect-factorized local operators are generally complex. In fact, as we will see in Section~\ref{Sec:OPE}, if a local operator is holomorphically-defect-factorized through an oriented line ($\cL \neq \ocL$), then its two-point function with itself vanishes, so it cannot be real.\footnote{Throughout this paper, $\cL = \cL'$ means that they are in the same isomorphism class.
}
In the concrete example of the free compact boson theory, the exponential operators, which are complex, are holomorphically-defect-factorized through ${\rm U}(1)$ symmetry defects. By contrast, the cosine and sine operators, which are real combinations of exponential operators, are themselves not holomorphically-defect-factorized by Definition~\ref{Hypo}.\footnote{One could define a relaxed notion of factorization by allowing finite sums of holomorphically-defect-factorized operators. We do not do so here.
}

\begin{defn}[Holomorphic-defect-factorization prerequisite]
\label{Prereq}
A local operator $\cO$ of weight $(h, \bar h)$ is said to satisfy the holomorphic-defect-factorization prerequisite if there exists a simple topological defect line $\cL$ such that the defect Hilbert space $\cH_\cL$ contains a defect operator of weight $(h, 0)$, and the dual defect Hilbert space $\cH_{\ocL}$ contains one of weight $(0, \bar h)$.
\end{defn}

Holomorphic factorization further implies the following statement about the analyticity of general correlators, with the special case of sphere four-point functions rigorously proven in \cite{Maldacena:2015iua}:\footnote{This implication is due to the anonymous JHEP referee.
}
\begin{prop}[Analyticity]
\label{Analyticity}
An $n$-point correlation function involving holomorphically-defect-factorized local operators at $(z_i, \bar z_i)$ admits an analytic continuation for $z_i$ and $\bar z_i$ being independent complex variables on a branched cover of $\bC^{2n}$, where branch points can only occur when two defect operators collide.
\end{prop}

\subsection{Operator product expansion}
\label{Sec:OPE}

Holomorphic-defect-factorization provides a new perspective on the local operator production expansion (OPE). The OPE between two holomorphically-defect-factorized $\cO_1$ and $\cO_2$ follows from performing an $F$-move on $\cL_1$ and $\cL_2$ and expressing the $\cO_1 \times \cO_2$ OPE as a sum of products of $\cD_1 \times \cD_2$ and $\ocD_1 \times \ocD_2$ OPEs,
\ie
\label{OPE}
&
\cO_1(z_1, \bar z_1) \, \cO_2(z_2, \bar z_2) 
=
\sqrt{\textstyle \prod_{i=1}^2 {\vev{\cL_i}_{\bR^2}}} ~
\begin{gathered}
\begin{tikzpicture}[scale=1]
\draw [line,->-=.6] (0,0) \dotsol{left}{$\cD_1(z_1)$} -- node [above] {$\cL_1$} node [below] {} (1,0) \dotemp{right}{$\ocD_1(\bar z_1)$};
\draw [line,->-=.6] (0,-1) \dotsol{left}{$\cD_2(z_2)$} -- node [above] {$\cL_2$} node [below] {} (1,-1) \dotemp{right}{$\ocD_2(\bar z_2)$};
\end{tikzpicture}
\end{gathered} 
\\
&\hspace{-.15in} =
\sqrt{\textstyle \prod_{i=1}^2 {\vev{\cL_i}_{\bR^2}}} \,
\sum_{\cL} 
\begin{gathered}
\begin{tikzpicture}[scale=1]
\draw [line,->-=.3,-<-=.8] (-.5,0) ++(90:.5) \dotsol{left} {$\cD_1(z_1)$} arc (90:-90:.5) \dotsol{left}{$\cD_2(z_2)$};
\draw [line,->-=.6] (0,0) -- node [above] {$\cL$} node [below] {} (1,0);
\draw [line,-<-=.3,->-=.8] (1.5,0) ++(90:.5) \dotemp{right}{$\ocD_1(\bar z_1)$} arc (90:270:.5) \dotemp{right}{$\ocD_2(\bar z_2)$};
\end{tikzpicture}
\end{gathered}
\circ
(F^{\cL_1, \ocL_1, \ocL_2}_{\ocL_2})_{\cI, \ocL}~(1_{\cL_1, \ocL_1, \cI}, 1_{\cI, \ocL_2, \cL_2}) \, .
\fe

Suppose $\cO_2$ is the hermitian conjugate of $\cO_1$, {\it i.e.} $\cO_2=\cO_1^\dagger$, and take the vacuum expectation value. Holomorphy forces
\ie
\cL = \cI \, , \quad \cL_2 = \ocL_1 \, , \quad \cD_2 = \cD_1^\dag \, , \quad \ocD_2 = \ocD_1^\dag \, ,
\fe
which gives
\ie\label{eqn:OOtoDDDD}
\la \cO_1(0) \, \cO_1^\dagger(1) \ra ~=~ \la\hspace{-4pt}
\begin{gathered}
\begin{tikzpicture}[scale=1]
\draw [line,->-=.6] (0,0) \dotsol{left}{$\cD_1(0)$} -- node [above] {$\cL_1$} node [below] {} (1,0) \dotsol{right}{$\cD_1^\dag(1)$};
\end{tikzpicture}
\end{gathered}
\hspace{-4pt}\ra
\times
\la\hspace{-4pt}
\begin{gathered}
\begin{tikzpicture}[scale=1]
\draw [line,->-=.6] (0,0) \dotemp{left}{$\ocD_1^\dag(0)$} -- node [above] {$\cL_1$} node [below] {} (1,0) \dotemp{right}{$\ocD_1(1)$};
\end{tikzpicture}
\end{gathered}
\hspace{-4pt}\ra \, .
\fe
This shows that the OPE formula \eqref{OPE} has the correct normalization factor.

To proceed, define the three-point defect correlation functionals\footnote{The notation $'$ means moving an operators to the other patch of the sphere while taking into account the conformal factors.
}
\ie
\label{Defect3pt}
C_{\cD_1, \cD_2, \cD_3} \ = \ 
\la
\begin{gathered}
\begin{tikzpicture}[scale=.75]
\draw [line,-<-=.55] (0,0) -- (0,1) \dotsol{above}{$\cD_1(0)$};
\draw [line,-<-=.55] (0,0) -- (-.87,-.5) \dotsol{below left}{$\cD_2(1)$};
\draw [line,-<-=.55] (0,0) -- (.87,-.5) \dotsol{below right}{$\cD_3'(\infty)$};
\end{tikzpicture}
\end{gathered} 
\ra
~:~
V_{\cL_1, \cL_2, \cL_3} \to \bC
\, ,
\\
C_{\ocD_1, \ocD_3, \ocD_2} \ = \ 
\la
\begin{gathered}
\begin{tikzpicture}[scale=.75]
\draw [line,->-=.55] (0,0) -- (0,1) \dotemp{above}{$\ocD_1(0)$};
\draw [line,->-=.55] (0,0) -- (-.87,-.5) \dotemp{below left}{$\ocD_3'(\infty)$};
\draw [line,->-=.55] (0,0) -- (.87,-.5) \dotemp{below right}{$\ocD_2(1)$};
\end{tikzpicture}
\end{gathered} 
\ra
~:~
V_{\ocL_1, \ocL_3, \ocL_2} \to \bC
\, .
\fe
A central formula is a relation between them and the three-point coefficient $C_{\cO_1, \cO_2, \cO_3}$ of 
local operators,
\ie
\label{3pt}
C_{\cO_1, \cO_2, \cO_3} ~=~
\sqrt{\textstyle \prod_{i=1}^3{\vev{\cL_i}_{\bR^2}}} \times
( C_{\cD_1, \cD_2, \cD_3} \otimes C_{\ocD_1, \ocD_3, \ocD_2} ) \circ
\Theta_{\cL_1, \cL_2, \cL_3} \, ,
\fe
where the bi-vector $\Theta_{\cL_1, \cL_2, \cL_3}$ has multiple equivalent expressions
\ie
\label{Theta}
\Theta_{\cL_1, \cL_2, \cL_3} ~&=~ \frac{1}{{\vev{\cL_3}_{\bR^2}}} \times 
(F^{\cL_1, \ocL_1, \ocL_2}_{\ocL_2})_{\cI, \cL_3}\,(1_{\cL_1, \ocL_1, \cI}, 1_{\cI, \ocL_2, \cL_2})
\\
~&=~ \frac{1}{{\vev{\cL_1}_{\bR^2}}} \times 
(F^{\cL_2, \ocL_2, \ocL_3}_{\ocL_3})_{\cI, \cL_1}\,(1_{\cL_2, \ocL_2, \cI}, 1_{\cI, \ocL_3, \cL_3})
\\
~&=~ \frac{1}{{\vev{\cL_2}_{\bR^2}}} \times 
(F^{\cL_3, \ocL_3, \ocL_1}_{\ocL_1})_{\cI, \cL_2}\,(1_{\cL_3, \ocL_3, \cI}, 1_{\cI, \ocL_1, \cL_1}) \, .
\fe
The formula \eqref{3pt} can be derived by starting with
\ie
\la \cO_1(z_1, \bar z_1) \, \cO_2(z_2, \bar z_2) \, \cO_3(z_3, \bar z_3) \ra
~=~
\sqrt{\textstyle \prod_{i=1}^3 {\vev{\cL_i}_{\bR^2}}}
~\times~
\la\,
\begin{gathered}
\begin{tikzpicture}[scale=1]
\draw [line,->-=.6] (0,0) \dotsol{left}{$\cD_1(z_1)$} -- node [above] {$\cL_1$} node [below] {} (1,0) \dotemp{right}{$\ocD_1(\bar z_1)$};
\draw [line,->-=.6] (0,-1) \dotsol{left}{$\cD_2(z_2)$} -- node [above] {$\cL_2$} node [below] {} (1,-1) \dotemp{right}{$\ocD_2(\bar z_2)$};
\draw [line,->-=.6] (0,-2) \dotsol{left}{$\cD_3(z_3)$} -- node [above] {$\cL_3$} node [below] {} (1,-2) \dotemp{right}{$\ocD_3(\bar z_3)$};
\end{tikzpicture}
\end{gathered}
\,\ra \, ,
\fe
performing an OPE via \eqref{OPE}, and then performing an $F$-move on a trivial line connecting $\cL$ and $\cL_3$ to arrive at
\ie\label{eqn:3ptDDDD}
&
\sqrt{\textstyle \prod_{i=1}^3 {\vev{\cL_i}_{\bR^2}}}
~\times~
\sum_{\cL, \cL'}
\,
\la\,
\begin{gathered}
\begin{tikzpicture}
\begin{scope}
\draw [line,->-=.3,-<-=.8] (0,0) ++(90:.5) \dotsol{left}{$\cD_1(z_1)$} arc (90:-90:.5) \dotsol{left}{$\cD_2(z_2)$};
\draw [line,->-=.51,-<-=.8] (0,-.67) ++ (-90:.83) \dotsol{left}{$\cD_3(z_3)$} arc (-90:53:.83);
\end{scope}
\begin{scope}[xscale = -1, shift = {(-3,0)}]
\draw [line,-<-=.3,->-=.8] (0,0) ++(90:.5) \dotemp{right}{$\ocD_1(\bar z_1)$} arc (90:-90:.5) \dotemp{right}{$\ocD_2(\bar z_2)$};
\draw [line,-<-=.51,->-=.8] (0,-.67) ++ (-90:.83) \dotemp{right}{$\ocD_3(\bar z_3)$} arc (-90:53:.83);
\end{scope}
\draw [line,->-=.55] (.83,-.67) -- (1.5,-.67) node [above] {$\cL'$} -- (2.17,-.67);
\node at (1,-.25) {$\cL$};
\node at (2,-.25) {$\cL$};
\end{tikzpicture}
\end{gathered}
\,\ra 
\\
&\hspace{.5in} \circ
(F^{\cL, \ocL, \ocL_3}_{\ocL_3})_{\cI, \ocL'}~(1_{\cL, \ocL, \cI}, 1_{\cI, \ocL_3, \cL_3})
\otimes
(F^{\cL_1, \ocL_1, \ocL_2}_{\ocL_2})_{\cI, \ocL}~(1_{\cL_1, \ocL_1, \cI}, 1_{\cI, \ocL_2, \cL_2})
\, .
\fe
If we take the vacuum expectation value, then holomorphy forces $\cL' = \cI$ and $\cL = \ocL_3$, and gives \eqref{3pt} with $\Theta_{\cL_1, \cL_2, \cL_3}$ written in its first expression in \eqref{Theta}. Analogous derivations by first taking the $\cO_2 \times \cO_3$ or the $\cO_1 \times \cO_3$ OPE arrive at the other two expressions for $\Theta_{\cL_1, \cL_2, \cL_3}$. Note that the equivalence of the three expressions for $\Theta_{\cL_1, \cL_2, \cL_3}$ is a purely fusion categorical property.

In Appendix~\ref{Sec:Crossing}, we show that given \eqref{3pt}, the crossing symmetry of holomorphically-defect-factorized local operators follows from the crossing symmetry of holomorphic defect operators.

\subsection{Closedness, uniqueness, and commutativity}

Using the above formulation of local OPE in terms of TDL fusion and defect OPE, we can argue for the following properties of holomorphic-defect-factorization.

\begin{prop}[Closedness of factorized operators]
\label{Closed}
If two local operators $\cO_1$ and $\cO_2$ are both holomorphically-defect-factorized, $\cO_i = \cD_i \stackrel{\cL_i}{\text{---}} \ocD_i$, then all operators in the $\cO_1 \times \cO_2$ operator product expansion (OPE) are holomorphically-defect-factorized.
\end{prop}
This proposition obviously follows from \eqref{OPE}.

\begin{prop}[Closedness of factorizing topological defect lines]
\label{ClosedTDL}
The set of factorizing topological defect lines is closed under fusion.
\end{prop}
Given Proposition~\ref{Closed}, it suffices to argue that every TDL $\cL$ appearing in the fusion of two factorizing TDLs $\cL_1$ and $\cL_2$ is factorizing. This can be shown by considering
\ie
\begin{gathered}
\begin{tikzpicture}[scale=.75]
\draw [line,->-=.55] (0,0) -- (0,1) node [above] {$\cL$};
\draw [line,-<-=.55] (0,0) -- (-.87,-.5) \dotsol{below left}{$\cD_1(z_1)$};
\draw [line,-<-=.55] (0,0) -- (.87,-.5) \dotsol{below right}{$\cD_2(z_2)$};
\end{tikzpicture}
\end{gathered}
\qquad
\begin{gathered}
\begin{tikzpicture}[scale=.75]
\draw [line,->-=.55] (0,0) -- (0,1) node [above] {$\ocL$};
\draw [line,->-=.55] (0,0) -- (-.87,-.5) \dotsol{below left}{$\ocD_2(\bar z_2)$};
\draw [line,->-=.55] (0,0) -- (.87,-.5) \dotsol{below right}{$\ocD_1(\bar z_1)$};
\end{tikzpicture}
\end{gathered}
\fe
and taking the $\cD_1 \, \cD_2$ and $\ocD_1 \, \ocD_2$ OPEs.

\begin{prop}[Uniqueness of factorization]
\label{eqn:Uniqueness}
The holomorphic-defect-factorization \eqref{Coincidence} of a local operator $\cO$ is unique (up to isomorphism) if existent.
\end{prop} 
Suppose $\cO = \cD \stackrel{\cL}{\text{---}} \ocD = \cD' \stackrel{\cL'}{\text{---}} \ocD'$, by taking the operator product expansion \eqref{OPE} of $\cO = \cD \stackrel{\cL}{\text{---}} \ocD$ with its hermitian conjugate $\cO^\dag = (\cD')^\dag \stackrel{\ocL'}{\text{---}} (\ocD')^\dag$, one would conclude that the fusion $\cL \, \ocL'$ produces the trivial TDL, which implies that $\cL = \cL'$. Furthermore, the orthonormality of defect operators implies $\cD = \cD'$ and $\ocD = \ocD'$.

\begin{prop}[Uniqueness of holomorphic defect operator]
Every topological defect line hosts at most one holomorphic defect operator that is highest-weight with respect to the maximally extended chiral algebra. 
\end{prop}
Suppose a topological defect line $\cL$ hosts a set of holomorphic defect highest-weight operators (with respect to the maximally extended chiral algebra) $\cD_i$, chosen to be orthonormal, then the holomorphic defect OPE gives
\ie
\begin{gathered}
\begin{tikzpicture}[scale=1]
\draw [line,->-=.6] (0,0) \dotsol{left}{$\cD_i(z)$} -- node [above] {$\cL$} node [below] {} (1,0) \dotsol{right}{$\cD_j^\dag(0)$};
\end{tikzpicture}
\end{gathered}
\ = \
\sum_\Omega z^{h_\Omega - h_{\cD_i} - h_{\cD_j}} \, C_{\cD_i, \cD_j^\dag, \Omega}\, \Omega(0) \, .
\fe
where $\Omega$ are holomorphic local operators. All $\Omega$ must be chiral algebra descendants of the vacuum, because otherwise the chiral algebra would have been further extended. Then by associativity, $\cD_i$ and $\cD_j$ appear in each other's OPE with $\Omega$, {\it i.e.} they are in the same chiral algebra module. Thus, every topological defect line hosts at most one holomorphic defect highest-weight operator, and only the vacuum module appears in the holomorphic defect OPE. 

However, not every topological defect line hosts a holomorphic defect operator in its defect Hilbert space. A simple example is given by the charge conjugation symmetry defect line in the three-state Potts model.

\begin{prop}[Commutativity]
\label{Commute}
The fusion rule of factorizing topological defect lines is commutative.
\end{prop}
Let $\cO_1$ and $\cO_2$ be local operators holomorphically-defect-factorized through $\cL_1$ and $\cL_2$, respectively. The operator product expansions of $\cO_1(z,\bar z)\cO_2(0,0)$ and $\cO_2(z,\bar z)\cO_1(0,0)$ 
contain the same set of local operators that factorize through $\cL_1 \, \cL_2$ and $\cL_2 \, \cL_1$. By Propositions~\ref{ClosedTDL} and \ref{eqn:Uniqueness}, we must therefore have $\cL_1 \, \cL_2 = \cL_2 \, \cL_1$.

\subsection{Non-compact topological defect lines}
\label{Sec:Non-compact}

In the above, we have assumed that the category of TDLs is {\it semi-simple}. To incorporate non-semi-simple TDLs, the usual fusion categorical framework needs to be enlarged. To motivate, consider the Tambara-Yamagami categories \cite{tambara1998tensor} with $G = \bZ_n$, and embed $\bZ_n$ in ${\rm U}(1)$. Heuristically, the infinite $n$ limit should give rise to a Tambara-Yamagami category with $G = {\rm U}(1)$. Indeed as we will see below, a properly-normalized version of the non-invertible TDL produces upon self-fusion an {\it integral} over ${\rm U}(1)$ symmetry lines. While such a mathematical framework has not been fully developed, we nevertheless attempt to characterize the key properties of such TDLs. For the lack of a better name, we refer to this generalized structure as a {\it TDL category}.

\begin{defn}[Weak Cardy condition]
\label{WCC}
A topological defect line is said to satisfy the {\it weak Cardy condition} if its defect Hilbert space has a positive norm.
\end{defn}
In particular, the weak Cardy condition allows for continuous (delta-function normalizable) spectra inside the defect Hilbert space along with discrete (normalizable) states. By contrast, the usual Cardy condition requires the spectrum of the defect Hilbert space to be discrete.
\begin{defn}[Non-compactness]
\label{NCP}
A topological defect line with a continuum inside the defect Hilbert space is said to be non-compact.
\end{defn}

A TDL category contains TDLs satisfying the weak Cardy condition. Importantly, there exists a {\it basis} of TDLs, parameterized by variables taking both discrete and continuous values, such that every TDL can be expressed as a direct integral over the basis TDLs with positive measure (discrete TDLs correspond to delta-function measures). This basis must contain all the simple TDLs, and possibly 
some non-compact TDLs.

There are two ways to normalize a simple TDL. The standard way, which we call {\it Cardy normalization}, is to demand that the leading term (corresponding to the ground state) in the $q, \, \bar q$-expansion of its defect partition function $Z_{\cL\ocL}(\tau, \bar\tau)$ has unit coefficient. The alternative way, which we call {\it loop normalization} and denote the corresponding TDL by $\widetilde\cL$, is to normalize the cylinder loop expectation value $\langle\widetilde\cL\rangle_{{\rm S}^1\times\bR}$ to one. The two are related by $\widetilde\cL = \vev{\cL}_{{\rm S}^1\times\bR}^{-1} \, \cL$. For non-compact basis TDLs, Cardy normalization is not always well-defined, as the ground state in $H_{\cL\ocL}$ may sit at the bottom of a continuous spectrum; therefore, the only natural normalization is the loop normalization.

In terms of basis TDLs, the $F$-move could be defined in the same way as \eqref{F}, but with the sum replaced by an integral over the basis TDLs, and with the $F$-symbol becoming an integration measure. Accordingly, every appearance of ``~$\sum_{\cL}$~", for instance in \eqref{OPE} and \eqref{eqn:3ptDDDD}, should be interpreted as integrals over the basis TDLs.

We will encounter an example of such a TDL category in the free boson orbifold theory in Section~\ref{Sec:S1Z2}. An important lesson we learn from this example is that the more general TDL category (which contains non-compact TDLs) can arise as a limit of a sequence of semi-simple fusion categories. More precisely, a non-compact basis TDL $\widetilde\cL$ can arise as a limit of a sequence of simple TDLs $\cL_n$, such that when $\cL_n$ is Cardy normalized, the loop expectation value $\vev{\cL_n}_{{\rm S}^1\times\bR}$ diverges in the $n \to \infty$ limit, while at the same time the spacing in the spectrum of the defect Hilbert space $\cH_{\cL_n}$ diminishes. Hence, the sequence of defect Hilbert spaces $\cH_{\widetilde\cL_n}$ of the loop-normalized simple TDLs $\widetilde\cL_n = \vev{\cL_n}_{{\rm S}^1\times\bR}^{-1} \, \cL_n$ converges to a Hilbert space with continua in its spectrum. This limiting defect Hilbert space could thereby be identified as that of a non-compact TDL $\widetilde\cL = \lim_{n\to\infty} \widetilde\cL_n$. 

The general structure of the fusion of two non-compact basis TDLs $\widetilde\cL$ and $\widetilde\cL'$ can also be nicely understood from the limit of a sequence of fusions of simple TDLs $\widetilde\cL_n$ and $\widetilde\cL_n'$. The decomposition of the fusion product $\cL_n \, \cL_n'$ must either contain a simple TDL whose loop expectation value diverges in the $n \to \infty$ limit, or be a sum whose number of summands diverges in the $n \to \infty$ limit. In the latter case, we find that the decomposition of the fusion $\widetilde\cL \, \widetilde\cL'$ should contain a direct integral of simple TDLs.

We stress that an infinite direct sum of Cardy-normalized TDLs is unphysical because the defect partition function diverges, as we presently explain. By the modular S-transformation, the defect partition function is related to the twisted partition function, which is proportional to the cylinder loop expectation value $\vev~_{{\rm S}^1\times\bR}$. In a unitary compact theory, the $\vev~_{{\rm S}^1\times\bR}$ of every topological defect line is lower-bounded by one. An infinite sum of numbers lower-bounded by one produces infinity. Therefore, when such an infinity is formally encountered in taking the limit of theories or fusion categories, one should loop-normalize the simple TDLs, and interpret the limiting TDL as a non-compact TDL that has a continuous yet finite defect partition function.

To illustrate the ideas presented above, consider the Tambara-Yamagami categories \cite{tambara1998tensor} with $G = \bZ_n$. At finite $n$, the fusion rule is
\ie
\label{TY}
\cN^2 = \sum_{m = 0}^{n-1} \eta^m \, ,
\fe
where $\eta$ is the symmetry line corresponding to a generator of $\bZ_n$, and $\cN$ is the non-invertible TDL with $\vev{\cN}_{{\rm S}^1\times\bR} = \sqrt{n}$. The naive $n \to \infty$ limit produces an infinite sum on the right, and relatedly $\vev{\cN}_{{\rm S}^1\times\bR}$ diverges. Suppose the $\bZ_n$ symmetry is embedded in a ${\rm U}(1)$ whose elements are parameterized by $\theta \in [0, 2\pi)$. If we denote the U(1) symmetry lines by $\cL_\theta$, then the embedding map is
\ie
\eta^m \mapsto \cL_{2\pi \frac{m}{n}} \, .
\fe
By defining the loop-normalized
\ie
\widetilde\cN \equiv \frac{\cN}{\vev{\cN}_{{\rm S}^1\times\bR}} = \frac{\cN}{\sqrt{n}} \, ,
\fe
\eqref{TY} becomes
\ie
\widetilde\cN^2 = \frac{1}{n} \sum_{m = 0}^{n-1} \eta^m \, .
\fe
In the $n \to \infty$ limit, the sum becomes an integral
\ie
\widetilde\cN^2 = \int_0^{2\pi} \frac{d\theta}{2\pi} \, \cL_{\theta} \, .
\fe

In the holomorphic-defect-factorization of a local operator $\cO$, the factorizing TDL could be a non-compact basis TDL $\widetilde\cL$, and the defect operator $\cD$ could sit in a continuum in the defect Hilbert space $\cH_{\widetilde\cL}$. Note that while  the local operator $\cO$ is normalizable, the defect operator $\cD$ is delta-function normalizable. To make sense of $\cO = \cD \stackrel{\widetilde\cL}{\text{---}} \ocD$ as an operator equivalence inside correlation functions, the expectation value $\vev{~}$ should be defined with the additional prescription of appropriately removing ``$\D(0)$" factors.

Let us try to make precise the preceding paragraph by considering a sequence of local operators $\cO_n$ that factorize through a sequence of simple TDLs $\cL_n$. We write the holomorphic-defect-factorization in a slightly different form: 
\ie
\begin{gathered}
\begin{tikzpicture}
\draw \dott{right}{$\cO_n(z, \bar z)|_{\bar z = z^*}$};
\end{tikzpicture}
\end{gathered}
\ = \ 
\lim_{\bar z' \to \bar z = z^*}
\begin{gathered}
\begin{tikzpicture}
\draw [line,->-=.6] (0,0) \dotsol{left}{$\widetilde\cD_n(z)$} -- node [above] {$\widetilde\cL_n$} node [below] {} (1,0) \dotemp{right}{$\widetilde\ocD_n(\bar z')$};
\draw[](0,-.75);
\end{tikzpicture}
\end{gathered} \, ,
\fe
where $\widetilde\cL_n$ is loop-normalized, and the defect operators $\widetilde\cD_n, \, \widetilde\ocD_n$ are normalized as\footnote{Note that $\vev{\cL_n}_{\bR^2}\vev{\cL_n}_{\bR^2}^*=\vev{\cL_n}_{{\rm S}^1\times\bR}^2$.}
\ie
\widetilde\cD_n \equiv \vev{\cL_n}_{\bR^2}^{1\over 4}\vev{\cL_n}_{{\rm S}^1\times\bR}^{1\over 2} \, \cD_n \, , \quad \widetilde\ocD_n \equiv \vev{\cL_n}_{\bR^2}^{1\over 4} \vev{\cL_n}_{{\rm S}^1\times\bR}^{1\over 2}\, \ocD_n \, ,
\fe
in order to absorb all factors of $\vev{\cL_n}_{\bR^2}$ and $\vev{\cL_n}_{{\rm S}^1\times\bR}$. Under this normalization, the two-point functions \eqref{eqn:D2ptfns} become
\ie
\label{eqn:tD2ptfcns}
\la\hspace{-4pt}
\begin{gathered}
\begin{tikzpicture}[scale=1]
\draw [line,->-=.6] (0,0) \dotsol{left}{$\widetilde\cD_n(0)$} -- node [above] {$\widetilde\cL_n$} node [below] {} (1,0) \dotsol{right}{$\widetilde\cD_n^\dag(1)$};
\draw[](0,-.625);
\end{tikzpicture}
\end{gathered}
\hspace{-4pt}\ra
~=~
\la\hspace{-4pt}
\begin{gathered}
\begin{tikzpicture}[scale=1]
\draw [line,->-=.6] (0,0) \dotemp{left}{$\widetilde\ocD_n^\dag(0)$} -- node [above] {$\widetilde\cL_n$} node [below] {} (1,0) \dotemp{right}{$\widetilde\ocD_n(1)$};
\draw[](0,-.75);
\end{tikzpicture}
\end{gathered}
\hspace{-4pt}\ra
~=~ 
\vev{\cL_n}_{{\rm S}^1\times\bR}^{1\over 2}
\, ,
\fe
and the relation \eqref{eqn:OOtoDDDD} between the local operator two-point function and defect operator two-point functions becomes
\ie
\label{t2pt}
\la \cO_n(0) \, \cO_n^\dagger(1) \ra ~=~ \vev{\cL_n}_{{\rm S}^1\times\bR}^{-1} \times \la\hspace{-4pt}
\begin{gathered}
\begin{tikzpicture}[scale=1]
\draw [line,->-=.6] (0,0) \dotsol{left}{$\widetilde\cD_n(0)$} -- node [above] {$\widetilde\cL_n$} node [below] {} (1,0) \dotsol{right}{$\widetilde\cD_n^\dag(1)$};
\draw[](0,-.625);
\end{tikzpicture}
\end{gathered}
\hspace{-4pt}\ra
\times
\la\hspace{-4pt}
\begin{gathered}
\begin{tikzpicture}[scale=1]
\draw [line,->-=.6] (0,0) \dotemp{left}{$\widetilde\ocD_n^\dag(0)$} -- node [above] {$\widetilde\cL_n$} node [below] {} (1,0) \dotemp{right}{$\widetilde\ocD_n(1)$};
\draw[](0,-.75);
\end{tikzpicture}
\end{gathered}
\hspace{-4pt}\ra \, .
\fe
The $n \to \infty$ limit of correlators of local operators, such as the two-point function \eqref{t2pt}, is finite.

In the spectrum of the limiting non-compact TDL $\widetilde\cL = \lim_{n \to \infty} \widetilde\cL_n$, the limiting holomorphic defect operator $\widetilde \cD = \lim_{n \to \infty} \widetilde \cD_n$ is buried inside a continuum, say parameterized by $\mu$, and becomes delta-function normalizable,
\ie
\la\hspace{-4pt}
\begin{gathered}
\begin{tikzpicture}[scale=1]
\draw [line,->-=.6] (0,0) \dotsol{left}{$\widetilde\cD(\mu; 0)$} -- node [above] {$\widetilde\cL$} node [below] {} (1,0) \dotsol{right}{$\widetilde\cD^\dag(\nu; 1)$};
\draw[](0,-.625);
\end{tikzpicture}
\end{gathered}
\hspace{-4pt}\ra
~=~ \D(\mu - \nu) \, .
\fe
Likewise for the anti-holomorphic defect operator $\widetilde \ocD$. From this prespective, the diverging $\vev{\cL_n}_{{\rm S}^1\times\bR}^{1\over 2}$ on the right hand side of \eqref{eqn:tD2ptfcns} should be interpreted as a ``$\D(0)$'' factor. Moreover, In a correlator of local operators, such as the two-point function \eqref{t2pt}, a specific power of ``$\D(0)$" should be removed. The $\vev{\cL_n}_{{\rm S}^1\times\bR}^{-1}$ factor in \eqref{t2pt} transits to such a removal operation in the $n \to \infty$ limit, and \eqref{t2pt} schematically becomes
\ie
\la \cO(0) \, \cO^\dagger(1) \ra ~=~
``\frac{1}{\D(0)^2}" ~
\la\hspace{-4pt}
\begin{gathered}
\begin{tikzpicture}[scale=1]
\draw [line,->-=.6] (0,0) \dotsol{left}{$\widetilde\cD(0)$} -- node [above] {$\widetilde\cL$} node [below] {} (1,0) \dotsol{right}{$\widetilde\cD^\dag(1)$};
\draw[](0,-.625);
\end{tikzpicture}
\end{gathered}
\hspace{-4pt}\ra
\times
\la\hspace{-4pt}
\begin{gathered}
\begin{tikzpicture}[scale=1]
\draw [line,->-=.6] (0,0) \dotemp{left}{$\widetilde\ocD^\dag(0)$} -- node [above] {$\widetilde\cL$} node [below] {} (1,0) \dotemp{right}{$\widetilde\ocD(1)$};
\draw[](0,-.75);
\end{tikzpicture}
\end{gathered}
\hspace{-4pt}\ra \, .
\fe

\section{Holomorphically-defect-factorized local operators in the Regge limit}
\label{Sec:FRL}

\subsection{Action on local operators in the conformal Regge limit}

Suppose a local operator is holomorphically-defect-factorized, $\cO = \cD \stackrel{\cL}{\text{---}} \ocD$, then to study the action \eqref{Act} of $\cL$ on a particular local operator $\phi$, we can take the four-point function $\la \cO^\dag(0) \cO(z, \bar z) \phi(1) {\phi'}^\dag(\infty) \ra$ and send $z$ around 1 while keeping $\bar z$ fixed. This wraps $\cL$ around $\phi(1)$. By then sending $z, \, \bar z \to 0$ with ${z / \bar z}$ fixed and removing the leading singularity, we obtain $\la \widehat{\widetilde\cL}(\phi) \, \phi' \ra$, where $\widetilde\cL$ is loop-normalized such that $\widehat{\widetilde\cL}(1) = 1$. This limit is none other than the {\bf conformal Regge limit} \cite{Cornalba:2007fs,Costa:2012cb} of the four-point function. The following is a visual for when $\cL$ is simple:
\ie
\label{ReggeVisual}
&
\sqrt{\vev{\cL}_{\bR^2}} ~\times~
\begin{gathered}
\begin{tikzpicture}[scale=1]
\draw [line,opacity=0.2] (.75,0) circle (3 and 1.5);
\draw [line,-<-=.55] (0,.5) \dotemp{above}{$\ocD(\bar z)$} -- (0,-.5) \dotsol{below}{$\cD(z)$};
\dt{(-1,0)}{below}{$\cO^\dag(0)$};
\dt{(1,0)}{below}{$\phi(1)$};
\dt{(2.5,0)}{below}{${\phi'}^\dag(\infty)$};
\end{tikzpicture}
\end{gathered}
\\
\ &\to \
\sqrt{\vev{\cL}_{\bR^2}} ~\times~
\begin{gathered}
\begin{tikzpicture}[scale=1]
\draw [line,opacity=0.2] (1.75,0) circle (3.5 and 1.5);
\draw (.5,.5) \dotsol{right}{$\cD(z)$};
\draw [line, ->-=.6] (.5,.5) .. controls (.5,2) and (-1,0) .. (.5,0)
(.5,0) .. controls (1.75,0) and (3,2) .. (3,0)
(3,0) .. controls (3,-1) and (.5,-2) .. (.5,-.5) ;
\draw (.5,-.5) \dotemp{right}{$\stackrel{~}{\ocD(\bar z)}$};
\dt{(-.5,0)}{below}{$\cO^\dag(0)$};
\dt{(2,0)}{below}{$\phi(1)$};
\dt{(4,0)}{below}{${\phi'}^\dag(\infty)$};
\end{tikzpicture}
\end{gathered}
\\
& 
= \ \frac{1}{\sqrt{\vev{\cL}_{\bR^2}}} ~\times~ \sum_{\cL'} ~
\begin{gathered}
\begin{tikzpicture}[scale=1]
\draw [line,opacity=0.2] (1.75,0) circle (3.5 and 1.5);
\draw [line,-<-=0] (2,0) circle (1);
\draw (0,.5) \dotsol{above}{$\cD(z)$};
\draw [line,->-=.6] (0,0) -- (.55,0) node [above] {$\cL'$} -- (1,0);
\draw [line, ->-=.25, ->-=.8] (0,.5) .. controls (0,1.15) and (-.75,0) .. (0,0)
(0,0) .. controls (.75,0) and (0,-1.15) .. (0,-.5) ;
\draw (0,-.5) \dotemp{below}{$\ocD(\bar z)$};
\dt{(-1,0)}{below}{$\cO^\dag(0)$};
\dt{(2,0)}{below}{$\phi(1)$};
\dt{(4,0)}{below}{${\phi'}^\dag(\infty)$};
\end{tikzpicture}
\end{gathered}
~(1_{\cL, \ocL, \cI}, 1_{\cI, \cL, \ocL})
\\
& 
\sim \
\frac{1}{{\vev{\cL}_{\bR^2}}} \
\frac{1}{z^{2h} \bar z^{2\bar h}}
\ \times \
\begin{gathered}
\begin{tikzpicture}[scale=1]
\draw [line,opacity=0.2] (2.75,0) circle (3 and 1.5);
\draw [line,-<-=0] (2,0) circle (1);
\dt{(2,0)}{below}{$\phi(1)$};
\dt{(4,0)}{below}{${\phi'}^\dag(\infty)$};
\end{tikzpicture}
\end{gathered}
\, .
\fe
In the second-to-last line, we used the relation \eqref{FDim} between the planar loop expectation value and the $F$-symbol $(F^{\cL,\ocL,\cL}_\cL)_{\cI, \cL'}~(1_{\cL, \ocL, \cI}, 1_{\cI, \cL, \ocL})$. In the last line, we kept the leading term in the $z, \, \bar z \to 0$ limit corresponding to the domination of $\cL' = \cI$, and used \eqref{FDim} to rewrite the $F$-symbol as an inverse planar loop expectation value. 

Normally, continuing $z$ and $\bar z$ independently takes a correlator off the Euclidean plane. However, if one of the operators is holomorphically-defect-factorized, then the correlator has a new interpretation as a Euclidean correlator involving not only local operators, but also defect operators joined by topological defect lines.

\subsection{Holomorphic-defect-factorization criterion in the torus Regge limit}
\label{Sec:TorusRegge}

To study the action of $\cL$ on all local operators at once, one can consider the torus two-point function $\la \cO(z, \bar z) \cO(0) \ra_{T^2(\tau, \bar\tau)}$. By sending $z \to z+1$ (spatial translation) with $\bar z$ fixed and then $z, \, \bar z \to 0$ with $z/\bar z$ fixed while removing the leading singularity, one obtains the torus partition function $Z^{\widetilde\cL}(\tau, \bar\tau)$ with the loop-normalized $\widetilde\cL$ wrapped along the spatial direction. The following is a visual for when $\cL$ is simple:
\ie
\label{SpatialTorus}
& \sqrt{\vev{\cL}_{\bR^2}} ~\times~
\begin{gathered}
\begin{tikzpicture}[scale=1]
\draw [line,-<-=.55] (0,.5) \dotemp{above}{$\ocD(\bar z)$} -- (0,-.5) \dotsol{below}{$\cD(z)$};
\node at (-1.4,-1.2) {$\cO(0)$};
\node at (-.25,0) {${\cal L}$};
\dt{(-2,-1.5)}{}{};
\draw [line] (-2,-1.5) -- (1,-1.5) -- (1,1.5) -- (-2,1.5) -- (-2,-1.5);
\end{tikzpicture}
\end{gathered}
\quad \to ~
\sqrt{\vev{\cL}_{\bR^2}} ~\times~
\begin{gathered}
\begin{tikzpicture}[scale=1]
\draw [line] (.5,.5) \dotemp{above}{$\ocD(\bar z)$};
\draw [line] (-1,.5) \dotsol{below}{$~\cD(z)$};
\draw [line,-<-=.13,->-=.77] (.5,.47) .. controls (.5,-.5) .. (1,-.5)
(-1,.5) .. controls (-1,1) and (-1.5,1) .. (-1.5,.5)
(-1.5,.5) .. controls (-1.5,-.5) .. (-2,-.5);
\node at (-1.4,-1.2) {$\cO(0)$};
\dt{(-2,-1.5)}{}{};
\draw [line] (-2,-1.5) -- (1,-1.5) -- (1,1.5) -- (-2,1.5) -- (-2,-1.5);
\end{tikzpicture}
\end{gathered}
\\
& \qquad \to 
\frac{1}{\sqrt{\vev{\cL}_{\bR^2}}} \times
\begin{gathered}
\begin{tikzpicture}[scale=1]
\draw [line,->-=.67] (-1,.5) .. controls (-1,1) and (-1.5,1) .. (-1.5,.5)
(-1.5,.5) .. controls (-1.5,0) and (.5,-.3) .. (.5,.5);
\draw [line,-<-=.51] (-2,-.5) -- (1,-.5);
\draw [line] (-2,-1.5) -- (1,-1.5) -- (1,1.5) -- (-2,1.5) -- (-2,-1.5);
\draw [line,->-=.6] (0,.04) -- (0,-.125) node [right] {$\cL'$} -- (0,-.5);
\dt{(-2,-1.5)}{}{};
\node at (-.5,-.75) {${\cal L}$};
\draw [line] (.5,.5) \dotemp{above}{$\ocD(\bar z)$};
\draw [line] (-1,.5) \dotsol{right}{$\cD(z)$};
\node at (-1.4,-1.2) {$\cO(0)$};
\end{tikzpicture}
\end{gathered}
\quad \sim ~
\frac{1}{\vev{\cL}_{\bR^2}}
\
\frac{e^{2i \pi h}}{z^{2h} \bar z^{2\bar h}}
\ \times \
\begin{gathered}
\begin{tikzpicture}[scale=1]
\draw [line,-<-=.51] (-2,0) -- (1,0);
\draw [line] (-2,-1.5) -- (1,-1.5) -- (1,1.5) -- (-2,1.5) -- (-2,-1.5);
\node at (-.5,-.25) {${\cal L}$};
\end{tikzpicture}
\end{gathered}
\, .
\fe
In the last step, we kept the dominant $\cL' = \cI$ contribution, and performed a $2\pi$ angle rotation of $\cL$ at $\cD$ to return to the original configuration, thereby creating the extra $e^{2i \pi h}$ phase. With the $e^{2i \pi h}$ phase stripped off, we call this the {\bf spatial torus Regge limit}.

The modular S transform of $Z^{\widetilde\cL}(\tau, \bar\tau)$ gives the defect partition function $Z_{\widetilde\cL}(\tau, \bar\tau)$, {\it i.e.} the torus partition function with $\widetilde\cL$
wrapped along the temporal direction. The latter could be obtained directly from $\la \cO(z, \bar z) \cO(0) \ra_{{\rm T}^2(\tau, \bar\tau)}$ by sending $z \to z-\tau$ (temporal translation) with 
$\bar z$ fixed and then $z, \, \bar z \to 0$ with $z/\bar z$ fixed while removing the leading singularity. The following is a visual for when $\cL$ is simple:
\ie
\sqrt{\vev{\cL}_{\bR^2}} ~\times~
\begin{gathered}
\begin{tikzpicture}[scale=1]
\draw [line,-<-=.55] (0,.5) \dotemp{above}{$\ocD(\bar z)$} -- (0,-.5) \dotsol{below}{$\cD(z)$};
\node at (-1.4,-1.2) {$\cO(0)$};
\dt{(-2,-1.5)}{}{};
\draw [line] (-2,-1.5) -- (1,-1.5) -- (1,1.5) -- (-2,1.5) -- (-2,-1.5);
\end{tikzpicture}
\end{gathered}
\quad \to \quad \dotsb \quad \sim \quad
\frac{1}{{\vev{\cL}_{\bR^2}}}
\
\frac{e^{2i \pi h}}{z^{2h} \bar z^{2\bar h}}
\ \times \
\begin{gathered}
\begin{tikzpicture}[scale=1]
\draw [line,->-=.55] (-.5,-1.5) -- (-.5,1.5);
\draw [line] (-2,-1.5) -- (1,-1.5) -- (1,1.5) -- (-2,1.5) -- (-2,-1.5);
\end{tikzpicture}
\end{gathered}
\, .
\fe
With the $e^{2i \pi h}$ phase stripped off, we call this the {\bf temporal torus Regge limit}.

As we have seen, the conformal and torus Regge limits naturally produce correlators with loop-normalized TDLs. When $\cL$ is simple, we expect that multiplication with $\vev{\cL}_{\bR^2}$ gives the more standard Cardy-normalized torus partition function, which has a $q, \, \bar q$ expansion with positive integer coefficients (Cardy condition). This requirement presents a nontrivial criterion for the factorization of the local operator $\cO$ through a simple TDL.

\begin{defn}[Strong holomorphic-defect-factorization criterion]
\label{Strong}
Given a local operator $\cO$ in a unitary conformal field theory, if the torus two-point function
$\la \cO(z, \bar z) \, \cO(0) \ra_{{\rm T}^2(\tau, \bar\tau)}$
in the temporal torus Regge limit has a $q, \, \bar q$ expansion with positive integer coefficients up to some overall number, then $\cO$ is said to satisfy the strong holomorphic-defect-factorization criterion.
\end{defn}

To incorporate holomorphic-defect-factorization through non-compact TDLs, the discreteness and integer-coefficient requirements need to be relaxed, hence the Cardy condition should be replaced by the weak Cardy condition of Definition~\ref{WCC}.

\begin{defn}[Weak holomorphic-defect-factorization criterion]
\label{Weak}
Given a local operator $\cO$ in a unitary conformal field theory, if the torus two-point function
$\la \cO(z, \bar z) \, \cO(0) \ra_{{\rm T}^2(\tau, \bar\tau)}$
in the temporal torus Regge limit is the Laplace transform of a non-negative density of states, then $\cO$ is said to satisfy the weak holomorphic-defect-factorization criterion.
\end{defn}

While the weak holomorphic-defect-factorization criterion is certainly natural in non-compact theories like Liouville or Toda \cite{Drukker:2010jp}, it also applies to compact theories. In particular, topological defect lines satisfying the weak criterion but not the strong criterion will arise in the free boson orbifold theory at irrational points in Section~\ref{Sec:S1Z2}.

\section{Lorentzian dynamics and holography}
\label{Sec:Lorentzian}

As discussed in Section \ref{Sec:FRL}, the conformal Regge limit \cite{Cornalba:2007fs,Costa:2012cb} of the four-point function of a pair of holomorphically-defect-factorized local operators $\cO = \cD \stackrel{\cL}{\text{---}} \ocD$ computes the matrix element of the map $\widehat\cL$ on the Hilbert space of local operators. Traditionally, the conformal Regge limit is interpreted as a limit of Lorentzian correlators, since analytically continuing $z$ around 1 while fixing $\bar z$ moves the local operator off the Euclidean plane onto the Lorentzian sheet. In holographic theories, the conformal Regge limit corresponds to the Regge limit of the bulk S-matrix --- the high energy limit with a fixed impact parameter. There is also a close connection to chaos \cite{Shenker:2013pqa,Roberts:2014ifa,Shenker:2014cwa,Maldacena:2015waa}, as the conformal Regge limit is equivalent to the late time limit of the out-of-time-ordered-correlator (OTOC) at finite temperature \cite{Murugan:2017eto}.

To be concrete, let us consider the Euclidean four-point function of a pair of hermitian conjugate operators $\cO, \, \cO^\dag$ with another pair of hermitian conjugate operators $\phi, \, \phi^\dag$ on the complex plane
\ie\label{eqn:4ptfonC}
G(z, \bar z)={\la \cO^\dag(z_1,\bar z_1)\cO(z_2,\bar z_2)\phi(z_3,\bar z_3)\phi^\dag(z_4,\bar z_4)\ra\over \la \cO^\dag(z_1,\bar z_1)\cO(z_2,\bar z_2)\ra\la \phi(z_3,\bar z_3)\phi^\dag(z_4,\bar z_4)\ra } \, ,
\fe
where the cross ratios are
\ie
z={z_{12}z_{34}\over z_{13}z_{24}} \, , \quad \bar z={\bar z_{12}\bar z_{34}\over \bar z_{13}\bar z_{24}} \, .
\fe
By conformal symmetry, the positions of the operators can be fixed to
\ie
z_1=-\rho \, ,\quad z_2=\rho \, ,\quad z_3=1 \, ,\quad z_4=-1 \, .
\fe
Then the cross ratios are related to the global variables $\rho, \, \bar\rho$ by \cite{Pappadopulo:2012jk}
\ie
\rho={z\over (1+\sqrt{1-z})^2} \, ,\quad\bar\rho={\bar z\over (1+\sqrt{1-\bar z})^2} \, .
\fe
Under the analytic continuation sending $z$ around 1 while fixing $\bar z$, the cross ratios become independent variables; on the Lorentzian sheet, they are both real. In the conformal Regge limit, $(1-z)\to e^{2\pi i}(1-z)$ with $\bar z$ fixed and then $z,\,\bar z\to 0$ with $z/\bar z$ fixed, $\rho$ and $\bar \rho$ scale as
\ie
\rho={4\over z}+\cO(z^0) \, ,\quad \bar\rho={\bar z\over 4}+\cO(\bar z^2) \, .
\fe
The analytic continuation and the conformal Regge limit could be equivalently described in the $\rho$-coordinate. One first write $\rho$ and $\bar \rho$ as
\ie
\rho = r e^{i\theta} \, ,\quad \bar\rho = r e^{-i\theta} \, .
\fe
In Euclidean signature, the distance from the origin $r$ and the angle $\theta$ are real. One then analytic continues the angle $\theta$ as $\theta = -i \zeta-\epsilon$, and arrives at the Rindler coordinates
\ie
\rho = r e^{\zeta-i\epsilon} \, ,\quad \bar\rho = r e^{-\zeta+i\epsilon} \, ,
\fe
where the $\zeta$ is the boost parameter (rapidity) of the $\cO, \, \cO^\dag$ operators relative to the $\phi, \, \phi^\dag$ operators.

The conformal Regge limit \cite{Cornalba:2007fs,Costa:2012cb} corresponds to the large boost limit where the pair of $\cO, \, \cO^\dag$ operators become time-like separated from the pair $\phi^\dag, \, \phi$, respectively, and $\cO$ and $\cO^\dag$ approach the light-cone of each other. Under the holographic duality, this limit can be interpreted as the high energy scattering of particles created by the operators $\phi$ and $\cO$ with a fixed finite impact parameter.

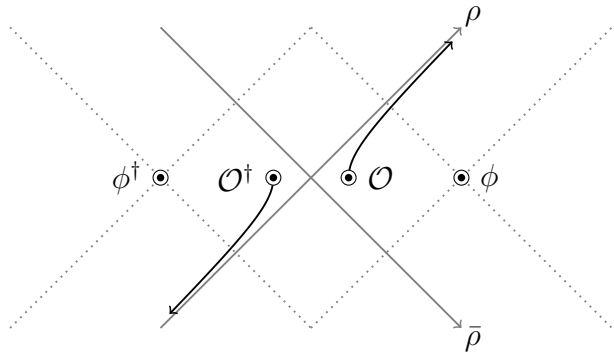
\begin{figure}
\centering
\begin{tikzpicture}[scale=1]
\draw [line,->,gray] (-2,2) -- (2,-2);
\draw [line,->,gray] (-2,-2) -- (2,2);
\begin{scope}[shift={(1,0)}]
\draw [line,dotted,gray] (-1,-2) -- (3,2);
\draw [line,dotted,gray] (-1,2) -- (3,-2);
\dt{(1,0)}{right}{$\phi$};
\end{scope}
\begin{scope}[shift={(-1,0)}]
\draw [line,dotted,gray] (-3,-2) -- (1,2);
\draw [line,dotted,gray] (-3,2) -- (1,-2);
\dt{(-1,0)}{left}{$\phi^\dag$};
\end{scope}
\draw[line,domain=0:2,smooth,variable=\t,->] plot ({cosh(\t )/2}, {sinh(\t)/2} );
\draw[line,domain=0:2,smooth,variable=\t,->] plot ({-cosh(\t )/2}, {-sinh(\t)/2} );
\dt{(.5,0)}{right}{$\cO$};
\dt{(-.5,0)}{left}{$~~\cO^\dag$};
\draw (2.15,2.15) node {$\rho$}; 
\draw (2.15,-2.15) node {$\bar \rho$}; 
\end{tikzpicture}
\caption{The conformal Regge limit depicted in the global $\rho$-coordinate. The dotted lines are the light-cones of the operators $\phi$ and $\phi^\dagger$.}
\label{Fig:globalC}
\end{figure}

\subsection{Opacity bound and spectral radius formula}

The four-point function in the conformal Regge limit has the expected behavior \cite{Costa:2012cb}\footnote{In the conformal Regge limit, our variables $\rho$ and $\bar\rho$ are related to the variables $\sigma$ and $\rho$ in (55) and (56) of \cite{Costa:2012cb} by
\ie
16{\bar \rho\over \rho}=\sigma^2 \, ,\quad \rho\bar\rho = e^{2\rho}\,
\fe
where the variables $\sigma$ and $\rho$ appearing on the right are the ones defined in \cite{Costa:2012cb}.
}
\ie\label{eqn:ReggeG}
G(z,\bar z)^{\circlearrowleft} \sim 1 - \# \left(\rho\over \bar \rho\right)^{{1\over 2}(j_0-1)} \, ,
\fe
where $G(z,\bar z)^{\circlearrowleft}$ denotes the four-point function after the continuation of $z$ around 1, and $j_0$ is the Regge intercept, {\it i.e.} the analytic continuation of the leading Regge trajectory $j(\Delta)$ to $\Delta=1$. In unitary theories, \cite{Caron-Huot:2017vep} used the Cauchy-Schwarz inequality\footnote{We thank Petr Kravchuk for a discussion.
}
\ie
\label{CauchySchwarz}
|G(z,\bar z)^{\circlearrowleft}| \le G(z,\bar z) \, , \quad 0 \le z, \, \bar z \le 1
\fe
to prove that the Regge intercept is bounded by
\ie
\label{ReggeBound}
j_0 \le 1 \, .
\fe

The Regge behavior of the Lorentzian four-point function can be separated into two distinct classes, transparent $j_0<1$ and opaque $j_0=1$ \cite{Caron-Huot:2020ouj}. When $j_0<1$, the Lorentzian four-point function factorizes into a product of two-point functions in the conformal Regge limit; holographically, the particle created by the operator $\phi$ and that by $\cO$ pass through each other without interacting in the high energy fixed impact parameter limit. By contrast, when $j_0=1$, the Lorentzian four-point function does not factorize, and the bulk scattering is nontrivial. 

The behavior in the Regge limit can be further subdivided into transparent, refractive, and opaque. If we define
\ie
{r[\cO,\phi]} \,\equiv\, \lim_{z\to 0,\,{z/\bar z}\,\text{fixed}} G(z,\bar z)^{\circlearrowleft} \, ,
\fe
then the four-point function is called transparent if ${r[\cO,\phi]} = 1$, refractive if $r[\cO,\phi]$ is a nontrivial phase, and opaque otherwise.\footnote{By the definition of \cite{Caron-Huot:2020ouj}, refractive scattering is opaque.
}
We then define the notion of {\it opacity}
\ie
\label{opacity}
\kappa[\cO,\phi] \equiv 1-{|r[\cO,\phi]|}
\fe
for the four-point function $G(z,\bar z)$. Note that while the four-point function has zero opacity when $|r[\cO,\phi]|=1$, there could still be nontrivial refraction that corresponds to a nontrivial phase of $r[\cO,\phi]$. The inequality \eqref{CauchySchwarz} shows that
\ie\label{eqn:opacitybound}
\kappa[\cO,\phi] \ge 0 \, .
\fe

If the operator $\cO$ is holomorphically-defect-factorized through a topological defect line $\cL$, then according to \eqref{ReggeVisual}, we have
\ie\label{eqn:reggeGeneral}
{r[\cO,\phi]} \, = \,
\frac{1}{\vev{\cL}_{\bR^2}} \frac{\big\la \phi^\dag(0) \, \widehat\cL(\phi)(1)\big\ra}{ \big\la \phi^\dag(0) \, \phi(1) \big\ra}
 \, .
\fe
When $\kappa[\cO,\phi]=0$ (in other words $|{r[\cO,\phi]}|=1$), the nontrivial Regge dynamics are encoded in the subleading $\sum_{\cL' \neq \cI}$ part of \eqref{ReggeVisual}. By contrast, when $\kappa[\cO,\phi] > 0$ or $|{r[\cO,\phi]}|\le 1$, the leading $\cL' = \cI$ piece is already nontrivial, and incorporates an $\cO(1)$ contribution from the second piece in \eqref{eqn:ReggeG}. The opacity bound \eqref{eqn:opacitybound} immediately implies a corollary about topological defect lines in unitary conformal field theory.
\begin{coro}
\label{SpectralRadius}
In a (1+1)$d$ unitary conformal field theory, the spectral radius $r_{\cL}$ of a factorizing topological defect line $\cL$, defined by
\ie
r_{\cL}\equiv \max_{\phi}\left\{\left|\frac{\big\la \phi^\dag(0) \, \widehat\cL(\phi)(1)\big\ra}{ \big\la \phi^\dag(0) \, \phi(1) \big\ra}\right|\right\}\,,
\fe
is equal to the loop expectation value of $\cL$, i.e.
\ie\label{eqn:SpectralRadius}
r_{\cL}=|\vev{\cL}_{\bR^2}|\,.
\fe
In other words, for any factorizing topological defect line $\cL$ and any local operator $\phi$,
\ie
\left|
\frac{1}{\vev{\cL}_{\bR^2}} \frac{\big\la \phi^\dag(0) \, \widehat\cL(\phi)(1)\big\ra}{ \big\la \phi^\dag(0) \, \phi(1) \big\ra}
\right|
\le 1 \, .
\fe
\end{coro}
In Appendix~\ref{Sec:Bounded}, we give complementing arguments for the spectral radius formula \eqref{eqn:SpectralRadius} without assuming that the TDL $\cL$ is factorizing, by use of the Perron-Frobenius theorem and its generalizations.

Finally, the spatial torus Regge limit \eqref{SpatialTorus} of the torus two-point function of $\cO$ conveniently packages the infinitely-boosted conformal Regge limit for all possible $\phi$.

\subsection{Aspects of chaos}

The relation between the conformal Regge limit and the chaos limit of the Lorentzian four-point function at finite temperature $T=\beta^{-1}$ could be seen by conformally mapping the complex plane to the cylinder ${\rm S}^1\times \bR$ by $z = e^{{2i \pi \over \beta}(\tau+ix)}$, where the ${\rm S}^1$ is the thermal circle with periodicity $\beta$ \cite{Roberts:2014ifa,Murugan:2017eto}. The Euclidean time $\tau$ could be further analytically continued to Lorentzian time $t$ by
\ie
&z_1=e^{{2\pi\over \beta}(t+i\epsilon_1)} \, ,&& z_2=e^{{2\pi\over \beta}(t+i\epsilon_2)} \, ,&& z_3=e^{{2\pi\over \beta}(x+i\epsilon_3)} \, ,&& z_4=e^{{2\pi\over \beta}(x+i\epsilon_4)} \, ,
\\
&\bar z_1=e^{-{2\pi\over \beta}(t+i\epsilon_1)} \, ,&& \bar z_2=e^{-{2\pi\over \beta}(t+i\epsilon_2)} \, ,&& \bar z_3=e^{{2\pi\over \beta}(x-i\epsilon_3)} \, ,&& \bar z_4=e^{{2\pi\over \beta}(x-i\epsilon_4)} \, .
\fe
The ordering of the operators in the correlator is specified by choosing $\epsilon_1<\epsilon_4<\epsilon_2<\epsilon_3$. At $t=0$, the operators are space-like separated and $\bar z_i = z_i^*$. When $t$ increases from $t = 0$ to $t > |x|$, the cross ratio $z$ moves across the branch cut at $[1,\infty)$ onto the second sheet, while $\bar z$ remains on the first sheet. In the late time limit $t \to\infty$, both $z$ and $\bar z$ approach 0 with their ratio ${z / \bar z} = e^{{4\pi\over \beta}x} +\cO(e^{-{2\pi\over \beta}t})$ fixed, which is precisely the conformal Regge limit.

The out-of-time-ordered correlator (OTOC) captures the perturbation caused by the operators $\phi$ on the later measurements $\cO$. The behavior of the four-point function in the conformal Regge limit \eqref{eqn:ReggeG} translates to the exponential time dependence of the OTOC at late time
\ie\label{eqn:OTOClatetime}
G(z,\bar z)^{\circlearrowleft} \sim 1 - \# e^{{2\pi\over \beta}\lambda t} \, .
\fe
The exponent $\lambda$ is related to the Regge intercept $j_0$ by $\lambda = j_0-1$, and bounded according to \eqref{ReggeBound} by $\lambda\le 0$. When $\lambda < 0$, the OTOC approaches the product of two-point functions signifying that the effect of the operators $\phi$'s on the measurements $\cO$'s exponentially decays at late time. When $\lambda = 0$, the effect of the operator $\phi$ could have finite imprint on the measurement $\cO$ at infinite time.

In a chaotic system, the effect of the operator $\phi$ on the measurement $\cO$ could grow exponentially during some intermediate time scale. At large central charge $c$ and the time scale $t \sim \beta\log c$, the OTOC is expected to behave as \cite{Roberts:2014ifa,Fitzpatrick:2016thx,Perlmutter:2016pkf,Liu:2018iki,Hampapura:2018otw,Chang:2018nzm}
\ie\label{eqn:chaosOTOC}
G(z,\bar z)^{\circlearrowleft}\sim 1 - \frac{\#}{c} e^{{2\pi\over \beta}\lambda_L t} \, .
\fe
The chaos exponent $\lambda_L$ could take positive values and bounded in unitary theories by \cite{Maldacena:2015waa}
\ie
\lambda_L\le 1 \, .
\fe
Probing the chaotic behavior \eqref{eqn:chaosOTOC} of the OTOC requires taking the limit $z\to 0$ while fixing $\bar z/z$ and $c \times z$. Such a limit could be similarly studied by applying the manipulations in \eqref{ReggeVisual} to large $c$ theories. One would need to include subleading terms that involve lasso diagrams \cite{Chang:2018iay}.

\section{Rational conformal field theory}
\label{Sec:Rational}

\subsection{Holomorphic-defect-factorization and Lorentzian dynamics}

The holomorphic-defect-factorization prerequisite (Definition~\ref{Prereq}) is the existence of holomorphic and anti-holomorphic defect operators of suitable weights in some defect Hilbert spaces, so that holomorphic-defect-factorization is at all possible.

The local operators transform as bi-modules of the holomorphic and anti-holomorphic chiral algebras. The highest-weight operators in the bi-modules are labeled by $\cO_{i, j}$, where the indices $i$ and $ j$ label the irreducible modules of the holomorphic and anti-holomorphic chiral algebras. Modular invariance further constrains the set of $\cO_{i, j}$ that appear in the theory, and the holomorphic-defect-factorization prerequisite is satisfied by the existence of the Verlinde lines \cite{Verlinde:1988sn,Petkova:2000ip,Drukker:2010jp,Gaiotto:2014lma}.

In a diagonal modular invariant rational conformal field theory, the partition function of local operators is
\ie
Z(\tau, \bar\tau)=\sum_{i}\chi_i(\tau) \bar\chi_i(\bar\tau) \, .
\fe
The Verlinde line $\cL_k$ acts the local operator $\cO_{i, i}$ by
\ie\label{eqn:VerlindeLonO}
\widehat\cL_k(\cO_{i, i})={S_{ki}\over S_{0i}} \, \cO_{i, i} \, ,
\fe
where $S_{ki}$ is the modular S matrix, and $i=0$ denotes the vacuum module. The partition function twisted 
by the Verlinde line $\cL_k$ is
\ie
Z^{\cL_k}(\tau, \bar\tau) = {\rm Tr}_{\cH} \, \widehat\cL \, q^{L_0-\frac{c}{24}} \bar q^{\bar L_0-\frac{\bar c}{24}} = \sum_{i}{S_{ki}\over S_{0i}} \,\chi_i(\tau) \bar\chi_i(\bar\tau) \, .
\fe
The partition function for the defect Hilbert space ${\cal H}_{\cL_k}$ is obtained by a modular S transform. The result is
\ie
Z_{\cL_k}(\tau, \bar\tau) = \sum_{i,j} N^j_{ki} \, \chi_i(\tau) \bar\chi_j(\bar\tau) \, ,
\fe
where the fusion coefficients $N^j_{ki}$ are non-negative integers given by the Verlinde formula \cite{Verlinde:1988sn},
\ie
N^j_{ki} = \sum_{\ell}{S_{k\ell}S_{i\ell}S_{j\ell}^*\over S_{0\ell}} \, .
\fe
The holomorphic-defect-factorization prerequisite is satisfied because $N^0_{ki} = \D_{ki}$ and $N^j_{k0} = \D^j_k$. In other words, for any admissible highest-weight operator $\cO_{i,i}$ with weight $(h_i,h_i)$, the defect Hilbert space of the Verlinde line $\cL_i$ contains one defect highest-weight operator of weight $(h_i, 0)$ and another one of weight $(0, h_i)$. 

When there exists a permutation automorphism $\zeta$ of the irreducible modules of the chiral algebra, satisfying
\ie
\zeta(0)=0 \, ,\quad S_{\zeta(i)\zeta(j)}=S_{ij} \, ,\quad T_{\zeta(i)\zeta(j)}=T_{ij} \, ,
\fe
there is a modular invariant partition function
\ie
Z(\tau, \bar\tau) = \sum_{i}\chi_i(\tau) \bar\chi_{\zeta(i)}(\bar\tau) \, .
\fe
The topological defects lines in such theories were classified by Petkova and Zuber \cite{Petkova:2000ip}. The Verlinde line $\cL_k$ acts on the local operator $\cO_{i, \zeta(i)}$ by
\ie\label{eqn:VerlindeLonO_2}
\widehat\cL_k(\cO_{i, \zeta(i)}) = {S_{ki}\over S_{0i}} \, \cO_{i, \zeta(i)} \, .
\fe
After similar manipulations as before, we find the partition function for the defect Hilbert space ${\cal H}_{\cL_k}$,
\ie
Z_{\cL_k}(\tau, \bar\tau) = \sum_{i,j} N^{\zeta^{-1}(j)}_{ki} \, \chi_i(\tau) \bar\chi_j(\bar\tau) \, .
\fe
The holomorphic-defect-factorization prerequisite in this case follows from $N^0_{ki} = \D_{ki}$ and $N^{\zeta^{-1}(j)}_{k0} = \D^j_{\zeta(k)}$. In other words, for any admissible highest-weight operator $\cO_{i,\zeta(i)}$ with weight $(h_i,h_{\zeta(i)})$, the defect Hilbert space of the Verlinde line $\cL_i$ contains one defect highest-weight operator of weight $(h_i, 0)$ and another one of weight $(0, h_{\zeta(i)})$. 

Diagonal or not, the defect Hilbert space ${\cal H}_{\cL_k}$ projected onto the subspace of holomorphic operators (resp. anti-holomorphic operators) is an irreducible module of the holomorphic (resp. anti-holomorphic) chiral algebra, encapsulated in the equations
\ie
\lim_{\bar q \to 0} \, \bar q\,^{-{c\over 24}} Z_{\cL_k}(\tau,\bar\tau)
=\chi_k(\tau) \, ,\quad 
\lim_{ q \to 0} \, q^{-{c\over 24}} Z_{\cL_k}(\tau,\bar\tau)
=\bar\chi_{\zeta(k)}(\bar\tau) \, .
\fe
The diagonal case is when the permutation map $\zeta$ is the identity map. 

As proven by Moore and Seiberg \cite{Moore:1988ss}, every rational theory has a maximally extended chiral algebra with respect to which the theory is either diagonal or permutation modular invariant. And since all operators in the same chiral algebra module can be factorized through the same topological defect line, the preceding discussion covers all possibilities. 

The full set of topological defect lines that not necessarily commutes with the maximally extended chiral algebra is vast, even in rational theories. The fact that all local operators are factorized through Verlinde lines, {\it i.e.} TDLs that commute with the maximally extended chiral algebra, suggests the following proposition.\footnote{We thank Zohar Komargodski and Kantaro Ohmori for highlighting this implication.
}
\begin{prop}
In rational conformal field theory, if $\cL$ is a topological defect line whose defect Hilbert space $\cH_\cL$ contains a holomorphic defect operator, and whose dual defect Hilbert space $\cH_\ocL$ contains an anti-holomorphic defect operator, then $\cL$ and $\ocL$ are Verlinde lines.
\end{prop}
Every $\cL$ satisfying the assumed property produces a local operator by holomorphic-defect-factorization, and this map is injective by Proposition~\ref{eqn:Uniqueness}, but as discussed in rational conformal field theory all local operators are factorized through Verlinde lines. 

Let us comment on the Lorentzian dynamics of rational conformal field theory. Using \eqref{eqn:VerlindeLonO} and \eqref{eqn:VerlindeLonO_2} for the action of $\cL_k$ on local operators, the infinite boost limit \eqref{eqn:reggeGeneral} is given by the modular S matrix as
\ie\label{eqn:rforRCFT}
{r[\cO_{k,\zeta(k)},\cO_{i,\zeta(i)}]} & \, = \, \displaystyle {S_{00}S_{ki}\over S_{0k}S_{0i}} \, .
\fe
The diagonal case ($\zeta$ being the trivial permutation) reproduces the result of \cite{Gu:2016hoy} derived from the monodromy properties of the chiral algebra blocks, or equivalently from a bulk perspective (reviewed in Section~\ref{Sec:Bulk}) by use of the braiding of anyons. However, we emphasize that the derivation of our formula \eqref{eqn:reggeGeneral} only involves the $F$-symbols alone, and hence applies beyond rationality.

\subsection{Example: Ising conformal field theory}
\label{Sec:Ising}

The Ising conformal field theory has three local operators, the identity $1$, the energy operator $\varepsilon$, and the spin operator $\sigma$. It has three topological defect lines, the trivial $\cI$, the $\bZ_2$ symmetry defect line $\eta$, and the non-invertible Kramers-Wannier duality line $\cN$ \cite{Frohlich:2004ef,Frohlich:2006ch,Frohlich:2009gb}. The fusion rule is
\ie
\eta^2 = \cI \, , \quad \cN^2 = \cI + \eta \, , \quad \eta \, \cN = \cN \, .
\fe
The local operators are holomorphically-defect-factorized as follows:
\ie
\varepsilon = \psi \stackrel{\eta}{\text{---}} \bar\psi \, , \quad
\sigma = \tau \stackrel{\cN}{\text{---}} \bar\tau \, ,
\fe
where $\psi$ is a weight $(\frac12, 0)$ free fermion, and $\tau$ is a weight $(\frac{1}{16}, 0)$ defect operator. Consider the vector of holomorphic-defect four-point functions
\ie
{\bf f}(z) \ = \
\begin{pmatrix} 
\begin{gathered}
\begin{tikzpicture}[scale=.5]
\draw [line,dashed] (-1,0) -- (-.5,0) node [above] {$\cI$} -- (0,0);
\draw [line] (-1,0) -- (-1.5,.87) \dotsol{above left}{$\tau$};
\draw [line] (-1,0) -- (-1.5,-.87) \dotsol{below left}{$\tau$};
\draw [line] (0,0) -- (.5,-.87) \dotsol{below right}{$\tau$};
\draw [line] (0,0) -- (.5,.87) \dotsol{above right}{$\tau$};
\end{tikzpicture}
\end{gathered}
\\
\begin{gathered}
\begin{tikzpicture}[scale=.5]
\draw [line] (-1,0) -- (-.5,0) node [above] {$\eta$} -- (0,0);
\draw [line] (-1,0) -- (-1.5,.87) \dotsol{above left}{$\tau$};
\draw [line] (-1,0) -- (-1.5,-.87) \dotsol{below left}{$\tau$};
\draw [line] (0,0) -- (.5,-.87) \dotsol{below right}{$\tau$};
\draw [line] (0,0) -- (.5,.87) \dotsol{above right}{$\tau$};
\end{tikzpicture}
\end{gathered}
\end{pmatrix} 
=
\begin{pmatrix}
\cF{\scriptsize
\begin{bmatrix}
\frac{1}{16} & \frac{1}{16}
\\
\frac{1}{16} & \frac{1}{16}
\end{bmatrix}
}_0^{\frac12}(z)
\\
\\
C_{\tau, \tau, \psi}^2 \times
\cF{\scriptsize
\begin{bmatrix}
\frac{1}{16} & \frac{1}{16}
\\
\frac{1}{16} & \frac{1}{16}
\end{bmatrix}
}_{\frac12}^{\frac12}(z)
\end{pmatrix} 
\, .
\fe
Under crossing, the (properly normalized) Virasoro blocks transform as
\ie
\begin{pmatrix}
\cF{\scriptsize
\begin{bmatrix}
\frac{1}{16} & \frac{1}{16}
\\
\frac{1}{16} & \frac{1}{16}
\end{bmatrix}
}_{0}^{\frac12}(1-z)
\\
\\
\cF{\scriptsize
\begin{bmatrix}
\frac{1}{16} & \frac{1}{16}
\\
\frac{1}{16} & \frac{1}{16}
\end{bmatrix}
}_{\frac12}^{\frac12}(1-z)
\end{pmatrix} 
\ = \
\frac{1}{\sqrt2}
\begin{pmatrix}
1 & \frac12
\\
2 & -1
\end{pmatrix}
\begin{pmatrix}
\cF{\scriptsize
\begin{bmatrix}
\frac{1}{16} & \frac{1}{16}
\\
\frac{1}{16} & \frac{1}{16}
\end{bmatrix}
}_{0}^{\frac12}(z)
\\
\\
\cF{\scriptsize
\begin{bmatrix}
\frac{1}{16} & \frac{1}{16}
\\
\frac{1}{16} & \frac{1}{16}
\end{bmatrix}
}_{\frac12}^{\frac12}(z)
\end{pmatrix} \, .
\fe

A gauge choice means that canonical junction vectors have been chosen, so all correlation functionals can be turned into correlation functions by the implicit insertion of canonical junction vectors. Henceforth defect thee-point correlation functionals become simply defect three-point coefficients. Suppose we adopt the gauge choice of \cite{Chang:2018iay} where the nontrivial $F$-symbols are
\ie
(F^{\eta,\,\cN,\,\eta}_\cN)_{\cN,\,\cN} = -1 \, , \quad F^{\cN,\,\cN,\,\cN}_\cN = 
\frac{1}{\sqrt2}
\begin{pmatrix}
1 & 1
\\
1 & -1
\end{pmatrix} \, .
\fe
The crossing equation
\ie
{\bf f}(1-z) = \frac{1}{\sqrt2}
\begin{pmatrix}
1 & 1
\\
1 & -1
\end{pmatrix} {\bf f}(z)
\fe
becomes simply
\ie
\begin{pmatrix}
1
\\
& C_{\tau, \tau, \psi}^2
\end{pmatrix}
\times
\frac{1}{\sqrt2}
\begin{pmatrix}
1 & \frac12
\\
2 & -1
\end{pmatrix}
\ = \
\frac{1}{\sqrt2}
\begin{pmatrix}
1 & 1
\\
1 & -1
\end{pmatrix} 
\times
\begin{pmatrix}
1
\\
& C_{\tau, \tau, \psi}^2
\end{pmatrix}
\, ,
\fe
which gives $C_{\tau, \tau, \psi}^2 = \frac12$. The formula \eqref{3pt} and \eqref{Theta} give the three-point coefficient
\ie
C_{\sigma, \sigma, \varepsilon} =
{\vev{\cN}_{\bR^2}} \sqrt{{\vev{\eta}_{\bR^2}}} \,
\, C_{\tau, \tau, \psi} \, C_{\bar\tau, \bar\tau, \bar\psi} \, \Theta_{\eta, \eta, \cN}
=
\sqrt2 \times 1 \times \sqrt{\frac12} \times \sqrt{\frac12} \times \sqrt{\frac12} = \frac12 \, ,
\fe
up to a sign that can be absorbed into a redefinition of $\psi$ and $\bar\psi$.

Alternatively, one may choose a gauge in which the $F$-symbols are identical to the crossing matrix of Virasoro blocks,
\ie
 F^{\cN,\,\cN,\,\cN}_\cN = 
\frac{1}{\sqrt2}
\begin{pmatrix}
1 & \frac12
\\
2 & -1
\end{pmatrix} \, ,
\fe
trivializing the defect three-point coefficients. The formula \eqref{3pt} becomes
\ie
C_{\sigma, \sigma, \varepsilon} =
{\vev{\cN}_{\bR^2}} \sqrt{{\vev{\eta}_{\bR^2}}} \, \Theta_{\cN, \cN, \eta}
=
\sqrt2 \, \Theta_{\cN, \cN, \eta} \, ,
\fe
and in this gauge \eqref{Theta} is computed to be (using the first expression)
\ie
\Theta_{\cN, \cN, \eta} ~&=~ \frac{1}{{\vev{\eta}_{\bR^2}}} \times 
(F^{\cN, \cN, \cN}_{\cN})_{\cI, \eta} = \frac{1}{2\sqrt2} \, ,
\fe
giving the same result $C_{\sigma, \sigma, \varepsilon} = \frac12$. However, in this gauge, many previously trivial ($=1$) $F$-symbols have become nontrivial. For instance,
\ie
(F^{\cN, \cN, \eta}_{\eta})_{\cI, \cN} = \frac{1}{2} \, .
\fe
The trivialization of defect three-point coefficients is at the cost of complicating the $F$-symbols.

Next let us study the emergence of the Kramers-Wannier duality line $\cN$ from Lorentzian dynamics. The torus two-point function of the spin operator $\sigma$ is \cite{DiFrancesco:1987ez}
\ie
\la \sigma(z, \bar z) \sigma(0) \ra_{{\rm T}^2(\tau, \bar\tau)} = \left|\frac{\partial_z \theta_1(0|\tau)}{\theta_1(z|\tau)}\right|^{\frac14} \sum_{\nu = 2}^4 \left|\frac{\theta_\nu(\frac{z}{2}|\tau)}{\eta(\tau)}\right| \, ,
\fe
normalized such that in the limit $z, \, \bar z \to 0$,
\ie
\la \sigma(z, \bar z) \sigma(0) \ra_{{\rm T}^2(\tau, \bar\tau)} \to |z|^{-\frac14} Z(\tau, \bar\tau) \, , 
\fe
where $Z(\tau, \bar\tau)$ is the torus partition function
\ie
Z(\tau, \bar\tau) = \sum_{\nu = 2}^4 |\theta_\nu(0|\tau)| \, .
\fe

Consider the torus Regge limits.
\paragraph
{\bf Spatial torus Regge limit.}
Under $z \to z+1$, 
\ie
& \la \sigma(z, \bar z) \sigma(0) \ra_{{\rm T}^2(\tau, \bar\tau)} \to
\la \sigma(z+1, \bar z) \sigma(0) \ra_{{\rm T}^2(\tau, \bar\tau)} 
\\
&= e^{\frac{i\pi}{8}} \left|\frac{\partial_z \theta_1(0|\tau)}{\theta_1(z|\tau)}\right|^{\frac14} \frac{- \theta_1(\frac{z}{2}|\tau) \theta_2(\frac{\bar z}{2}|\bar\tau) + \theta_4(\frac{z}{2}|\tau) \theta_3(\frac{\bar z}{2}|\bar\tau) + \theta_3(\frac{z}{2}|\tau) \theta_4(\frac{\bar z}{2}|\bar\tau)}{|\eta(\tau)|} \, .
\fe
Then
\ie
\lim_{z, \bar z \to 0} e^{-\frac{i\pi}{8}} |z|^{\frac14} \la \sigma(z+1, \bar z) \sigma(0) \ra_{{\rm T}^2(\tau, \bar\tau)} &= \frac{\theta_4(\frac{z}{2}|\tau) \theta_3(\frac{\bar z}{2}|\bar\tau) + \theta_3(\frac{z}{2}|\tau) \theta_4(\frac{\bar z}{2}|\bar\tau)}{|\eta(\tau)|}
= \frac{Z^N(\tau, \bar\tau)}{\sqrt2} \, .
\fe
\paragraph
{\bf Temporal torus Regge limit.}
Under $z \to z+\tau$, 
\ie
& \la \sigma(z, \bar z) \sigma(0) \ra_{{\rm T}^2(\tau, \bar\tau)} \to
\la \sigma(z+\tau, \bar z) \sigma(0) \ra_{{\rm T}^2(\tau, \bar\tau)} 
\\
&= e^{\frac{i\pi}{8}} \left|\frac{\partial_z \theta_1(0|\tau)}{\theta_1(z|\tau)}\right|^{\frac14} \frac{\theta_3(\frac{z}{2}|\tau) \theta_2(\frac{\bar z}{2}|\bar\tau) + \theta_2(\frac{z}{2}|\tau) \theta_3(\frac{\bar z}{2}|\bar\tau) + i \, \theta_1(\frac{z}{2}|\tau) \theta_4(\frac{\bar z}{2}|\bar\tau)}{|\eta(\tau)|} \, .
\fe
Then
\ie
\lim_{z, \bar z \to 0} e^{-\frac{i\pi}{8}} |z|^{\frac14} \la \sigma(z+1, \bar z) \sigma(0) \ra_{{\rm T}^2(\tau, \bar\tau)} &= \frac{\theta_3(\frac{z}{2}|\tau) \theta_2(\frac{\bar z}{2}|\bar\tau) + \theta_2(\frac{z}{2}|\tau) \theta_3(\frac{\bar z}{2}|\bar\tau)}{|\eta(\tau)|}
= \frac{Z_N(\tau, \bar\tau)}{\sqrt2} \, .
\fe
In the above we used some identities \eqref{JacobiShifts} for the Jacobi theta functions. Noting that ${\vev{\cL}_{\bR^2}} = \sqrt2$, we recover the expected twisted torus partition functions $Z^\cL(\tau, \bar\tau)$ and $Z_\cL(\tau, \bar\tau)$.

\subsection{Bulk perspective}
\label{Sec:Bulk}

The holomorphic part of a rational conformal field theory (RCFT) is the boundary edge theory of a bulk topological quantum field theory (TQFT) \cite{Witten:1988hf,Elitzur:1989nr,Moore:1989vd,Moore:1989yh}. A celebrated example is Witten's correspondence between Wess-Zumino-Witten (WZW) models and Chern-Simons theory \cite{Witten:1988hf}. The states of the latter quantized on any spatial slice $\cM_2$ correspond to the chiral algebra blocks of the WZW on $\cM_2$. General RCFTs are dual to more general topological orders, such as Dijkgraaf-Witten theories, or abstract sets of anyons described by modular tensor categories.

A TQFT on $\cM_2 \times [0,1]$ corresponds to a diagonal RCFT on $\cM_2$ \cite{Fuchs:2002cm,Fuchs:2003id,Fuchs:2004dz,Fuchs:2004xi,Fjelstad:2005ua}. The holomorphic degrees of freedom live on one boundary, the anti-holomorphic ones live on the other, connected through the bulk by anyons. From this point of view, the Verlinde lines in a diagonal RCFT are the two-dimensional avatars of anyons in the TQFT, and the holomorphic-defect-factorization of local operators becomes evident,
\ie\label{eqn:bulkfactorizationO}
\cO(z, \bar z) \ = \
\begin{gathered}
\begin{tikzpicture}[scale=1]
\begin{scope}[scale=2]
\draw [line, opacity=.2] (0,0) -- (2,0) -- (3,1) -- (1,1) -- (0,0);
\end{scope}
\begin{scope}[scale=2, shift={(0,-1.5)}]
\draw [line, opacity=.2] (0,0) -- (2,0) -- (3,1) -- (1,1) -- (0,0);
\end{scope}
\draw [line,->-=.55] (3,-2) \dotsol{below}{$\cD(z)$} -- (3,-.5) node [right] {$\cL$} -- (3,1) \dotemp{above}{$\ocD(\bar z)$};
\end{tikzpicture}
\end{gathered} \, ,
\fe
where we abused the notation by labeling the anyon also by $\cL$. The meaning of treating $z , \, \bar z$ as independent complex variables is also clear, and the nontrivial monodromies of blocks correspond to the braiding of anyons. The action of a Verlinde line $\cL'$ on a holomorphically-defect-factorized local operator $\cO = \cD \stackrel{\cL}{\text{---}} \ocD$ could be realized as the linking of the anyon lines $\cL'$ and $\cL$ in the three-dimensional bulk. For example, the action of the Verlinde line $\cL_k$ on local operator $\cO_{i,i}$ is realized as
\ie\label{eqn:bulkTDLaction}
\widehat\cL_k(\cO_{i, i;\alpha}) \ = \
\begin{gathered}
\begin{tikzpicture}[scale=0.8]
\begin{scope}[scale=2]
\draw [line, opacity=.2] (0,0) -- (2,0) -- (3,1) -- (1,1) -- (0,0);
\end{scope}
\begin{scope}[scale=2, shift={(0,-1.5)}]
\draw [line, opacity=.2] (0,0) -- (2,0) -- (3,1) -- (1,1) -- (0,0);
\end{scope}
\draw [line] (3,-2) \dotsol{below}{$\cD(z)$} -- (3,-1.05);
\draw [line,->-=.2] (3,-.75) -- (3,-.5) node [right] {$\cL_i$} -- (3,1) \dotemp{above}{$\ocD(\bar z)$};
\draw [line,->-=.52] (3,-.6) ++(85:1 and .5) arc (85:-265:1 and .4);
\node at (1.6,-.5) {$\cL_k$};
\end{tikzpicture}
\end{gathered}
 \ = \ {S_{ki}\over S_{0i}}\,\times \, \begin{gathered}
\begin{tikzpicture}[scale=0.8]
\begin{scope}[scale=2]
\draw [line, opacity=.2] (0,0) -- (2,0) -- (3,1) -- (1,1) -- (0,0);
\end{scope}
\begin{scope}[scale=2, shift={(0,-1.5)}]
\draw [line, opacity=.2] (0,0) -- (2,0) -- (3,1) -- (1,1) -- (0,0);
\end{scope}
\draw [line,->-=.55] (3,-2) \dotsol{below}{$\cD(z)$} -- (3,-.5) node [right] {$\cL_i$} -- (3,1) \dotemp{above}{$\ocD(\bar z)$};
\end{tikzpicture}
\end{gathered} \, ,
\fe
where we apply braiding to unlink the $\cL$ and $\cL'$ and use the relation between braiding and the modular S matrix. The result agrees with the action \eqref{eqn:VerlindeLonO}. Since topological defect lines in general conformal field theory need not admit braiding, we refrain from using braiding in the following.

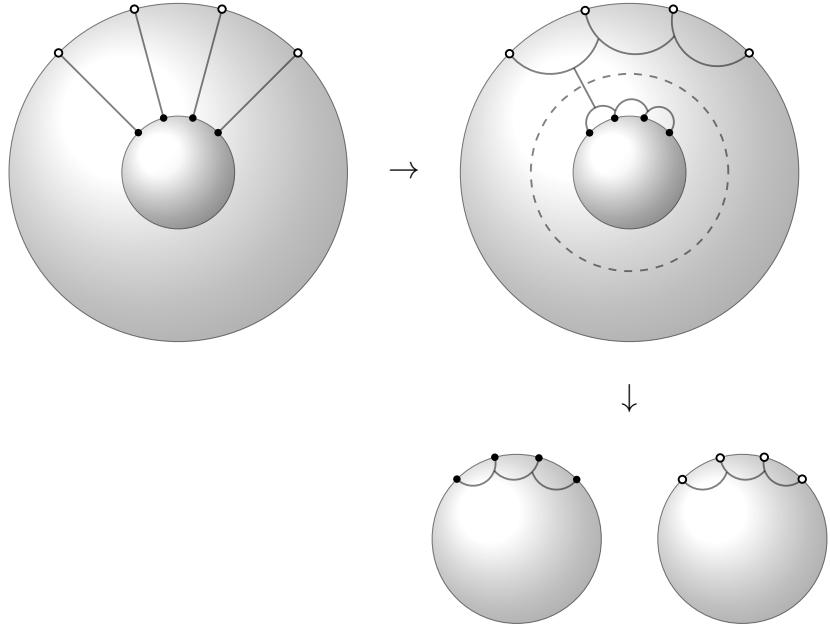
\begin{figure}
\centering
\begin{tikzpicture}[scale=.75]
\def\ri{1}
\def\rm{1.75}
\def\ro{3}
\begin{scope}
\draw[ball color=white] (0,0) circle [radius=\ri];
\draw[ball color=white, opacity=.5] (0,0) circle [radius=\ro];
\def\ang{45}
\draw[line, opacity=.5] ({\ri * cos(\ang)}, {\ri * sin(\ang)}) \dotsol{}{} -- ({\ro * cos(\ang)}, {\ro * sin(\ang)}) \dotemp{}{};
\def\ang{75}
\draw[line, opacity=.5] ({\ri * cos(\ang)}, {\ri * sin(\ang)}) \dotsol{}{} -- ({\ro * cos(\ang)}, {\ro * sin(\ang)}) \dotemp{}{};
\def\ang{105}
\draw[line, opacity=.5] ({\ri * cos(\ang)}, {\ri * sin(\ang)}) \dotsol{}{} -- ({\ro * cos(\ang)}, {\ro * sin(\ang)}) \dotemp{}{};
\def\ang{1\ro5}
\draw[line, opacity=.5] ({\ri * cos(\ang)}, {\ri * sin(\ang)}) \dotsol{}{} -- ({\ro * cos(\ang)}, {\ro * sin(\ang)}) \dotemp{}{};
\end{scope}
\node at (4,0) {$\to$};
\begin{scope}[xshift = 8cm]
\draw[ball color=white] (0,0) circle [radius=\ri];
\draw[ball color=white, opacity=.5] (0,0) circle [radius=\ro];
\draw[line, dashed, opacity=.5] (0,0) circle [radius=\rm];
\begin{scope}[rotate=-30]
\draw [line, opacity=.5] (0,{\ro / cos(15)}) ++(-15:{\ro * tan(15)}) \dotemp{}{} arc (-15:-165:{\ro * tan(15)}) \dotemp{}{};
\end{scope}
\begin{scope}[rotate=-2]
\draw [line, opacity=.5] (0,{1 * \ro}) ++(-38:{.3 * \ro}) arc (-38:-170:{.3 * \ro}) \dotemp{}{};
\end{scope}
\begin{scope}[rotate=28]
\draw [line, opacity=.5] (0,{1 * \ro}) ++(-46:{.3 * \ro}) arc (-46:-171:{.3 * \ro}) \dotemp{}{};
\end{scope}
\begin{scope}[rotate=-30]
\draw [line, opacity=.5] (0,{\ri / cos(15)}) ++(-15:{\ri * tan(15)}) \dotsol{}{} arc (-15:195:{\ri * tan(15)}) \dotsol{}{};
\end{scope}
\begin{scope}[rotate=-2]
\draw [line, opacity=.5] (0,{1 * \ri}) ++(16:{.3 * \ri}) arc (16:189:{.3 * \ri}) \dotsol{}{};
\end{scope}
\begin{scope}[rotate=28]
\draw [line, opacity=.5] (0,{1 * \ri}) ++(16:{.3 * \ri}) arc (16:189:{.3 * \ri}) \dotsol{}{};
\end{scope}
\begin{scope}[rotate=28]
\draw [line, opacity=.5] (0,1.3) -- (0,2.1);
\end{scope}
\end{scope}
\node at (8,-4) {$\downarrow$};
\begin{scope}[xshift = 6cm, yshift = -6.5cm, xscale = -.5, yscale = .5]
\draw[ball color=white, opacity=.5] (0,0) circle [radius=\ro];
\begin{scope}[rotate=-30]
\draw [line, opacity=.5] (0,{\ro / cos(15)}) ++(-15:{\ro * tan(15)}) \dotsol{}{} arc (-15:-165:{\ro * tan(15)}) \dotsol{}{};
\end{scope}
\begin{scope}[rotate=-2]
\draw [line, opacity=.5] (0,{1 * \ro}) ++(-38:{.3 * \ro}) arc (-38:-170:{.3 * \ro}) \dotsol{}{};
\end{scope}
\begin{scope}[rotate=28]
\draw [line, opacity=.5] (0,{1 * \ro}) ++(-46:{.3 * \ro}) arc (-46:-171:{.3 * \ro}) \dotsol{}{};
\end{scope}
\end{scope}
\begin{scope}[xshift = 10cm, yshift = -6.5cm, scale = .5]
\draw[ball color=white, opacity=.5] (0,0) circle [radius=\ro];
\begin{scope}[rotate=-30]
\draw [line, opacity=.5] (0,{\ro / cos(15)}) ++(-15:{\ro * tan(15)}) \dotemp{}{} arc (-15:-165:{\ro * tan(15)}) \dotemp{}{};
\end{scope}
\begin{scope}[rotate=-2]
\draw [line, opacity=.5] (0,{1 * \ro}) ++(-38:{.3 * \ro}) arc (-38:-170:{.3 * \ro}) \dotemp{}{};
\end{scope}
\begin{scope}[rotate=28]
\draw [line, opacity=.5] (0,{1 * \ro}) ++(-46:{.3 * \ro}) arc (-46:-171:{.3 * \ro}) \dotemp{}{};
\end{scope}
\end{scope}
\end{tikzpicture}
\caption{The conformal block decomposition of the four-point function of the holomorphically-defect-factorized local operator \eqref{eqn:bulkfactorizationO}.}
\label{Fig:3D}
\end{figure}

Consider ${\rm S}^2\times [0,1]$, and insert four anyon lines $\cL_1,\,\dotsc,\,\cL_4$ at $z_1,\,\dotsc,\, z_4\in{\rm S}^2$ extended from one ${\rm S}^2$ boundary to the other, as shown in the upper left picture of Figure \ref{Fig:3D}. This configuration gives a state in the Hilbert space $\widehat{\cal H}_{{\rm S}^2;z_i,\cL_i}\times \widehat{\cal H}_{{\rm S}^2;\bar z_i,\ocL_i}$, where $\widehat{\cal H}_{{\rm S}^2;z_i,\cL_i}$ is the Hilbert space of the holomorphic chiral algebra blocks of the RCFT, and $\widehat{\cal H}_{{\rm S}^2;\bar z_i,\ocL_i}$ the anti-holomorphic \cite{Witten:1988hf}. 

We now argue that this state corresponds to a crossing symmetric four-point function of local operators $\cO_1,\,\dotsc,\,\cO_4$. First, we apply a sequence of $F$-moves on the anyons, to achieve the configuration on the upper right of Figure~\ref{Fig:3D}. Next, we cut the space along the spherical surface represented by the dashed line. The cutting generates two new boundaries with opposite orientations that could be either ${\rm S}^2$ with one marked point or no marked point, which has a zero-dimensional or one-dimensional Hilbert space, respectively. Hence, the anyon that crosses the cutting surface must be a trivial line. By gluing this configuration with two solid ${\rm B}^3$ balls with opposite orientations along the cutting surface, we obtain the configuration on the bottom right of Figure~\ref{Fig:3D}, where the left (resp. right) connected component gives a state in the Hilbert space $\widehat{\cal H}_{{\rm S}^2;z_i,\cL_i}$ (resp. $\widehat{\cal H}_{{\rm S}^2;\bar z_i,\ocL_i}$). They correspond to the holomorphic and anti-holomorphic blocks of the chiral algebra. The total configuration is a finite sum over the holomorphically factorized products and gives the conformal block decomposition.

\section{Free boson theory}
\label{Sec:Free}

Are operators holomorphically-defect-factorized in irrational theories? This section examines the $c = 1$ free boson theory whose moduli space contains both rational and irrational points.

\subsection{Toroidal branch}

As we presently explain, all local operators in the compact boson theory are holomorphically-defect-factorized through the U(1) symmetry defect lines, which are Wilson lines of the background U(1) gauge field.

The ${\rm U(1)}_{\rm m} \times {\rm U(1)}_{\rm w}$ momentum and winding symmetry Wilson lines can be explicitly represented by
\ie
\label{ExplicitU1Line}
\cL_{(\theta_{\rm m}, \theta_{\rm w})} \ = \ :\exp\left[\frac{i}{2\pi} (\theta_{\rm m} r + \frac{\theta_{\rm w}}{r}) \int dz \, \partial X_L(z) - \frac{i}{2\pi} (\theta_{\rm m} r - \frac{\theta_{\rm w}}{r}) \int d\bar z \, \bar\partial X_R(\bar z) \right]: \, .
\fe
Integer spectral flow gives an equivalence relation
\ie
\cL_{(\theta_{\rm m}, \theta_{\rm w})} \sim \cL_{(\theta'_m, \theta'_w)} \, , \quad \theta'_m - \theta_{\rm m} \, , \ \theta'_w - \theta_{\rm w} \in 2 \pi \bZ \, .
\fe
The flavored torus partition function of $\cL_{(\theta_{\rm m}, \theta_{\rm w})}$ is
\ie
Z^{\cL_{(\theta_{\rm m}, \theta_{\rm w})}}(\tau, \bar\tau) = \frac{1}{|\eta(\tau)|^2} \sum_{m, w \in \bZ} e^{i \theta_{\rm m} m + i \theta_{\rm w} w} q^{\frac{p_{\rm L}^2}{4}} \bar q^{\frac{p_{\rm R}^2}{4}} \, , \quad p_{L,R} = \frac{m}{r} \pm wr \, ,
\fe
whose modular S transform gives the defect partition function 
\ie
\label{DefectZToroidal}
Z_{\cL_{(\theta_{\rm m}, \theta_{\rm w})}}(\tau, \bar\tau) = \frac{1}{|\eta(\tau)|^2} \sum_{m, w \in \bZ} q^{\frac{p_{\rm L}^2}{4}} \bar q^{\frac{p_{\rm R}^2}{4}} \, , \quad p_{L,R} = \frac{m + \theta_{\rm w}/2\pi}{r} \pm (w + \theta_{\rm m}/2\pi)r \, .
\fe
In fact, a defect operator can be explicitly identified by taking the representation \eqref{ExplicitU1Line} of $\cL_{(\theta_{\rm m}, \theta_{\rm w})}$ and integrating by parts. Doing so in different spectral flow frames gives different defect operators that belong to the same defect Hilbert space of $\cL_{(\theta_{\rm m}, \theta_{\rm w})}$. We will see an example momentarily.

An exponential local operator
\ie
\label{ExpOp}
\cO_{m, w}(z, \bar z) \,=\, :e^{i p_{\rm L} X_L(z) + i p_{\rm R} X_R(\bar z)}: \, , \quad p_{\rm L} = \frac{m}{r} + wr \, , \quad p_{\rm R} = \frac{m}{r} - wr
\fe
is holomorphically-defect-factorized through a particular symmetry Wilson line $\cL$, which has two useful representations (among infinitely many) 
\ie
\cL_{-\pi(\frac{m}{r^2}+w, m+w r^2)} \sim \cL_{\pi(-\frac{m}{r^2}+w, m-w r^2)} \, ,
\fe
which are equivalent under $(w, m)$ units of spectral flow. Using the first representation, the defect partition function \eqref{DefectZToroidal} involves the sum
\ie
\sum_{m', w' \in \bZ} q^{\frac{1}{4}\left( \frac{m' - m}{r} + (w' - w)r \right)^2} \bar q^{\frac{1}{4}\left( \frac{m'}{r} - w'r \right)^2} \, .
\fe
The term $m' = w' = 0$ corresponds to the unique holomorphic-defect current algebra primary $\cD_{m, w}$, whereas the term $m' = m, \, w' = w$ corresponds to the unique anti-holomorphic one $\ocD_{m, w}$. These two defect current algebra primaries can be explicitly obtained via integration by parts. Using the first representation
\ie
\cL_{-\pi(\frac{m}{r^2}+w, m+w r^2)} \ = \ :\exp\left[-i (\frac{m}{r}+wr) \int_{z_1}^{z_2} dz \, \partial X(z) \right]: \, ,
\fe
integration by parts gives a holomorphic defect operator on one end
\ie
\label{LeftU1Defect}
\cD_{m,w}(z_1) =
:e^{i (\frac{m}{r}+wr) X_L(z_1)}: \, , \quad h = \frac{m^2}{r^2} + w^2 r^2 + 2mw \, , \quad \bar h = 0 \, .
\fe
Using the second representation
\ie
\cL_{\pi(-\frac{m}{r^2}+w, m-w r^2)} \ = \ :\exp\left[i (\frac{m}{r}-wr) \int_{\bar z_1}^{\bar z_2} d\bar z \, \bar\partial X(\bar z) \right]: \, ,
\fe
integration by parts gives an anti-holomorphic defect operator on the other end
\ie
\label{RightU1Defect}
\ocD_{m,w}(\bar z_2) =
:e^{i (\frac{m}{r}-wr) X_R(\bar z_2)}: \, , \quad h = 0 \, , \quad \bar h = \frac{m^2}{r^2} + w^2 r^2 - 2mw \, .
\fe
The exponential local operator is holomorphically-defect-factorized as
\ie
\cO_{m, w} = \cD_{m, w} \stackrel{\cL}{\text{---}} \ocD_{m, w} \, .
\fe

Let us check that the torus Regge limits are consistent with the above analysis. The torus two-point function of the exponential local operator \eqref{ExpOp} with its conjugate is 
\ie
\vev{\cO_{m,w}(z,\bar z) \cO_{-m,-w}(0)}_{{\rm T}^2(\tau, \bar\tau)} &=
\left(\partial_z \theta_1(0|\tau)\over \theta_1(z|\tau)\right)^{{1\over 2}p^2_{L,m,w}}\left(\partial_{\bar z} \theta_1(0|\bar \tau)\over \theta_1(\bar z|\bar \tau)\right)^{{1\over 2}p^2_{R,m,w}}
\\
&\quad\times{1\over |\eta(\tau)|^2}\sum_{m',w'}\Theta_{m,m',w,w'}(\tau,\bar\tau,z,\bar z) \, ,
\fe
where
\ie
\Theta_{m,m',w,w'}(\tau,\bar\tau,z,\bar z)=q^{{1\over 4}p_{L,m',w'}^2}\bar q^{{1\over 4}p_{R,m',w'}^2}e^{i \pi \left(p_{L,m',w'}p_{L,m,w}z-p_{R,m',w'}p_{R,m,w}\bar z\right)} \, .
\fe
Consider the spatial torus Regge limit. Under $z\to z+1$, we find
\ie
\vev{\cO_{m,w}(z+1,\bar z) \cO_{-m,-w}(0)}_{{\rm T}^2(\tau, \bar\tau)} &=
e^{{1\over 2}i \pi p_{L,m,w}^2} \, \left(\partial_z \theta_1(0|\tau)\over \theta_1(z|\tau)\right)^{{1\over 2}p^2_{L,m,w}} \left(\partial_{\bar z} \theta_1(0|\bar \tau)\over \theta_1(\bar z|\bar \tau)\right)^{{1\over 2}p^2_{R,m,w}}
\\
&\quad\times{1\over |\eta(\tau)|^2} \sum_{m',w'} e^{i \pi p_{L,m',w'}p_{L,m,w}} \, \Theta_{m,m',w,w'}(\tau,z) \, ,
\fe
where we have used
\ie
\theta_1(z+1|\tau) = -\theta_1(z|\tau) \, .
\fe
In the further $z, \bar z \to 0$ limit, we find
\ie
& \vev{\cO_{m,w}(z+1,\bar z) \cO_{-m,-w}(0)}_{{\rm T}^2(\tau, \bar\tau)} 
\\
&\to e^{{1\over 2}i \pi p_{L,m,w}^2} z^{-\frac12 p_{L,m,w}^2} \bar z^{-\frac12 p_{R,m,w}^2} \sum_{m',w'} e^{i \pi p_{L,m',w'}p_{L,m,w}} \, \frac{q^{\frac14 p_{L,m,w}^2} \bar q^{\frac14 p_{R,m,w}^2}}{|\eta(\tau)|^2} \, .
\fe
Stripping off the leading $z, \, \bar z$ divergence and the overall $e^{\frac12 i \pi p_{L,m,w}^2}$ factor corresponding to $e^{- 2 i \pi h}$ phase in \eqref{SpatialTorus}, the exponential operators $\cO_{m', w'}$ are transformed by the phases
\ie
e^{i \pi p_{L,m',w'} p_{L,m,w}} = e^{i \pi \left( \frac{m'}{r} + w'r \right) \left( \frac{m}{r} + wr \right)} 
= e^{i \pi \left[ m' \left( \frac{m}{r^2} + w \right) + w' \left( m + wr^2 \right) \right]} 
= e^{i (m' \theta_{\rm m} + w' \theta_{\rm w})} \, .
\fe
A modular S transform recovers the expected defect partition function \eqref{DefectZToroidal}. We could have also directly taken the temporal torus Regge limit to arrive at \eqref{DefectZToroidal}.

Let us comment on the Lorentzian dynamics. For the four-point function of exponential operators, \eqref{eqn:reggeGeneral} gives the conformal Regge limit at infinite boost
\ie
\label{rExp}
r[\cO_{m,w},\cO_{m',w'}]=e^{i \pi \left[ m' \left( \frac{m}{r^2} + w \right) + w' \left( m + wr^2 \right) \right]} \, .
\fe
Suppose one is interested in the four-point function of real operators, {\it i.e.} the cosine and sine operators
\ie
&\cO^{\cos}_{m,n}(z, \bar z)={1\over \sqrt{2}}(\cO_{m,w}(z, \bar z)+\cO_{-m,-w}(z, \bar z))=\sqrt{2}\cos(p_{\rm L} X_L(z)+p_{\rm R} X_R(\bar z)) \, ,
\\
&\cO^{\sin}_{m,n}(z, \bar z)={1\over \sqrt{2}i}(\cO_{m,w}(z, \bar z)-\cO_{-m,-w}(z, \bar z))=\sqrt{2}\sin(p_{\rm L} X_L(z)+p_{\rm R} X_R(\bar z)) \, ,
\fe
suitable combinations of \eqref{rExp} give
\ie
r[\cO^{\cos}_{m,w},\cO^{\cos}_{m',w'}] &= -r[\cO^{\cos}_{m,w},\cO^{\sin}_{m',w'}] = r[\cO^{\sin}_{m,w},\cO^{\sin}_{m',w'}]
\\
&=\cos\pi \left[ m' \left( \frac{m}{r^2} + w \right) + w' \left( m + wr^2 \right) \right] \, .
\fe

\subsection{Orbifold branch}
\label{Sec:S1Z2}

The ${\rm S}^1/\bZ_2$ partition function is
\ie
\label{ZOrbifold}
Z_{{\rm S}^1_r/\bZ_2}(\tau, \bar\tau) &= \frac12 \left( Z_{{\rm S}^1_r}(\tau, \bar\tau) + \frac{|\theta_3(\tau) \theta_4(\tau)|}{|\eta(\tau)|^2} + \frac{ |\theta_2(\tau) \theta_4(\tau)| + |\theta_2(\tau) \theta_3(\tau)| }{|\eta(\tau)|^2} \right) 
\\
&= \frac12 Z_{{\rm S}^1_r}(\tau, \bar\tau) + \left| \frac{\eta(\tau)}{\theta_2(\tau)} \right| + \left| \frac{\eta(\tau)}{\theta_3(\tau)} \right| + \left| \frac{\eta(\tau)}{\theta_4(\tau)} \right|
\, .
\fe
The first two terms enumerate the untwisted sector, and the latter two enumerate the twisted sector which is universal and independent of the radius $r$. At $c=1$, an irreducible module with primary weight $h = n^2$ for $n \in \bZ_{\ge0}$ has a null state at level $2n+1$, and $h = (n + \frac12)^2$ for $n \in \bZ_{\ge0}$ at level $2n+2$, so the degenerate characters are
\ie
\chi_{h = n^2}(\tau) = \frac{q^{n^2} - q^{(n+1)^2}}{\eta(\tau)} \, , \quad 
\chi_{h = (n+\frac12)^2}(\tau) = \frac{q^{(n+\frac12)^2} - q^{(n+\frac32)^2}}{\eta(\tau)} \, .
\fe
The untwisted sector can be written as
\ie
\label{S1Z2Untwisted}
Z_{{\rm S}^1_r/\bZ_2}^{\rm untwisted}(\tau, \bar\tau) &= \sum_{\substack{n, \bar n \in \bZ_{\ge0} \\ n - \bar n \in 2\bZ}} \chi_{h = n^2}(\tau) \bar\chi_{\bar h = \bar n^2}(\bar \tau) + \frac12 \left( Z_{{\rm S}^1_r}(\tau, \bar\tau) - \frac{1}{|\eta(\tau)|^2} \right) \, ,
\fe
where the first piece enumerates the degenerate Verma modules that are universal on the orbifold branch, and the second piece enumerates the rest including the cosine operators
\ie
\cO_{m, w}(z, \bar z) = \cos(p_{\rm L} X + p_{\rm R} \bar X) \, , \quad p_{L,R} = \frac{m}{r} \pm wr \, , \quad (m, w) \neq (0, 0) \, .
\fe
At irrational $r$ (not $r^2$), all cosine operators are non-degenerate, but at rational $r$, some cosine operators become degenerate. For the simplicity of discussion, we ignore the subtlety at rational $r$, and always refer to the states counted by the first piece in \eqref{S1Z2Untwisted} as degenerate Verma modules, and to the states counted by the second piece as cosine operators and their descendants.

On the orbifold branch there is a universal $D_4$ symmetry, as reviewed in Appendix~\ref{Sec:D4}. The five order-two elements correspond to the symmetry lines
\ie
\eta_{\rm m} \, , \quad \eta_{\rm w} \, , \quad \eta \equiv 
\eta_{\rm m} \, \eta_{\rm w} \, \eta_{\rm m} \, \eta_{\rm w} \, , \quad \eta_m' \equiv \eta_{\rm w} \, \eta_{\rm m} \, \eta_{\rm w} \, , \quad \eta_w' \equiv \eta_{\rm m} \, \eta_{\rm w} \, \eta_{\rm m} \, .
\fe
From the orbifolding perspective, $\eta_{\rm m}$ and $\eta_{\rm m}'$ descend from the momentum $\bZ_2$ symmetry line in the ${\rm S}^1$ theory, $\eta_{\rm w}$ and $\eta_{\rm w}'$ from the winding $\bZ_2$, and $\eta$ is the emergent $\bZ_2$ symmetry line that assigns $+1$ charge to the untwisted sector and $-1$ charge to the twisted sector.\footnote{That a single $\bZ_2$ symmetry line in the ${\rm S}^1$ theory descends to multiple symmetry lines in the ${\rm S}^1/\bZ_2$ orbifold theory is due to the non-uniqueness of symmetry action on the twisted sector.
}

At arbitrary radius $r$, there is a continuous family of unoriented topological defect lines --- which we call {\it cosine} lines --- with ${\vev{\cL}_{\bR^2}} = 2$. They descend from the orientation-reversal-invariant combinations of the ${\rm U(1)}_{\rm m} \times {\rm U(1)}_{\rm w}$ symmetry Wilson lines in the ${\rm S}^1$ theory,\footnote{This combination is not simple before orbifold, but can become simple after.
}
\ie
\cL_{(\theta_{\rm m}, \theta_{\rm w})}^{{\rm S}^1/\bZ_2} = \cL_{(\theta_{\rm m}, \theta_{\rm w})}^{{\rm S}^1} + \cL_{-(\theta_{\rm m}, \theta_{\rm w})}^{{\rm S}^1} \, . 
\fe
Cosine lines are labeled by a pair of quantum numbers $(\theta_{\rm m}, \theta_{\rm w})$, which not only have periodicity $(2\pi, 0)$ and $(0, 2\pi)$ due to integer spectral flow, but are also identified under $(\theta_{\rm m}, \theta_{\rm w}) \to - (\theta_{\rm m}, \theta_{\rm w})$. The fusion of cosine lines gives
\ie
\cL_{(\theta_{\rm m}, \theta_{\rm w})} \, \cL_{(\theta_{\rm m}', \theta_{\rm w}')} = \cL_{(\theta_{\rm m}+\theta_{\rm m}', \theta_{\rm w}+\theta_{\rm w}')} + \cL_{(\theta_{\rm m}-\theta_{\rm m}', \theta_{\rm w}-\theta_{\rm w}')} \, .
\fe
They act on the nontrivial cosine operators by
\ie
\label{S1Z2}
\widehat\cL_{(\theta_{\rm m}, \theta_{\rm w})} (\cO_{m', w'}) = 2 \cos( m' \theta_{\rm m} + w' \theta_{\rm w} ) \, \cO_{m', w'} \, ,
\fe
on the degenerate Verma modules by a factor of 2, and annihilate the twisted sector states. For any pair of positive integers $(N_m, N_w)$, there is a subring generated by finitely many objects
\ie
\left\{ \cL_{(\theta_{\rm m}, \theta_{\rm w})} \mid \theta_{\rm m} \in \bZ/N_m \, , \, \theta_{\rm w} \in \bZ/N_w \right\} \, .
\fe

A cosine line can be either simple or non-simple; in the latter case it must be the direct sum of two symmetry lines.\footnote{The quantum dimension of a cosine line is $\vev{\cL_{(\theta_{\rm m}, \theta_{\rm w})}}_{{\rm S}^1 \times \bR} = |\vev{\cL_{(\theta_{\rm m}, \theta_{\rm w})}}_{\bR^2}| = 2$. In a compact conformal field theory, every topological defect line has quantum dimension $\ge 1$, and $= 1$ if and only if the topological defect line is a symmetry line \cite{Chang:2018iay}. Since the quantum dimension is additive under direct sum, the claim follows.
} 
Because the orbifold theory does not have any continuous symmetry except at ${\rm S}^1_{r = 1}/\bZ_2 = {\rm S}^1_{r=2}$, generic cosine lines are simple. However, for
\ie
\label{NonsimpleTheta}
(\theta_{\rm m}, \theta_{\rm w}) = (0,0) \, , \, (\pi, 0) \, , \, (0, \pi) \, , \, (\pi, \pi) \, ,
\fe
because the original $\cL_{(\theta_{\rm m}, \theta_{\rm w})}^{{\rm S}^1}$ was already unoriented, one expects $\cL_{(\theta_{\rm m}, \theta_{\rm w})}^{{\rm S}^1/\bZ_2}$ to be non-simple.\footnote{See \cite{Chang:2018iay} for this phenomenon in the $\bZ_2$ orbifold relation between tetra-critical Ising and three-state Potts.
}
They are the following direct sums of $D_4$ symmetry lines:
\ie
&\cL_{(0, 0)}^{{\rm S}^1/\bZ_2} = \cI + \eta \, , 
\quad \cL_{(\pi, 0)}^{{\rm S}^1/\bZ_2} = \eta_{\rm m} + \eta_{\rm m}' \, , \quad \cL_{(0, \pi)}^{{\rm S}^1/\bZ_2} = \eta_{\rm w} + \eta_{\rm w}' \, , 
\\
&\cL_{(\pi, \pi)}^{{\rm S}^1/\bZ_2} = \eta_{\rm m} \, \eta_{\rm w} + \eta_{\rm w} \, \eta_{\rm m} \, .
\fe
In the rest of this section, the label ${\rm S}^1/\bZ_2$ will be suppressed.

The torus partition function twisted by $\cL_{(\theta_{\rm m}, \theta_{\rm w})}$ (in the temporal direction) is
\ie
& Z^{\cL_{(\theta_{\rm m}, \theta_{\rm w})}}(\tau, \bar\tau)
= 
\left(\sum_{\substack{m \in \bZ \\ w\in\bZ_{>0}}}+\sum_{\substack{m \in \bZ_{>0} \\ w=0}}\right) \, 2 \cos(\theta_{\rm m} m + \theta_{\rm w} w) \, \chi_{ \frac{p_{\rm L}^2}{4}}(\tau) \overline{\chi_{  \frac{p_{\rm R}^2}{4}}(\tau)}
\\
&
\hspace{2.25in} + 
2\sum_{n, \bar n \in \bZ_{\ge0} }{1+(-1)^{n+\bar n}\over 2} \chi_{h = n^2}(\tau) \bar\chi_{\bar h = \bar n^2}(\bar \tau)
\\
&= \frac{1}{|\eta(\tau)|^2} \sum_{m, w \in \bZ} e^{i \theta_{\rm m} m + i \theta_{\rm w} w} \, {q^{\frac{p_{\rm L}^2}{4}} \bar q^{\frac{p_{\rm R}^2}{4}}} + \frac{|\theta_3(\tau) \theta_4(\tau)|}{|\eta(\tau)|^2} \, , \quad p_{L,R} = \frac{m}{r} \pm wr \, .
\fe
The defect partition function of $\cL_{(\theta_{\rm m}, \theta_{\rm w})}$ is obtained by a modular S transform to be
\ie
\label{DefectZ}
Z_{\cL_{(\theta_{\rm m}, \theta_{\rm w})}}(\tau, \bar\tau) = \sum_{m, w \in \bZ} \frac{q^{\frac{p_{\rm L}^2}{4}} \bar q^{\frac{p_{\rm R}^2}{4}}}{|\eta(\tau)|^2} + \frac{|\theta_2(\tau) \theta_3(\tau)|}{|\eta(\tau)|^2}
\, , \quad p_{L,R} = \frac{m + \frac{\theta_{\rm w}}{2\pi}}{r} \pm (w + {\textstyle \frac{\theta_{\rm m}}{2\pi}})r \, ,
\fe
where
\ie
\sqrt{\frac{\theta_2(\tau) \theta_3(\tau)}{2}} = \sum_{n \in \bZ_{\ge0}} q^{\frac{(2n+1)^2}{16}}
\fe
is a $q$-series with positive integer coefficients.

Consider a cosine line $\cL_{\pi ({m \over r^2} + w, m + w r^2)}$ with $m, w \in \bZ$. Its defect partition function involves the sum
\ie
\sum_{m', w' \in \bZ} q^{\frac14\left(\frac{m' + m}{r} + (w' + w)r\right)^2} \bar q^{\frac14\left(\frac{m'}{r} - w' r\right)^2} \, .
\fe
The $m' = w' = 0$ term corresponds to a holomorphic defect primary $\cD_{m,w}$ of weight $h = \frac14 \left( \frac{m}{r} + wr \right)^2$, and the $m' = -m, \, w' = -w$ terms corresponds to an anti-holomorphic defect primary $\ocD_{m,w}$ of weight $\bar h = \frac14 \left( \frac{m}{r} - wr \right)^2$. The cosine operator is holomorphically-defect-factorized as
\ie
\cO_{m, w} = \cD_{m, w} \stackrel{\cL}{\text{---}} \ocD_{m, w} \, , \quad \cL = \cL_{\pi ({m \over r^2} + w, m + w r^2)} \, .
\fe
In particular, $\cO_{m, w}$ has charge $2 \cos( \pi \frac{m^2}{r^2} + \pi w^2 r^2 )$ under the line $\cL_{\pi ({m \over r^2} + w, m + w r^2)}$ it factorizes through.

What about operators in the twisted sector? Consider the twisted sector ground states ${\cal E}_i$ of weight $(\frac{1}{16}, \frac{1}{16})$, where $i = 1, 2$ label the two fixed points. When rational, by the discussion in Section~\ref{Sec:Rational}, in some (possibly complex) basis, they must be holomorphically-defect-factorized. Let ${\cal E}$ denote an operator in such a basis, then
\ie
{\cal E} = \cD \stackrel{\cL_{{\cal E}}}{\text{---}} \ocD \, ,
\fe
and the defect partition function of $\cL_{{\cal E}}$ is obtainable from a limit of the twist field two-point function on the torus. In Appendix~\ref{Sec:RationalOrbifold}, we examine special rational points on the orbifold branch and identify $\cL_{{\cal E}}$ as Verlinde lines. However, we can characterize $\cL_{{\cal E}}$ in a more universal fashion by computing the torus two-point function of twist fields in the temporal torus Regge limit. This computation is carried out in Appendix~\ref{Sec:ReggeS1Z2}, using the formulae of \cite{Hamidi:1986vh,Dixon:1986qv,Miki:1987mp,Dijkgraaf:1987vp} for general correlators in orbifolds. Moreover, the fusion rule of $\cL_{\cal E}$ or $\widetilde\cL_{\cal E}$ with its orientation reversal is computed in Appendix \ref{LLfusion}.

Interestingly, we find a clear distinction between rational and irrational theories:
\begin{enumerate}
\item If $r^2 = u/v$ is rational with $u, v$ coprime, then the strong holomorphic-defect-factorization criterion is satisfied, and the planar loop expectation value of $\cL_{{\cal E}}$ is ${\vev{\cL_{{\cal E}}}_{\bR^2}} = \sqrt{uv}$. The loop-normalized defect partition function is given in \eqref{eqn:rDPF}. 

When $u$ is even, the fusion rule is
\ie\label{eqn:LLrationaleven}
\cL_{{\cal E}} \, \ocL_{{\cal E}} = \cI+\eta_{\rm m}+\sum_{\substack{m=0\\m\in2\bZ}}^{2u-2} \sum^{v-1}_{\substack{w=2\\m\in2\bZ}} \cL_{-\pi(\frac{m}{r^2}+w, m+w r^2)}+\sum_{\substack{m=2\\m\in2\bZ}}^{u-2} \cL_{-\pi(\frac{m}{r^2}, m)}\, .
\fe
When $u$ and $v$ are both odd, the fusion rule is
\ie\label{eqn:LLrationalodd}
\cL_{{\cal E}} \, \ocL_{{\cal E}} = \cI+\sum_{\substack{m=0\\m\in2\bZ}}^{2u-2} \sum^{v-1}_{\substack{w=2\\m\in2\bZ}} \cL_{-\pi(\frac{m}{r^2}+w, m+w r^2)}+\sum_{\substack{m=2\\m\in2\bZ}}^{u-1} \cL_{-\pi(\frac{m}{r^2}, m)}\, .
\fe
~
\item If $r^2$ is irrational, then the strong holomorphic-defect-factorization criterion fails, but the weak criterion is satisfied. More precisely, the so-obtained loop-normalized defect partition function is
\ie\label{eqn:IRDPF}
Z_{\widetilde\cL_{\cal E}}
(\tau, \bar\tau)
&= \frac{1}{|\eta(\tau)|^2} \sum_{n=0}^\infty \int_0^\infty dp \left( q^{\frac{(2n+1)^2}{16}} \bar q^{\frac{p^2}{4}} + q^{\frac{p^2}{4}}\bar q^{\frac{(2n+1)^2}{16}} \right) \, ,
\fe
which in fact does not depend on $r$. The defect spectrum $\cH_{\widetilde\cL_{\cal E}}$ is continuous, hence $\widetilde\cL_{\cal E}$ is a non-compact TDL.

The loop-normalized torus partition function with $\widetilde\cL_{{\cal E}}$ wrapped along the spatial direction is
\ie
\label{ZLS1Z2Spatial}
Z^{\widetilde\cL_{{\cal E}}}(\tau, \bar\tau) &= \sum_{\substack{n, \bar n = 0 \\ n - \bar n \in 2\bZ}}^\infty \, (-)^n \chi_{h=n^2}(\tau) \bar\chi_{\bar h = \bar n^2}(\bar\tau) \, ,
\fe
indicating that $\widetilde\cL_{{\cal E}}$ annihilates all non-degenerate modules, and acts on the degenerate modules by a sign. 

The fusion rule is 
\ie\label{eqn:LLirrational}
\widetilde\cL_{\cal E} \, \widetilde\ocL_{\cal E}={1\over 2}\int^{2\pi}_0{d\theta_{\rm w}d\theta_{\rm m}\over (2\pi)^2}\cL_{(\theta_{\rm m}, \theta_{\rm w})} \, .
\fe

\end{enumerate}

The TDL $\widetilde\cL_{\cal E}$ in the theory with irrational $r^2$ is a non-compact TDL characterized in Section~\ref{Sec:Non-compact}, and belongs to a more general {\it TDL category}. In fact, many of the structures of this more general category could be understood by taking limits of fusion categories. For any irrational $r^2$, consider a sequence of coprime integers $(u_n, v_n)$ for $n=1,\,2,\,\cdots$, such that in the $n\to\infty$ limit, $u_n / v_n$ converges to $r^2$. The TDL $\widetilde\cL_{{\cal E}}$ in irrational theory could be obtained by taking the $n \to \infty$ limit of the $\widetilde\cL_{{\cal E}}$ in rational theory. Both the defect partition function for $\widetilde\cL_{{\cal E}}$ and the fusion rule for $\widetilde\cL_{\cal E} \, \widetilde\ocL_{\cal E}$ in the irrational theories indeed arise as limits of those in rational theories, as we presently explain.

First, in Appendix~\ref{Sec:ReggeS1Z2}, we find that the sequence of defect partition functions $Z_{\cL_{{\cal E}},\,{\rm S}^1_r/\bZ_2}$ for $r^2={u_n / v_n}$ in the $n\to\infty$ limit reproduces the defect partition function \eqref{eqn:IRDPF} at irrational points. Note that there are infinitely many different sequences of coprime integers $(u_n, v_n)$ whose ratios $u_n / v_n$ converge to the same irrational number. At first sight, it is not obvious that the corresponding sequences of the defect partition functions all converge to the same result. However, as we find in \eqref{eqn:rDPF}, the defect partition function $Z_{\cL_{{\cal E}},\,{\rm S}^1_r/\bZ_2} $ depends only on the product $uv$. Hence, the $n\to\infty$ limit coincides with the $uv\to\infty$ limit, and the limits of all possible sequences agree. Furthermore, the result does not depend on $r$. 

Second, consider the sequence of fusion rules \eqref{eqn:LLrationaleven} or \eqref{eqn:LLrationalodd} with $r^2 = {u_n / v_n}$. Divide by $u_n v_n$ on both sides of the fusion rule, and change the Cardy normalized $\cL_{\cal E}$ to loop-normalized $\widetilde\cL_{\cal E}$. Now, in the $n \to \infty$ limit, the sequence of fusion rules converges to the fusion rule \eqref{eqn:LLirrational} for the non-compact TDL $\widetilde\cL_{\cal E}$ at irrational points.

To end, let us remark on the Lorentzian dynamics of twist fields. According to \eqref{ZLS1Z2Spatial}, at irrational points, the Lorentzian four-point function exhibits transparent behavior for degenerate primaries with even $n$, refractive behavior for degenerate primaries with odd $n$, and opaque behavior for all non-degenerate primary $\phi$ (we have $r = 1, -1, 0$ in the three cases, respectively).

\section{Summary and discussion}
\label{Sec:Conclusion}

In this paper, we explicated the following aspects of two-dimensional conformal field theory.
\begin{enumerate}
\item We presented a purely Euclidean portrayal of treating the coordinates $z, \, \bar z$ of a local operator as independent complex variables. The local operator can often be factorized into a pair of holomorphic and anti-holomophic defect operators, connected by a topological defect line.
\item We proposed that local operators can be factorized not only through {\it simple} topological defect lines, but also through {\it non-compact} topological defect lines that have continua in their defect spectra. We extended the categorical framework to include such topological defect lines.
\item Based on factorization, we derived relations among correlation functions of local operators, correlation functionals of defect operators, and the $F$-symbols characterizing the splitting and joining of topological defect lines.
\item We proposed a procedure for discovering topological defect lines. This point warrants further remarks. A topological defect line is traditionally characterized by a map on local operators satisfying certain conditions --- including but not limited to the commutativity with the Virasoro algebra and the consistency of the defect partition function obtained by the modular S transform. From this perspective, a topological defect line is a solution to a set of consistency conditions, rather than something computed directly from the data of local operators. In this paper, by considering the conformal Regge limit, we have shown how the four-point function or torus two-point function directly generates the defining data for topological defect lines.
\item We characterized aspects of the conformal Regge limit by fundamental properties of topological defect lines. In particular, whether the bulk scattering is transparent, refractive or opaque \cite{Caron-Huot:2020ouj} is dictated by the action of topological defect lines on local operators. The proof of the unitarity bound on the opacity by \cite{Caron-Huot:2017vep} gave us Corollary~\ref{SpectralRadius}, which says that the spectral radius of any factorizing topological defect line is always given by the loop expectation value. We also give a complementing argument for the spectral radius formula \eqref{eqn:SpectralRadius}, with additional caveats but without assuming that the topological defect line is factorizing, by utilizing the representation theory of the fusion rule.
\item Applying our procedure for discovering topological defect lines, we obtained a unified description of the topological defect line through which the twist field factorizes in the $c = 1$ free boson orbifold theory. The result at irrational points suggests the existence of non-compact topological defect lines even in compact theories.
\end{enumerate}

Consider a local operator $\cO$ that is holomorphically-defect-factorized through a topological defect line $\cL$. As shown in Table~\ref{cases}, there are three logical possibilities regarding the finiteness of highest-weight operators (with respect to the maximally extended chiral algebra) in the $\cO \times \cO$ OPE and whether a Cardy-normalized $\cL$ is well-defined, such that the fusion $\cL \, \cL$ gives a direct sum. Most of our examples, including all local operators in rational theories and the exponential or cosine operators in the $c=1$ free boson theory, fall into Scenario (a). The twist field in the free boson orbifold theory falls into scenario (c). We are not aware of any realization of Scenario (b).

\begin{table}[t]
\centering
\begin{tabular}{c|c|c}
 & $\cL \, \cL$ finite direct sum & Cardy-normalized $\cL$ not well-defined
\\\hline
$\cO \times \cO$ finite & (a) & (b)
\\
$\cO \times \cO$ not finite & impossible & (c)
\end{tabular}
\caption{For a local operator $\cO$ that is holomorphically-defect-factorized through a topological defect line $\cL$, the logical possibilities regarding the finiteness of highest-weight operators (with respect to the maximally extended chiral algebra) in the $\cO \times \cO$ OPE and whether a Cardy-normalized $\cL$ is well-defined, such that the fusion $\cL \, \cL$ gives a direct sum.}
\label{cases}
\end{table}

Does every conformal field theory admit a (generally complex) basis of local operators in which every local operator is holomorphically-defect-factorized?  The answer is negative in the strong sense of Definition~\ref{Strong}, since it is violated at irrational points in the free boson orbifold theory, where the topological defect line through which the twist field hypothetically factorizes exhibits a continuous spectrum in the defect Hilbert space, violating the usual Cardy condition. However, in the more general weak sense of Definition~\ref{Weak} that allows factorization through non-compact topological defect lines, the posed question becomes more intriguing. For irrational theories embedded in a conformal manifold with ``dense enough'' rational points, such non-compact topological defect lines may be regarded as the limit of sequences of Verlinde lines.\footnote{See \cite{Benjamin:2020flm} and references within for discussions on the degree of prevalence of rational points on conformal manifolds.
}
Under this generalized notion, we conjecture that every conformal field theory has a holomorphically-defect-factorized basis of local operators.

The close connection between the opacity bound and the spectral radius formula illuminates a virtue of this conjecture. The Perron-Frobenius theorem allows us to prove the spectral radius formula for simple lines. Moreover, as noted in Appendix~\ref{Sec:Bounded}, generalizations of the Perron-Frobenius theorem to integral bounded operators extends the scope of the spectral radius formula to non-compact topological defect lines. These arguments complement the proof using the opacity bound of \cite{Caron-Huot:2017vep}. Finally, we comment that a similar bound on the four-point function \eqref{eqn:4ptfonC} in the light-cone limit, $(1-z)\to e^{2\pi i}(1-z)$ with $\bar z$ fixed and then $\bar z\to 0$ with $z$ fixed, was derived from causality constraints in \cite{Hartman:2015lfa}.

\section*{Acknowledgements}

We are grateful to David Simmons-Duffin, Hirosi Ooguri and Xi Yin for discussions and helpful comments. CC thanks the hospitality of National Taiwan University. CC is partly supported by National Key R\&D Program of China (NO. 2020YFA0713000). YL is supported by the Simons Collaboration Grant on the Nonperturbative Bootstrap, the Sherman Fairchild Foundation, and by the U.S. Department of Energy, Office of Science, Office of High Energy Physics, under Award Number DE-SC0011632.

\appendix

\section{Defect crossing implies local crossing}
\label{Sec:Crossing}

The crossing symmetry of a four-point function of holomorphic defect operators is the $F$-move
\ie
&
\la\,
\begin{gathered}
\begin{tikzpicture}[scale=.75]
\draw [line,->-=.56] (-1,0) -- (-.5,0) node [above] {$\cL_5$} -- (0,0);
\draw [line,-<-=.55] (-1,0) -- (-1.5,.87) \dotsol{above left}{$\cD_1(0)$};
\draw [line,-<-=.55] (-1,0) -- (-1.5,-.87) \dotsol{below left}{$\cD_2(z)$};
\draw [line,-<-=.55] (0,0) -- (.5,-.87) \dotsol{below right}{$\cD_3(1)$};
\draw [line,-<-=.55] (0,0) -- (.5,.87) \dotsol{above right}{$\cD_4'(\infty)$};
\end{tikzpicture}
\end{gathered}
\,\ra
\quad = \quad
\la\,
\begin{gathered}
\begin{tikzpicture}[scale=.75]
\draw [line,-<-=.56] (0,-1) -- (0,-.5) node [right] {$\cL_6$} -- (0,0);
\draw [line,->-=.55] (0,-1) -- (.87,-1.5) \dotsol{below right}{$\cD_3(1)$};
\draw [line,->-=.55] (0,-1) -- (-.87,-1.5) \dotsol{below left}{$\cD_2(z)$};
\draw [line,->-=.55] (0,0) -- (-.87,.5) \dotsol{above left}{$\cD_1(0)$};
\draw [line,->-=.55] (0,0) -- (.87,.5) \dotsol{above right}{$\cD_4'(\infty)$};
\end{tikzpicture}
\end{gathered}
\,\ra
~\circ~
(F^{\cL_1, \cL_2, \cL_3}_{\ocL_4})_{\cL_5, \cL_6}
\fe
decomposed into properly normalized $s$- and $t$-channel Virasoro blocks times defect three-point correlation functionals (bi-covectors),\footnote{A standard normalization for a block is to require unit coefficient for the leading coefficient in the cross ratio expansion
\ie
\cF{\scriptsize
\begin{bmatrix}
h_{\cD_1} & h_{\cD_4}
\\
h_{\cD_2} & h_{\cD_3}
\end{bmatrix}
}_{h_{\cD_5}}^{c}(z)
= z^{h_{\cD_5} - h_{\cD_1} - h_{\cD_2}} \left( 1 + \cO(z) \right) \, .
\fe
} 
\ie
\label{DefectCrossing}
\sum_{\cD_5} \, 
&
\cF{\scriptsize
\begin{bmatrix}
h_{\cD_1} & h_{\cD_4}
\\
h_{\cD_2} & h_{\cD_3}
\end{bmatrix}
}_{h_{\cD_5}}^{c}(z)
\times C_{\cD_1, \cD_2, \cD_5^\dag} \otimes C_{\cD_5, \cD_3, \cD_4}
\\
&= \ \sum_{\cL_6} \, \sum_{\cD_6} \, 
\cF{\scriptsize
\begin{bmatrix}
h_{\cD_2} & h_{\cD_1}
\\
h_{\cD_3} & h_{\cD_4}
\end{bmatrix}
}_{h_{\cD_6}}^{c}(1-z)
\times C_{\cD_2, \cD_3, \cD_6^\dag} \otimes C_{\cD_1, \cD_6, \cD_4} \circ (F^{\cL_1, \cL_2, \cL_3}_{\ocL_4})_{\cL_5, \cL_6} \, ,
\fe
where $c$ is the holomorphic central charge. The sums $\sum_{\cD_5}$ and $\sum_{\cD_6}$ are over holomorphic Virasoro primaries in the defect Hilbert spaces $\cH_{\cL_5}$ and $\cH_{\cL_6}$. When the theory has an extended chiral algebra, one could decompose the defect four-point function with respect to the extended chiral algebra. The crossing equation takes the same form as \eqref{DefectCrossing}, but with $\cF$ representing the chiral algebra blocks that may depend on other quantum numbers beside $h$, and the sums $\sum_{\cD_5}$ and $\sum_{\cD_6}$ are over holomorphic highest-weight operators of the chiral algebra in the defect Hilbert spaces $\cH_{\cL_5}$ and $\cH_{\cL_6}$.

In rational conformal field theory, the defect Hilbert space of a simple topological defect line $\cL$ projected onto the subspace of holomorphic operators is an irreducible module of the maximally extended chiral algebra. In other words, there is a single highest-weight defect operator $\cD_i$ for each $\cH_{\cL_i}$, and hence, each defect four-point correlation functional is equal to a single chiral algebra block composed with the appropriate three-point defect correlation functionals. One can always trivialize the defect three-point correlation functionals by a special choice of basis junction vectors. This has two complementary ramifications. First, the formula \eqref{3pt} for the three-point coefficients of local primary operators now only involves fusion categorical quantities, 
and the holomorphic defect four-point crossing equation \eqref{DefectCrossing} reads simply
\ie
\cF{\scriptsize
\begin{bmatrix}
h_{\cD_1} & h_{\cD_4}
\\
h_{\cD_2} & h_{\cD_3}
\end{bmatrix}
}_{h_{\cD_5}}^{c}(z)
= \ \sum_{\cL_6} \, 
\cF{\scriptsize
\begin{bmatrix}
h_{\cD_2} & h_{\cD_1}
\\
h_{\cD_3} & h_{\cD_4}
\end{bmatrix}
}_{h_{\cD_6}}^{c}(1-z)
\times (F^{\cL_1, \cL_2, \cL_3}_{\ocL_4})_{\cL_5, \cL_6} \, .
\fe
Hence the nontrivial dynamical data is solved if one could determine the explicit values of the $F$-symbols in this special basis that trivializes the defect three-point correlation functionals. However, actually finding such a basis requires knowing the explicit blocks, for which one must resort to solving the null state decoupling equation \cite{Knizhnik:1984nr} or the Wronskian method \cite{Mukhi:2017ugw}. Moreover, as demonstrated in the example of Ising in Section~\ref{Sec:Ising}, the $F$-symbols in such a basis are rather complicated.

The four-point function of holomorphically-defect-factorized local operators can be evaluated as follows. In the $s$-channel,
\ie
\label{SChannel}
& \la \cO_1(z_1, \bar z_1) \, \cO_2(z_2, \bar z_2) \, \cO_3(z_3, \bar z_3) \, \cO_4(z_4, \bar z_4) \ra
\\
&= \, \sqrt{\textstyle \prod_{i=1}^4 {\vev{\cL_i}_{\bR^2}}}
~
\sum_{\cL, \cL', \cL''}
\,
\la\,
\begin{gathered}
\begin{tikzpicture}
\begin{scope}
\draw [line,->-=.3,-<-=.8] (0,0) ++(90:.5) \dotsol{left}{$\cD_1(z_1)$} arc (90:-90:.5) \dotsol{left}{$\cD_2(z_2)$};
\draw [line,->-=.51,-<-=.8] (0,-.67) ++ (-90:.83) \dotsol{left}{$\cD_3(z_3)$} arc (-90:53:.83);
\draw [line,->-=.55,-<-=.85] (0,-1.4) ++ (-90:1.13) \dotsol{left}{$\cD_4(z_4)$} arc (-90:43:1.13);
\end{scope}
\begin{scope}[xscale = -1, shift = {(-3.5,0)}]
\draw [line,-<-=.3,->-=.8] (0,0) ++(90:.5) \dotemp{right}{$\ocD_1(\bar z_1)$} arc (90:-90:.5) \dotemp{right}{$\ocD_2(\bar z_2)$};
\draw [line,-<-=.51,->-=.8] (0,-.67) ++ (-90:.83) \dotemp{right}{$\ocD_3(\bar z_3)$} arc (-90:53:.83);
\draw [line,-<-=.55,->-=.85] (0,-1.4) ++ (-90:1.13) \dotemp{right}{$\ocD_4(z_4)$} arc (-90:43:1.13);
\end{scope}
\draw [line,->-=.55] (1.13,-1.4) -- (1.75,-1.4) node [above] {$\cL''$} -- (2.37,-1.4);
\node at (1,-.25) {$\cL$};
\node at (2.5,-.25) {$\cL$};
\node at (1.35,-.75) {$\cL'$};
\node at (2.25,-.75) {$\cL'$};
\end{tikzpicture}
\end{gathered}
\,\ra
\\
&\hspace{.5in} \circ
\Big[(F^{\cL', \ocL', \ocL_4}_{\ocL_4})_{\cI, \ocL''}~(1_{\cL', \ocL', \cI}, 1_{\cI, \ocL_4, \cL_4})
\otimes
(F^{\cL, \ocL, \ocL_3}_{\ocL_3})_{\cI, \ocL'}~(1_{\cL, \ocL, \cI}, 1_{\cI, \ocL_3, \cL_3})
\\
&\hspace{3in} \otimes
(F^{\cL_1, \ocL_1, \ocL_2}_{\ocL_2})_{\cI, \ocL}~(1_{\cL_1, \ocL_1, \cI}, 1_{\cI, \ocL_2, \cL_2})\Big]
\\
~
\\
&= \, \frac{\sqrt{\textstyle \prod_{i=1}^4 {\vev{\cL_i}_{\bR^2}}}}{{\vev{\cL_4}_{\bR^2}}} \sum_\cL \, 
\la\,
\begin{gathered}
\begin{tikzpicture}[scale=.75]
\draw [line,->-=.6] (-1,0) -- (-.5,0) node [above] {$\cL$} -- (0,0);
\draw [line,-<-=.55] (-1,0) -- (-1.5,.87) \dotsol{above left}{$\cD_1(z_1)$};
\draw [line,-<-=.55] (-1,0) -- (-1.5,-.87) \dotsol{below left}{$\cD_2(z_2)$};
\draw [line,-<-=.55] (0,0) -- (.5,-.87) \dotsol{below right}{$\cD_3(z_3)$};
\draw [line,-<-=.55] (0,0) -- (.5,.87) \dotsol{above right}{$\cD_4(z_4)$};
\end{tikzpicture}
\end{gathered}
\,\ra
\la\,
\begin{gathered}
\begin{tikzpicture}[scale=.75]
\draw [line,->-=.6] (-1,0) -- (-.5,0) node [above] {$\cL$} -- (0,0);
\draw [line,->-=.55] (-1,0) -- (-1.5,.87) \dotemp{above left}{$\ocD_4(\bar z_4)$};
\draw [line,->-=.55] (-1,0) -- (-1.5,-.87) \dotemp{below left}{$\ocD_3(\bar z_3)$};
\draw [line,->-=.55] (0,0) -- (.5,-.87) \dotemp{below right}{$\ocD_2(\bar z_2)$};
\draw [line,->-=.55] (0,0) -- (.5,.87) \dotemp{above right}{$\ocD_1(\bar z_1)$};
\end{tikzpicture}
\end{gathered}
\,\ra
\\
&\hspace{.5in} \circ\Big[
(F^{\cL, \ocL, \ocL_3}_{\ocL_3})_{\cI, \cL_4}~(1_{\cL, \ocL, \cI}, 1_{\cI, \ocL_3, \cL_3})
\otimes
(F^{\cL_1, \ocL_1, \ocL_2}_{\ocL_2})_{\cI, \ocL}~(1_{\cL_1, \ocL_1, \cI}, 1_{\cI, \ocL_2, \cL_2})\Big]
\, .
\fe
By performing block expansions on the defect four-point functions, and using \eqref{3pt} and \eqref{Theta}, we recover with the usual $s$-channel conformal block expansion for local operators,
\ie
\label{SLocal}
& \la \cO_1(z_1, \bar z_1) \, \cO_2(z_2, \bar z_2) \, \cO_3(z_3, \bar z_3) \, \cO_4(z_4, \bar z_4) \ra 
\\
&= \,
\frac{\sqrt{\prod_{i=1}^4 {\vev{\cL_i}_{\bR^2}}}}{{\vev{\cL_4}_{\bR^2}}} \, \sum_\cL \, 
\sum_{\cD \in \cH_\ocL^{\bar h = 0}} \sum_{\ocD \in \cH_\cL^{h = 0}} \,
\cF{\scriptsize
\begin{bmatrix}
h_{\cD_1} & h_{\cD_4}
\\
h_{\cD_2} & h_{\cD_3}
\end{bmatrix}
}_{h_{\cD}}^{c}(z)
\,
\cF{\scriptsize
\begin{bmatrix}
\bar h_{\ocD_1} & \bar h_{\ocD_4}
\\
\bar h_{\ocD_2} & \bar h_{\ocD_3}
\end{bmatrix}
}_{\bar h_{\ocD}}^{\bar c}(\bar z)
\\
&\hspace{.5in}\times C_{\cD_1, \cD_2, \cD^\dag} \otimes C_{\cD_3, \cD_4, \cD} \otimes C_{\ocD_1, \ocD^\dag, \ocD_2} \otimes C_{\ocD_3, \ocD, \ocD_4}
\\
&\hspace{1in}\circ\Big[
(F^{\cL, \ocL, \ocL_3}_{\ocL_3})_{\cI, \cL_4}~(1_{\cL, \ocL, \cI}, 1_{\cI, \ocL_3, \cL_3})
\otimes
(F^{\cL_1, \ocL_1, \ocL_2}_{\ocL_2})_{\cI, \ocL}~(1_{\cL_1, \ocL_1, \cI}, 1_{\cI, \ocL_2, \cL_2})\Big] 
\\
&= \,
\sum_\cO \, 
\cF{\scriptsize
\begin{bmatrix}
h_{\cO_1} & h_{\cO_4}
\\
h_{\cO_2} & h_{\cO_3}
\end{bmatrix}
}_{h_{\cO}}^{c}(z)
\,
\cF{\scriptsize
\begin{bmatrix}
\bar h_{\cO_1} & \bar h_{\cO_4}
\\
\bar h_{\cO_2} & \bar h_{\cO_3}
\end{bmatrix}
}_{\bar h_{\cO}}^{\bar c}(\bar z)
\, C_{\cO_1, \cO_2, \cO} \, C_{\cO_3, \cO_4, \cO} \, .
\fe
Similarly, in the $t$-channel,
\ie
\label{TChannel}
& \la \cO_1(z_1, \bar z_1) \, \cO_2(z_2, \bar z_2) \, \cO_3(z_3, \bar z_3) \, \cO_4(z_4, \bar z_4) \ra
\\
&= \, \sqrt{\textstyle \prod_{i=1}^4 {\vev{\cL_i}_{\bR^2}}}
~
\sum_{\cL, \cL', \cL''}
\,
\la\,
\begin{gathered}
\begin{tikzpicture}
\begin{scope}[yscale = -1]
\draw [line,->-=.3,-<-=.8] (0,0) ++(90:.5) \dotsol{left}{$\cD_3(z_3)$} arc (90:-90:.5) \dotsol{left}{$\cD_2(z_2)$};
\draw [line,->-=.51,-<-=.8] (0,-.67) ++ (-90:.83) \dotsol{left}{$\cD_1(z_1)$} arc (-90:53:.83);
\draw [line,->-=.54,-<-=.82] (0,.26) ++ (90:1.24) \dotsol{left}{$\cD_4(z_4)$} arc (90:-48:1.24);
\end{scope}
\begin{scope}[xscale = -1, yscale = -1, shift = {(-3.5,0)}]
\draw [line,-<-=.3,->-=.8] (0,0) ++(90:.5) \dotemp{right}{$\ocD_3(\bar z_3)$} arc (90:-90:.5) \dotemp{right}{$\ocD_2(\bar z_2)$};
\draw [line,-<-=.51,->-=.8] (0,-.67) ++ (-90:.83) \dotemp{right}{$\ocD_1(\bar z_1)$} arc (-90:53:.83);
\draw [line,-<-=.54,->-=.82] (0,.26) ++ (90:1.24) \dotemp{right}{$\ocD_4(z_4)$} arc (90:-48:1.24);
\end{scope}
\draw [line,->-=.55] (1.24,-.26) -- (1.75,-.26) node [above] {$\cL''$} -- (2.25,-.26);
\node at (.85,0) {$\cL$};
\node at (2.65,0) {$\cL$};
\node at (1.35,.5) {$\cL'$};
\node at (2.25,.5) {$\cL'$};
\end{tikzpicture}
\end{gathered}
\,\ra 
\\
&\hspace{.5in}
\circ
\Big[(F^{\cL', \ocL', \ocL_4}_{\ocL_4})_{\cI, \ocL''}~(1_{\cL', \ocL', \cI}, 1_{\cI, \ocL_4, \cL_4})
\otimes
(F^{\cL_1, \ocL_1, \ocL}_{\ocL})_{\cI, \ocL'}~(1_{\cL_1, \ocL_1, \cI}, 1_{\cI, \ocL, \cL})
\\
&\hspace{3in}
\otimes
(F^{\cL_2, \ocL_2, \ocL_3}_{\ocL_3})_{\cI, \ocL}~(1_{\cL_2, \ocL_2, \cI}, 1_{\cI, \ocL_3, \cL_3})\Big]
\\
~
\\
&= \, 
\frac{\sqrt{\prod_{i=1}^4 {\vev{\cL_i}_{\bR^2}}}}{{\vev{\cL_4}_{\bR^2}}} \, \sum_\cL \, 
\la\,
\begin{gathered}
\begin{tikzpicture}[scale=.75]
\draw [line,->-=.6] (0,-1) -- (0,-.5) node [right] {$\cL$} -- (0,0);
\draw [line,-<-=.55] (0,-1) -- (.87,-1.5) \dotsol{below right}{$\cD_3(z_3)$};
\draw [line,-<-=.55] (0,-1) -- (-.87,-1.5) \dotsol{below left}{$\cD_2(z_2)$};
\draw [line,-<-=.55] (0,0) -- (-.87,.5) \dotsol{above left}{$\cD_1(z_1)$};
\draw [line,-<-=.55] (0,0) -- (.87,.5) \dotsol{above right}{$\cD_4(z_4)$};
\end{tikzpicture}
\end{gathered}
\,\ra
\la\,
\begin{gathered}
\begin{tikzpicture}[scale=.75]
\draw [line,-<-=.6] (0,-1) -- (0,-.5) node [right] {$\cL$} -- (0,0);
\draw [line,->-=.55] (0,-1) -- (.87,-1.5) \dotemp{below right}{$\ocD_2(\bar z_2)$};
\draw [line,->-=.55] (0,-1) -- (-.87,-1.5) \dotemp{below left}{$\ocD_3(\bar z_3)$};
\draw [line,->-=.55] (0,0) -- (-.87,.5) \dotemp{above left}{$\ocD_4(\bar z_4)$};
\draw [line,->-=.55] (0,0) -- (.87,.5) \dotemp{above right}{$\ocD_1(\bar z_1)$};
\end{tikzpicture}
\end{gathered}
\,\ra
\\
&\hspace{.5in}\circ
\Big[(F^{\cL_1, \ocL_1, \ocL}_{\ocL})_{\cI, \cL_4}~(1_{\cL_1, \ocL_1, \cI}, 1_{\cI, \ocL, \cL})
\otimes
(F^{\cL_2, \ocL_2, \ocL_3}_{\ocL_3})_{\cI, \ocL}~(1_{\cL_2, \ocL_2, \cI}, 1_{\cI, \ocL_3, \cL_3})\Big]
\, .
\fe
Hence, by \eqref{3pt} and \eqref{Theta}, we recover the usual $t$-channel conformal block expansion for local operators
\ie
\label{TLocal}
& \la \cO_1(z_1, \bar z_1) \, \cO_2(z_2, \bar z_2) \, \cO_3(z_3, \bar z_3) \, \cO_4(z_4, \bar z_4) \ra
\\
~
\\
&= \,
\frac{\sqrt{\prod_{i=1}^4 {\vev{\cL_i}_{\bR^2}}}}{{\vev{\cL_4}_{\bR^2}}} \, \sum_\cL \, 
\sum_{\cD \in \cH_\ocL^{\bar h = 0}} \sum_{\ocD \in \cH_\cL^{h = 0}} \, 
\cF{\scriptsize
\begin{bmatrix}
h_{\cD_2} & h_{\cD_1}
\\
h_{\cD_3} & h_{\cD_4}
\end{bmatrix}
}_{h_{\cD}}^{c}(1-z)
\,
\cF{\scriptsize
\begin{bmatrix}
\bar h_{\ocD_2} & \bar h_{\ocD_1}
\\
\bar h_{\ocD_3} & \bar h_{\ocD_4}
\end{bmatrix}
}_{\bar h_{\ocD}}^{\bar c}(1-\bar z)
\\
&\hspace{.5in}\times C_{\cD_2, \cD_3, \cD^\dag} \otimes C_{\cD_1, \cD, \cD_4} \otimes C_{\ocD_2, \ocD^\dag, \ocD_3} \otimes C_{\ocD_1, \ocD_4, \ocD}
\\
&\hspace{1in}\circ\Big[
(F^{\cL_1, \ocL_1, \ocL}_{\ocL})_{\cI, \cL_4}~(1_{\cL_1, \ocL_1, \cI}, 1_{\cI, \ocL, \cL})
\otimes
(F^{\cL_2, \ocL_2, \ocL_3}_{\ocL_3})_{\cI, \ocL}~(1_{\cL_2, \ocL_2, \cI}, 1_{\cI, \ocL_3, \cL_3}) \Big] 
\\
&= \, \sum_\cO \, 
\cF{\scriptsize
\begin{bmatrix}
h_{\cO_2} & h_{\cO_1}
\\
h_{\cO_3} & h_{\cO_4}
\end{bmatrix}
}_{h_{\cO}}^{c}(1-z)
\,
\cF{\scriptsize
\begin{bmatrix}
\bar h_{\cO_2} & \bar h_{\cO_1}
\\
\bar h_{\cO_3} & \bar h_{\cO_4}
\end{bmatrix}
}_{\bar h_{\cO}}^{\bar c}(1-\bar z)
\, C_{\cO_2, \cO_3, \cO} \, C_{\cO_1, \cO, \cO_4} \, .
\fe

Let us perform two more $F$-moves on the last line of the $s$-channel expression \eqref{SChannel} to arrive at
\ie
& \dotsb \, = \,
\frac{\sqrt{\textstyle \prod_{i=1}^4 {\vev{\cL_i}_{\bR^2}}}}{{\vev{\cL_4}_{\bR^2}}} \sum_{\cL, \cL', \cL''} \, 
\la\,
\begin{gathered}
\begin{tikzpicture}[scale=.75]
\draw [line,-<-=.6] (0,-1) -- (0,-.5) node [right] {$\cL'$} -- (0,0);
\draw [line,-<-=.55] (0,-1) -- (.87,-1.5) \dotsol{below right}{$\cD_3(z_3)$};
\draw [line,-<-=.55] (0,-1) -- (-.87,-1.5) \dotsol{below left}{$\cD_2(z_2)$};
\draw [line,-<-=.55] (0,0) -- (-.87,.5) \dotsol{above left}{$\cD_1(z_1)$};
\draw [line,-<-=.55] (0,0) -- (.87,.5) \dotsol{above right}{$\cD_4(z_4)$};
\end{tikzpicture}
\end{gathered}
\,\ra
\la\,
\begin{gathered}
\begin{tikzpicture}[scale=.75]
\draw [line,-<-=.6] (0,-1) -- (0,-.5) node [right] {$\cL''$} -- (0,0);
\draw [line,->-=.55] (0,-1) -- (.87,-1.5) \dotsol{below right}{$\cD_2(\bar z_2)$};
\draw [line,->-=.55] (0,-1) -- (-.87,-1.5) \dotsol{below left}{$\cD_3(\bar z_3)$};
\draw [line,->-=.55] (0,0) -- (-.87,.5) \dotsol{above left}{$\cD_4(\bar z_4)$};
\draw [line,->-=.55] (0,0) -- (.87,.5) \dotsol{above right}{$\cD_1(\bar z_1)$};
\end{tikzpicture}
\end{gathered}
\,\ra
\\
&\hspace{.5in} \circ\Big[(F^{\ocL_1, \ocL_2, \ocL_3}_{\cL_4})_{\ocL, \cL'}
\otimes 
(F^{\cL_4, \cL_3, \cL_2}_{\ocL_1})_{\ocL, \cL''}\Big]
\\
&\hspace{.5in} \circ
\Big[(F^{\cL, \ocL, \ocL_3}_{\ocL_3})_{\cI, \cL_4}~(1_{\cL, \ocL, \cI}, 1_{\cI, \ocL_3, \cL_3})
\otimes
(F^{\cL_1, \ocL_1, \ocL_2}_{\ocL_2})_{\cI, \ocL}~(1_{\cL_1, \ocL_1, \cI}, 1_{\cI, \ocL_2, \cL_2})\Big]
\, .
\fe
Compared to the last line of the $t$-channel expression \eqref{TChannel}, we see that crossing symmetry of holomorphically-defect-factorized local operators is a consequence of
\begin{enumerate}
\item Crossing symmetry \eqref{DefectCrossing} of holomorphic defect operators, and
\item The fusion categorical identity
\ie
\label{FIdentity}
& \sum_\cL \Big[(F^{\ocL_1, \ocL_2, \ocL_3}_{\cL_4})_{\ocL, \cL'}
\otimes 
(F^{\cL_4, \cL_3, \cL_2}_{\ocL_1})_{\ocL, \cL''}\Big]
\\
&\qquad
\circ
\Big[(F^{\cL, \ocL, \ocL_3}_{\ocL_3})_{\cI, \cL_4}~(1_{\cL, \ocL, \cI}, 1_{\cI, \ocL_3, \cL_3})
\otimes
(F^{\cL_1, \ocL_1, \ocL_2}_{\ocL_2})_{\cI, \ocL}~(1_{\cL_1, \ocL_1, \cI}, 1_{\cI, \ocL_2, \cL_2})\Big]
\\
&=
\D_{\cL', \ocL''}
\,
(F^{\cL_1, \ocL_1, \ocL'}_{\ocL'})_{\cI, \cL_4}~(1_{\cL_1, \ocL_1, \cI}, 1_{\cI, \ocL', \cL'})
\otimes
(F^{\cL_2, \ocL_2, \ocL_3}_{\ocL_3})_{\cI, \cL'}~(1_{\cL_2, \ocL_2, \cI}, 1_{\cI, \ocL_3, \cL_3})
\fe
for fusion categories that admit a gauge in which the cyclic permutation map is trivial. If not, the identity involves extra cyclic permutation maps/$F$-symbols.
\end{enumerate}

\section{Spectral radius formula from the Perron-Frobenius theorem}
\label{Sec:Bounded}

Consider a quantum field theory hosting a finite (sub)set of simple topological defect lines (TDLs) $\{ \cL_i \mid i = 1, \dotsc, n \}$ that generate a commutative ring $R$ under fusion and direct sum. Let the fusion coefficients be $N_{ij}^k$, and let ${\bf N}_i$ denote the matrix whose $(j, k)$ component is given by $N_{ij}^k$. Associativity implies that ${\bf N}_*$ furnishes a non-negative matrix representation of the fusion rule, called the regular representation {\bf reg}, which is the direct sum of irreducible complex representations, ${\bf reg} = \bigoplus_{a=1}^{n_r} {\bf r}_a$.\footnote{While the regular representation has integer entries, we purposefully omit the word integer as it serves no purpose here. Also, the irreducible complex representations comprising the regular representation are not necessarily non-negative in any basis.
}
We write ${\bf r} < {\bf reg}$ if ${\bf r} \in \{ r_1, \dotsc, r_{n_r} \}$. 

On a cylinder, a TDL wrapped on the spatial circle acts as an operator on the Hilbert space. If the theory is unitary and if there is a unique vacuum, then every TDL acts on the vacuum with a positive eigenvalue. In other words, the cylinder loop expectation value is positive, $\vev{\cL_i}_{{\rm S}^1 \times \bR} > 0$ for all $i = 1, \dotsc, n$. This set of numbers solves the abelianized fusion rule,
\ie
\label{AbelianizedFusion}
\vev{\cL_i}_{{\rm S}^1 \times \bR} \, \vev{\cL_j}_{{\rm S}^1 \times \bR} = \sum_k N_{ij}^k \, \vev{\cL_k}_{{\rm S}^1 \times \bR} \, ,
\fe
and furnishes a one-dimensional representation of $R$. The relation between $\vev{\cL_i}_{{\rm S}^1 \times \bR}$ and $\vev{\cL_i}_{\bR^2}$ was discussed in footnote~\ref{VEV}; in particular,
\ie
\vev{\cL_i}_{{\rm S}^1 \times \bR} = |\vev{\cL_i}_{\bR^2}| \, .
\fe
The abelianized fusion rule \eqref{AbelianizedFusion} can be interpreted as saying that $\vev{\cL_*}_{{\rm S}^1 \times \bR}$ is a simultaneous eigenvector of ${\bf N}_i$ with eigenvalue $\vev{\cL_i}_{{\rm S}^1 \times \bR}$. 

Consider the matrix ${\bf N}_i(\epsilon)={\bf N}_i+\epsilon \sum_{j}{\bf N}_j$ for $\epsilon>0$, which is irreducible (in the Perron-Frobenius sense\footnote{A matrix is called reducible if an off-diagonal block can be set to zero by a {\it permutation} of basis. A matrix that is not reducible is irreducible. A matrix $M$ is irreducible if and only if for any pair of matrix indices $(i,j)$, there exists a positive integer $n$ such that $(M^n)_{i,j}>0$.}) because for any pair of simple TDLs $(\cL_k, \, \cL_l)$ one can always find a (not necessary simple) TDL $\cL$ such that $\cL_l$ appears in the decomposition of the fusion $\cL \, \cL_k$. By the Perron-Frobenius theorem, $\vev{\cL_*}_{{\rm S}^1 \times \bR}$ is the unique positive eigenvector of ${\bf N}_i(\epsilon)$ (up to an overall multiplicative factor), and the spectral radius of ${\bf N}_i(\epsilon)$ is the Perron-Frobenius eigenvalue $\vev{\cL_i}_{{\rm S}^1 \times \bR}+\epsilon \sum_{j}\vev{\cL_j}_{{\rm S}^1 \times \bR}$. By taking the $\epsilon\to0$ limit, we find that $\vev{\cL_i}_{{\rm S}^1 \times \bR}$ is the spectral radius of ${\bf N}_i$, {\it i.e.} 
\ie
\left| \frac{v^\dag {\bf N}_i v}{v^\dag v} \right| \le \vev{\cL_i}_{{\rm S}^1 \times \bR} \quad \forall i = 1, \dotsc, n, \quad \forall v \in \bC^n \, .
\fe

\begin{prop}
In a (1+1)$d$ unitary quantum field theory on a cylinder with a unique vacuum, let $R$ be the fusion rule of a finite set of simple topological defect lines $\{ \cL_i \mid i = 1, \dotsc, n\}$. Denote by $\widehat\cL_i$ the operator corresponding to wrapping $\cL_i$ on the spatial circle. For any state $|\phi\rangle$ transforming in an irreducible representation ${\bf r} < {\bf reg}$ with respect to the ring $R$, the following inequality holds
\ie
\left| \frac{\vev{\phi^\dag \widehat\cL_i \phi}}{\vev{\phi^\dag \phi}} \right| \le \vev{\cL_i}_{{\rm S}^1 \times \bR} \, , \quad \forall i = 1, \dotsc, n \, .
\fe
In particular, if $R$ is a group, then because every irreducible representation $< {\bf reg}$, the above inequality holds for all $|\phi\rangle$.
\end{prop}

In conformal field theory, the simple factorizing TDLs generate a commutative ring (see Definition~\ref{Factorizing} and Proposition~\ref{Commute}).\footnote{The full set of factorizing TDLs generally contains non-compact ones. Here we focus on a commutative ring generated by the simple factorizing TDLs.
}
By further utilizing the state-operator map, we obtain a unitary bound on the opacity.
\begin{coro}
In (1+1)$d$ unitary conformal field theory, if a local operator $\cO$ is holomorphically-defect-factorized through a topological defect line $\cL$ that generates under fusion a finite sum of simple objects, and if $\phi$ (not necessarily holomorphically-defect-factorized) transforms in an irreducible representation ${\bf r} < {\bf reg}$, then in the infinite boost limit, the opacity given by \eqref{eqn:reggeGeneral} and \eqref{opacity} is bounded by $\kappa[\cO, \phi] = 1-|r[\cO, \phi]| \ge 0$. If $\cL$ is invertible, then there is no restriction on $\phi$.
\end{coro}

For Verlinde lines in rational conformal field theory, the simultaneous eigenvectors can be expressed in terms of the modular S-matrix by the Verlinde formula \cite{Verlinde:1988sn}
\ie
{\bf N}_i\cdot {\bf v}_m = {S_{im}\over S_{0m} } {\bf v}_m \, ,\quad ({\bf v}_m)_j={S_{jm}\over S_{0m}} \, .
\fe
The Perron-Frobenius eigenvector is the zeroth eigenvector ${\bf v}_0 = {{S_{*m} / S_{0m}}} = \vev{\cL_*}_{{\rm S}^1 \times \bR}$. The expression in \eqref{eqn:rforRCFT} is the ratio between the $k$-th eigenvalue and the Perron-Frobenius eigenvalue, so its absolute value is no more than one. This proves the spectral radius formula \eqref{eqn:SpectralRadius} for Verlinde lines in all unitary rational conformal field theories.

Finally, the Perron-Frobenius theorem have been generalized to integral bounded operators by several theorems: Jentzsch Theorem \cite{jentzsch1912integralgleichungen}, Schaefer Theorem \cite{ellis1977hh} and Zerner Theorem \cite{zerner1987quelques}; see {\it e.g.} \cite{kloeden2000generalization} for a summary of these theorems. One could include the non-compact factorizing TDLs into the set of basis TDLs, and apply these theorems to fusion rules involving direct integrals.

\section{Free boson orbifold theory}
\label{App:Free}

This appendix concerns the holomorphic-defect-factorization of twist fields on the orbifold branch of the $c = 1$ free boson theory. We first review the basic definition and properties of Riemann theta functions that are used to express general correlators, give the character decomposition of the torus partition function, and describe the universal $D_4$ symmetry. We then examine dual descriptions at special rational points, and cast the topological defect lines as Verlinde lines. Finally, we compute the torus Regge limit of twist fields, and determine the action of the factorizing topological defect line and its fusion properties.

\subsection{Riemann and Jacobi theta functions}

The Riemann theta function is defined as
\ie
\label{Riemann}
\theta\bk{\A}{\B}(z|\tau) &= \sum_{n \in \bZ^g} \exp( i \pi (n+\A) \cdot \tau \cdot (n+\A) + 2i \pi (n+\A) \cdot (z+\B) ) \, .
\fe
By definition, it changes characteristic under shifts in $z$:
\ie
\theta\bk{\A}{\B}(z+k|\tau) &= \theta\bk{\A}{\B+k}(z|\tau) \, ,
\\
\theta\bk{\A}{\B}(z+k\tau|\tau) &= \exp(-{{i\pi}k^2\tau}) \, \theta\bk{\A+k}{\B}(z|\tau) \, .
\fe
When $g = 1$ and $\A, \, \B$ take values in $\frac12 \bZ$, they are the Jacobi theta functions
\ie
& \theta_1(z|\tau) = - \theta\bk{\frac12}{\frac12}(z|\tau) \, , \quad
\theta_2(z|\tau) = \theta\bk{\frac12}{0}(z|\tau) \, ,
\\
& \theta_3(z|\tau) = \theta\bk{0}{0}(z|\tau) \, , \quad
\theta_4(z|\tau) = \theta\bk{0}{\frac12}(z|\tau) \, .
\fe
Thus
\ie
\label{JacobiShifts}
& \theta_1(z+\frac12|\tau) = - \theta_2(z|\tau) \, , \quad \theta_2(z+\frac12|\tau) = \theta_1(z|\tau) \, ,
\\
& \theta_3(z+\frac12|\tau) = \theta_4(z|\tau) \, , \quad \theta_4(z+\frac12|\tau) = \theta_3(z|\tau) \, ,
\\
& \theta_1(z+\frac\tau2|\tau) = e^{-\frac{i\pi}{4}\tau} \theta_4(z|\tau) \, , \quad \theta_2(z+\frac\tau2|\tau) = e^{-\frac{i\pi}{4}\tau} \theta_3(z|\tau) \, ,
\\
& \theta_3(z+\frac\tau2|\tau) = e^{-\frac{i\pi}{4}\tau} \theta_2(z|\tau) \, , \quad \theta_4(z+\frac\tau2|\tau) = - e^{-\frac{i\pi}{4}\tau} \theta_1(z|\tau) \, .
\fe

Next consider modular transformations, and restrict to $z = 0$ for simplicity. For Riemann theta functions,
\ie
& \theta\bk{\A}{\B-\A+\frac12}(0|\tau+1) = \varepsilon_T \, e^{i\pi\A(1-\A)} \, \theta\bk{\A}{\B}(0|\tau) \, ,
\\
& \theta\bk{-\B}{\A}(0|-\frac{1}{\tau}) = \varepsilon_S \, e^{-2i \pi \A \B} \sqrt\tau \, \theta\bk{\A}{\B}(0|\tau) \, .
\fe
For Jacobi theta functions,
\ie
& \theta_4(0|\tau+1) = \varepsilon_T \, \theta_3(0|\tau) \, ,
\quad
\theta_3(0|-\frac{1}{\tau}) = \varepsilon_S \, \sqrt\tau \, \theta_3(0|\tau) \, ,
\\
& \theta_3(0|\tau+1) = \varepsilon_T \, \theta_4(0|\tau) \, ,
\quad
\theta_2(0|-\frac{1}{\tau}) = \varepsilon_S \, \sqrt\tau \, \theta_4(0|\tau) \, ,
\\
& \theta_2(0|\tau+1) = \varepsilon_T \, e^{\frac{i\pi}{4}} \, \theta_2(0|\tau) \, ,
\quad
\theta_4(0|-\frac{1}{\tau}) = \varepsilon_S \, \sqrt\tau \, \theta_2(0|\tau) \, ,
\fe
with $\varepsilon_T = \varepsilon_S = 1$.

\subsection{Partition function and character decomposition}

The partition function of the free boson orbifold theory is
\ie
 Z(\tau, \bar\tau)
&= \frac{1}{2|\eta(\tau)|^2} \sum_{m, w \in \bZ} \, {q^{\frac{p_{\rm L}^2}{4}} \bar q^{\frac{p_{\rm R}^2}{4}}}  + \frac{|\theta_3(\tau) \theta_4(\tau)|}{2 |\eta(\tau)|^2} + \frac{|\theta_2(\tau) \theta_3(\tau)|}{2 |\eta(\tau)|^2}+ \frac{|\theta_2(\tau) \theta_4(\tau)|}{2 |\eta(\tau)|^2} \, .
\fe
Let us decompose it into irreducible Virasoro characters
\ie
\chi_h(\tau)=\begin{cases} 
\displaystyle
{q^{n^2}-q^{(n+1)^2}\over \eta(\tau)}& h=n^2,\,n\in\bZ \, ,
\\
~
\\
\displaystyle
{q^{(n+\frac12)^2}-q^{(n+\frac32)^2}\over \eta(\tau)}& h=(n+\frac12)^2,\,n\in\bZ \, ,
\\
~
\\
\displaystyle 
{q^h\over \eta(\tau)}& \text{otherwise} \, .
\end{cases}
\fe
At irrational $r^2$,
\ie
 Z(\tau, \bar\tau)&=\left(\sum_{\substack{m \in \bZ \\ w\in\bZ_{>0}}}+\sum_{\substack{m \in \bZ_{>0} \\ w=0}}\right) \,\chi_{ \frac{p_{\rm L}^2}{4}}(\tau) \overline{\chi_{  \frac{p_{\rm R}^2}{4}}(\tau)}+
\sum_{n, \bar n \in \bZ_{\ge0}}{1+(-1)^{n+\bar n}\over 2} \chi_{ n^2}(\tau) \overline{\chi_{ \bar n^2}(\tau)}
\\
&\hspace{1in}
+ \sum_{n,\bar n \in \bZ_{\ge0}} (1+\cos\tfrac{(n-\bar n)\pi}{2}-\sin\tfrac{(n+\bar n)\pi}{2})\chi_{{(2n+1)^2\over 16}}(\tau)\overline{\chi_{{(2\bar n+1)^2\over 16}}(\tau)} \, ,
\fe
where we have used the identities
\ie
&\sqrt{\frac{\theta_2(\tau) \theta_3(\tau)}{2}} = \sum_{n \in \bZ_{\ge0}} q^{\frac{(2n+1)^2}{16}} \, ,\quad \sqrt{\frac{\theta_2(\tau) \theta_4(\tau)}{2}} = \sum_{n \in \bZ_{\ge0}} \left(\cos\tfrac{n\pi}{2}-\sin\tfrac{n\pi}{2}\right)q^{\frac{(2n+1)^2}{16}} \, ,
\\
&\frac{1+|\theta_3(\tau) \theta_4(\tau)|}{2 |\eta(\tau)|^2}=\sum_{\substack{n, \bar n \in \bZ_{\ge0} \\ n - \bar n \in 2\bZ}} \chi_{ n^2}(\tau) \overline{\chi_{ \bar n^2}(\tau)} = \sum_{n, \bar n \in \bZ_{\ge0}}{1+(-1)^{n+\bar n}\over 2} \chi_{ n^2}(\tau) \overline{\chi_{ \bar n^2}(\tau)} \, .
\fe
At rational $r^2={u/v}$ with $u, v$ coprime, but irrational $r$,
\ie
\label{RationalDecomp}
 Z(\tau, \bar\tau)
&=\left(\sum_{\substack{m \in \bZ,\,w\in\bZ_{>0} \\  mv\neq\pm wu}}+\sum_{\substack{m \in \bZ_{>0} \\ w=0}}\right) \,\chi_{\frac{p_{\rm L}^2}{4}}(\tau) \overline{\chi_{  \frac{p_{\rm R}^2}{4}}(\tau)}+
\sum_{n, \bar n \in \bZ_{\ge0}}{1+(-1)^{n+\bar n}\over 2} \chi_{ n^2}(\tau) \overline{\chi_{ \bar n^2}(\tau)}
\\
&\quad+ \sum_{\substack{m \in \bZ,\, w\in\bZ_{>0} \\  mv=wu}} \chi_{  \frac{p_{\rm L}^2}{4}}(\tau)\sum_{n=0}^\infty \overline{\chi_{n^2}(\tau)}+\sum_{n=0}^\infty \chi_{n^2}(\tau)  \sum_{\substack{m \in \bZ,\, w\in\bZ_{>0} \\  mv= -wu}} \overline{\chi_{  \frac{p_{\rm R}^2}{4}}(\tau)}
\\
&\quad+ \sum_{n,\bar n \in \bZ_{\ge0}} (1+\cos\tfrac{(n-\bar n)\pi}{2}-\sin\tfrac{(n+\bar n)\pi}{2})\chi_{{(2n+1)^2\over 16}}(\tau)\overline{\chi_{{(2\bar n+1)^2\over 16}}(\tau)} \, .
\fe
If $r$ itself is rational, then the characters with $p_{\rm L} \in \bZ$ or $p_{\rm R} \in \bZ$ are further reducible.

\subsection{$D_4$ symmetry}
\label{Sec:D4}

The momentum and winding $\bZ_2$ symmetry lines in the ${\rm S}^1$ theory descend to pairs of $\bZ_2$ symmetry lines $(\eta_{\rm m}, \, \eta_{\rm m}')$ and $(\eta_{\rm w}, \, \eta_{\rm w}')$, respectively, in the ${\rm S}^1/\bZ_2$ orbifold theory. Without loss of generality, $\eta_{\rm m}$ and $\eta_{\rm w}$ generate a $D_4$ symmetry. The emergent $\bZ_2$ symmetry that assigns $+1$ charge to the untwisted sector states and $-1$ charge to the twisted sector states corresponds to the symmetry line $\eta \equiv \eta_{\rm m} \, \eta_{\rm w} \, \eta_{\rm m} \, \eta_{\rm w}$. The five order-two elements act on the cosine operators and the twisted sector ground states associated to the two fixed points by 
\ie
& \widehat\eta_{\rm m}(\cO_{m,w})=(-1)^m\cO_{m,w} \, ,\quad && \widehat\eta_{\rm m}({\cal E}_1)={\cal E}_2 \, ,\quad && \widehat\eta_{\rm m}({\cal E}_2)={\cal E}_1 \, ,
\\
&\widehat\eta_{\rm w}(\cO_{m,w})=(-1)^w\cO_{m,w} \, ,\quad && \widehat\eta_{\rm w}({\cal E}_1)=-{\cal E}_1 \, ,\quad && \widehat\eta_{\rm w}({\cal E}_2)={\cal E}_2 \, ,
\\
&\widehat\eta(\cO_{m,w}) = \cO_{m,w} \, , \quad 
&&\widehat\eta({\cal E}_1)=-{\cal E}_1 \, , \quad 
&&\widehat\eta({\cal E}_2) = - {\cal E}_2 \, ,
\\
&\widehat\eta_{\rm m}'(\cO_{m,w})=(-1)^m\cO_{m,w} \, ,\quad && \widehat\eta_{\rm m}'({\cal E}_1)=-{\cal E}_2 \, ,\quad && \widehat\eta_{\rm m}'({\cal E}_2)=-{\cal E}_1\, ,
\\
&\widehat\eta_{\rm w}'(\cO_{m,w})=(-1)^w\cO_{m,w} \, ,\quad && \widehat\eta_{\rm w}'({\cal E}_1)={\cal E}_1 \, ,\quad && \widehat\eta_{\rm w}'({\cal E}_2)=-{\cal E}_2 \, .
\fe
The twisted partition functions are
\ie
 Z^{\eta_{\rm m}}(\tau, \bar\tau)
&= 
\left(\sum_{\substack{m \in \bZ \\ w\in\bZ_{>0}}}+\sum_{\substack{m \in \bZ_{>0} \\ w=0}}\right) \, (-1)^m\chi_{ \frac{p_{\rm L}^2}{4}}(\tau) \overline{\chi_{  \frac{p_{\rm R}^2}{4}}(\tau)}
 + 
\sum_{\substack{n, \bar n \in \bZ_{\ge0} \\ n - \bar n \in 2\bZ}} \chi_{h = n^2}(\tau) \bar\chi_{\bar h = \bar n^2}(\bar \tau) 
\\
&= \frac{1}{2|\eta(\tau)|^2} \sum_{m, w \in \bZ} (-1)^m \, {q^{\frac{p_{\rm L}^2}{4}} \bar q^{\frac{p_{\rm R}^2}{4}}} + \frac{|\theta_3(\tau) \theta_4(\tau)|}{2 |\eta(\tau)|^2} = Z^{\eta_{\rm m}'}(\tau, \bar\tau) \, ,
\\
 Z^{\eta_{\rm w}}(\tau, \bar\tau)
&= 
\left(\sum_{\substack{m \in \bZ \\ w\in\bZ_{>0}}}+\sum_{\substack{m \in \bZ_{>0} \\ w=0}}\right) \, (-1)^w\chi_{ \frac{p_{\rm L}^2}{4}}(\tau) \overline{\chi_{  \frac{p_{\rm R}^2}{4}}(\tau)}
 + 
\sum_{\substack{n, \bar n \in \bZ_{\ge0} \\ n - \bar n \in 2\bZ}} \chi_{h = n^2}(\tau) \bar\chi_{\bar h = \bar n^2}(\bar \tau) 
\\
&= \frac{1}{2|\eta(\tau)|^2} \sum_{m, w \in \bZ} (-1)^w \, {q^{\frac{p_{\rm L}^2}{4}} \bar q^{\frac{p_{\rm R}^2}{4}}} + \frac{|\theta_3(\tau) \theta_4(\tau)|}{2 |\eta(\tau)|^2} = Z^{\eta_{\rm w}'}(\tau, \bar\tau) \, ,
\\
 Z^{\eta}(\tau, \bar\tau)
&= 
\left(\sum_{\substack{m \in \bZ \\ w\in\bZ_{>0}}}+\sum_{\substack{m \in \bZ_{>0} \\ w=0}}\right) \, \chi_{ \frac{p_{\rm L}^2}{4}}(\tau) \overline{\chi_{  \frac{p_{\rm R}^2}{4}}(\tau)}+ 
\sum_{\substack{n, \bar n \in \bZ_{\ge0} \\ n - \bar n \in 2\bZ}} \chi_{h = n^2}(\tau) \bar\chi_{\bar h = \bar n^2}(\bar \tau) 
\\
&
\hspace{.75in} - \sum_{n,\bar n \in \bZ_{\ge0}} (1+\cos\tfrac{(n-\bar n)\pi}{2}-\sin\tfrac{(n+\bar n)\pi}{2})\chi_{{(2n+1)^2\over 16}}(\tau)\overline{\chi_{{(2\bar n+1)^2\over 16}}(\tau)} 
\\
&= \frac{1}{2|\eta(\tau)|^2} \sum_{m, w \in \bZ} \, {q^{\frac{p_{\rm L}^2}{4}} \bar q^{\frac{p_{\rm R}^2}{4}}} + \frac{|\theta_3(\tau) \theta_4(\tau)|}{2 |\eta(\tau)|^2} - \frac{|\theta_2(\tau) \theta_3(\tau)|}{2 |\eta(\tau)|^2}- \frac{|\theta_2(\tau) \theta_4(\tau)|}{2 |\eta(\tau)|^2} \, .
\fe
The partition functions of their defect Hilbert spaces are
\ie
Z_{\eta_{\rm m}}(\tau, \bar\tau)& = Z_{\eta_{\rm m}'}(\tau, \bar\tau) = \sum_{m, w \in \bZ} \frac{q^{\frac{p_{\rm L}^2}{4}} \bar q^{\frac{p_{\rm R}^2}{4}}}{2|\eta(\tau)|^2} +\frac{|\theta_2(\tau) \theta_3(\tau)|}{2 |\eta(\tau)|^2}
\, , \quad p_{L,R} = \frac{m }{r} \pm (w + {\textstyle \frac{1}{2}})r \, ,
\\
Z_{\eta_{\rm w}}(\tau, \bar\tau) &= Z_{\eta_{\rm w}'}(\tau, \bar\tau) = \sum_{m, w \in \bZ} \frac{q^{\frac{p_{\rm L}^2}{4}} \bar q^{\frac{p_{\rm R}^2}{4}}}{2|\eta(\tau)|^2} +\frac{|\theta_2(\tau) \theta_3(\tau)|}{2 |\eta(\tau)|^2}
\, , \quad p_{L,R} = \frac{m +{1\over 2}}{r} \pm w r \, ,
\\
 Z_{\eta}(\tau, \bar\tau)
&= \frac{1}{2|\eta(\tau)|^2} \sum_{m, w \in \bZ} \, {q^{\frac{p_{\rm L}^2}{4}} \bar q^{\frac{p_{\rm R}^2}{4}}}  - \frac{|\theta_3(\tau) \theta_4(\tau)|}{2 |\eta(\tau)|^2} + \frac{|\theta_2(\tau) \theta_3(\tau)|}{2 |\eta(\tau)|^2}- \frac{|\theta_2(\tau) \theta_4(\tau)|}{2 |\eta(\tau)|^2} 
\\
&=\frac{1}{{|\eta(\tau)|^2}}
\left( \sum_{\substack{m \in \bZ \\ w \in \bZ_{>0}}} + \sum_{\substack{m \in \bZ_{>0} \\ w = 0}} \right) \, {q^{\frac{p_{\rm L}^2}{4}} \bar q^{\frac{p_{\rm R}^2}{4}}}+
\sum_{\substack{n, \bar n \in \bZ_{\ge0} \\ n - \bar n \in 2\bZ+1}} \chi_{h = n^2}(\tau) \bar\chi_{\bar h = \bar n^2}(\bar \tau)
\\
&\quad+ \frac{|\theta_2(\tau) \theta_3(\tau)|}{2 |\eta(\tau)|^2}- \frac{|\theta_2(\tau) \theta_4(\tau)|}{2 |\eta(\tau)|^2} \, .
\fe

\subsection{Special rational points}
\label{Sec:RationalOrbifold}

Holomorphic-defect-factorization can be explicitly examined at special rational points via dual descriptions:
\ie
& \text{(a)} \quad \frac{{\rm S}^1_{r=1}}{\bZ_2} = {\rm S}^1_{r=2} \, ,
\\
& \text{(b)} \quad \frac{{\rm S}^1_{r=\sqrt{3}/\sqrt{2}}}{\bZ_2} = \frac{{\cal SM}(4, 6)}{(-1)^F}
\\
& \text{(c)} \quad \frac{{\rm S}^1_{r=\sqrt2}}{\bZ_2} = \text{Ising}^2 \, ,
\\
& \text{(d)} \quad \frac{{\rm S}^1_{r=\sqrt3}}{\bZ_2} = \frac{{\rm SU}(2)_4}{{\rm U(1)}} \, ,
\\
& \text{(e)} \quad \frac{{\rm S}^1_{r=\sqrt{6}}}{\bZ_2} = \frac{{\cal SM}(4, 6) \otimes (-1)^{\rm Arf}}{(-1)^F}
\\
& \text{(f)} \quad \frac{{\rm S}^1_{r=2\sqrt2}}{\bZ_2} = {\rm Sym}^2 \, \text{Ising} \, , \quad \dotsc \, ,
\fe
where ${\cal SM}$ denotes an $\cN = 1$ super-Virasoro minimal model. We adopt the notation for $D_4$ symmetry lines in Appendix~\ref{Sec:D4}, and denote the cosine lines by $\cL_{(\theta_{\rm m}, \theta_{\rm w})}$ as in Section~\ref{Sec:S1Z2}.
\begin{enumerate}[(a)]

\item In the ${{\rm S}^1_{r=1}}/{\bZ_2}$ theory, all local operators including the twist fields are exponential operators in the ${\rm S}^1_{r=2}$ description, factorized through ${\rm U(1)}$ symmetry lines. Complex combinations of the twist fields in ${\rm S}^1_{r=1}/\bZ_2$ correspond to the exponential operators with $m = \pm 1, \, w = 0$ in ${\rm S}^1_{r=2}$, and are factorized through the $\bZ_8$ symmetry lines $\cL_{(\pm\frac{\pi}{4}, \pi)}^{{\rm S}^1_{r=2}}$.

\item The twist fields of ${{\rm S}^1_{r=\sqrt3/\sqrt2}}/{\bZ_2}$ are the two weight $(\frac{1}{16}, \frac{1}{16})$ operators (one in the Neveu-Schwarz sector and one in the Ramond sector) in the bosonized ${\cal SM}(4,6)$ description. They are factorized through unoriented Verlinde lines with $\vev{\cL_{{\cal E}_i}}_{\bR^2} = \sqrt6$. The self-fusion of each line gives
\ie
\cL_{{\cal E}_1} \, \cL_{{\cal E}_1} = \cI + \eta_{\rm w} + \cL_{(\frac{2\pi}{3}, 0)} + \cL_{(\frac{2\pi}{3}, \pi)} \, ,
\quad
\cL_{{\cal E}_2} \, \cL_{{\cal E}_2} = \cI + \eta_{\rm w}' + \cL_{(\frac{2\pi}{3}, 0)} + \cL_{(\frac{2\pi}{3}, \pi)} \, .
\fe

\item The twist fields ${\cal E}_i$ at the two fixed points of ${{\rm S}^1_{r=\sqrt2}}/{\bZ_2}$ are linear combinations of the $\sigma_1$ and $\sigma_2$ operators in the $\text{Ising}^2$ description, 
\ie
{\cal E}_1 = {1\over 2}(\sigma_1+\sigma_2) \, ,\quad  {\cal E}_2 = {1\over 2}(\sigma_1-\sigma_2) \, .
\fe
They are factorized through the Kramers-Wannier duality lines $\cN_1$ and $\cN_2$ with $\vev{\cN_1}_{\bR^2} = \vev{\cN_2}_{\bR^2} = \sqrt2$. Fusion gives
\ie
\cN_i \, \cN_i = \cI + \eta_i \, , \quad i = 1, 2 \, ,
\fe
where $\eta_i$ is the Ising $\bZ_2$ symmetry line in each copy. They are identified inside the universal $D_4$ symmetry as
\ie
\eta_1 = \eta_{\rm m} \, , \quad \eta_2 = \eta_{\rm m}' \, .
\fe
Further note the identification between cosine lines and Verlinde lines
\ie
\cL_{(0,0)} = \cI + \eta \, , \quad
\cL_{(\pi,0)} = \eta_m + \eta_m'
\, , \quad \cL_{(\frac{\pi}{2}, \pi)} = 
\cN_1 \, \cN_2 \, ,
\fe
and $\eta_{\rm w}$ is the symmetry line that permutes the two copies of the Ising models.

\item The twist fields of ${{\rm S}^1_{r=\sqrt3}}/{\bZ_2}$ are the two weight $(\frac{1}{16}, \frac{1}{16})$ operators in the $\bZ_4$ parafermion theory. Complex combinations of the twist fields are factorized through oriented Verlinde lines $\cL_{\cal E}, \ocL_{\cal E}$ with $\vev{\cL_{\cal E}}_{\bR^2} = \sqrt3$. Their fusion with each other gives
\ie
\cL_{\cal E} \, \ocL_{\cal E} = \cI + \cL_{(\frac{2\pi}{3}, 0)} \, .
\fe

\item The twist fields of ${{\rm S}^1_{r=\sqrt6}}/{\bZ_2}$ are the two weight $(\frac{1}{16}, \frac{1}{16})$ operators (one in the Neveu-Schwarz sector and one in the Ramond sector) in the bosonization of the tensor product of ${\cal SM}(4,6)$ with the $(-1)^{\rm Arf}$ topological field theory. They are factorized through Verlinde lines with $\vev\cL_{\bR^2} = \sqrt6$.

\item The twist fields of ${{\rm S}^1_{r=2\sqrt2}}/{\bZ_2}$ in the language of ${\rm Sym}^2 \, \text{Ising}$ include the weight $(\frac{1}{16}, \frac{1}{16})$ operator $\frac{1}{\sqrt2}(\sigma_1 + \sigma_2)$ in the untwisted sector, and also the replica twist field ground state. The former is factorized through the Verlinde line $\cL$ ($\equiv \cN_1 + \cN_2$ before the symmetric product orbifold) with $\vev\cL_{\bR^2} = 2\sqrt2$.

\end{enumerate}
We observe a pattern: If $r^2 = u/v$ with $u, v$ coprime, then the twist fields in the ${{\rm S}^1_r}/{\bZ_2}$ theory are factorized through a topological defect line $\cL$ with ${\vev{\cL}_{\bR^2}} = \sqrt{uv}$. That this is true for all rational $r^2$ is proven in Appendix~\ref{Sec:ReggeS1Z2}.

\subsection{Torus Regge limit of the twist field two-point function}
\label{Sec:ReggeS1Z2}

The torus Regge limit computes the loop-normalized defect partition function $Z_{\widetilde\cL}(\tau, \bar\tau)$, as was explained in Section~\ref{Sec:TorusRegge}. If $Z_{\widetilde\cL}(\tau, \bar\tau)$ has a discrete expansion in $q, \, \bar q$, and if the coefficients are integers up to an overall multiplicative factor, then one can strip off the factor and obtain the Cardy-normalized defect partition function $Z_\cL(\tau, \bar\tau)$ with positive integer multiplicities. This overall factor is inverse ${\vev{\cL}_{\bR^2}}$ of the defect, so
\ie
Z_{\widetilde\cL}(\tau, \bar\tau) = \frac{Z_\cL(\tau, \bar\tau)}{{\vev{\cL}_{\bR^2}}} \, .
\fe

In the free boson orbifold theory, the holomorphic-defect-factorization of local operators in the untwisted sector can be figured out relatively easily using hints from the pre-orbifold free compact boson theory. However, the holomorphic-defect-factorization in the twisted sector is far from obvious. To characterize the factorizing TDL, we resort to the torus Regge limit. The covering space formalism for computing general correlators of orbifolds \cite{Dixon:1985jw,Dixon:1986jc} was developed in \cite{Hamidi:1986vh,Dixon:1986qv}, and applied to the $c = 1$ free boson theory in great detail in \cite{Miki:1987mp,Dijkgraaf:1987vp}. In particular, our notation and formulae follow \cite{Dijkgraaf:1987vp} closely.

In the free boson orbifold theory, the bosonic field $X$ is double valued. When computing the partition function on a Riemann surface, there are distinct topological sectors distinguished by whether $X$ flips sign around each nontrivial cycle. On a closed Riemann surface of genus $g$, those sectors are labeled by $\varepsilon_i \in \frac{1}{2} \bZ_2$ around $a$-cycles and $\D_i \in \frac{1}{2} \bZ_2$ around $b$-cycles, for $i = 1, \dotsc, g$. In a given sector described by $\varepsilon_i, \D_i$, the double-valued field $X$ on $\Sigma_g$ can be lifted to a single-valued field $X$ on a double-sheeted cover $\widetilde \Sigma_g$. The cover $\widetilde \Sigma_g$ is a replica-symmetric genus $2g$ Riemann surface, and its modulus is described by the period matrix $\Pi_{\varepsilon_i, \D_i}$ of Prym differentials (replica-symmetric holomorphic one-forms on $\widetilde \Sigma_g$). The modulus $\Pi_{\epsilon_i, \D_i}$ is fixed by the period matrix $\tau$ of $\Sigma_g$, the sector $\varepsilon_i, \D_i$, and the positions of twist fields; this relation will be explicitly given for $g = 1$ later.

Consider an orthonormal basis of twist field ground states, and let ${\cal E}$ be any of the two basis twist fields. The twist-field two-point function on a genus-$g$ Riemann surface $\Sigma_g$ is given in Dijkgraaf-Verilinde-Verlinde (5.13) to be
\ie
\la {\cal E}(z, \bar z) {\cal E}(0) \ra_{\Sigma_g(\tau, \bar\tau)} = 2^{-g} \sum_{\varepsilon_i, \D_i \in (\frac12 \bZ_2)^g} Z^{\rm cl}(r, \Pi_{\varepsilon_i, \D_i}, \overline\Pi_{\varepsilon_i, \D_i}) \, Z^{\rm qu}_{\varepsilon_i, \D_i}(\tau, \bar\tau) \, ,
\fe
where
\ie
Z^{\rm qu}_{\varepsilon, \D}(\tau, \bar\tau) = Z^{\rm qu}_0(\tau, \bar\tau) \left| c\bk{\varepsilon_i}{\D_i}(\tau) \right|^{-2} \, .
\fe
Let us explain the pieces comprising this formula.
\begin{enumerate}
\item $Z^{\rm cl}(r, \Pi_{\varepsilon_i, \D_i}, \overline\Pi_{\varepsilon_i, \D_i})$ is the classical contribution to the partition function. It is a solitonic sum over momentum and winding on the two-sheeted cover $\widetilde \Sigma_g$ of $\Sigma_g$,
\ie
& Z^{\rm cl}(r, \Pi_{\varepsilon_i, \D_i}, \overline\Pi_{\varepsilon_i, \D_i}) = \sum_{p, \bar p \in \Gamma_r} 
\exp\left[ \frac{i\pi}{2} (p \cdot \Pi_{\varepsilon_i, \D_i} \cdot p - \bar p \cdot \overline\Pi_{\varepsilon_i, \D_i} \cdot \bar p) \right] \, ,
\fe
where
\ie
& 
\Gamma_r = \left\{ \left( \frac{m_i}{r} + w_ir, \frac{m_i}{r} - w_ir \right) \mid m_i, w_i \in \bZ \right\} \, ,
\quad (i = 1, \dotsc, g) \, .
\fe
\item $Z^{\rm qu}_{\varepsilon_i, \D_i}(\tau, \bar\tau)$ is the quantum contribution to the partition function. And $Z^{\rm qu}_0(\tau, \bar\tau)$ is a common factor shared by all distinct topological sectors, that only depends on the period matrix $\tau$ of $\Sigma_g$.
\item Finally,
\ie
{c\bk{\varepsilon_i}{\D_i}(\tau)}^{-1} = E(z, 0)^{-\frac18} \frac{\theta\bk{\C_i+\frac{\varepsilon_i}{2}}{\D_i}(\frac12 \int_0^z \omega \mid 2\tau)}{\theta\bk{\C_i}{0}(0 \mid 2\Pi_{\varepsilon_i, \D_i})} \, ,
\fe
where $\C_i \in (\frac12 \bZ_2)^g$ is arbitrary, $\omega$ is the holomorphic one-form on $\Sigma_g$, and $E(z, 0)$ is the prime form, the closest thing to $z$ that respects the global structure of the Riemann surface. At short distances, $E(z, 0) \sim z$.
\end{enumerate}

We now specialize to $g=1$. The classical solitonic sum $Z^{\rm cl}(r, \Pi_{\varepsilon, \D}, \overline\Pi_{\varepsilon, \D})$ is just the free compact boson partition function with $\tau$ set to $\Pi_{\varepsilon, \D}$. The common factor in the quantum contributions to the partition function is
\ie
Z^{\rm qu}_0(\tau, \bar\tau) = \frac{1}{|\eta(\tau)|^2} \, .
\fe
The prime form on a torus is
\ie
E(z,q)=\frac{\theta\bk{\frac12}{\frac12}(z \mid \tau)}{\partial_z\theta\bk{\frac12}{\frac12}(z \mid \tau)\big|_{z=0}}
\fe
The Abel map is $\mathfrak{z} = \frac12 \int_0^z \omega$, where $\omega$ is the holomorphic one-form on $\Sigma_{g=1}$. The Schottky relation (the arbitrariness of $\C_i$ mentioned before)
\ie
\label{Schottky}
\frac{\theta\bk{0}{0}(\mathfrak{z} \mid 2\tau)}{\theta\bk{0}{0}(0 \mid 2\Pi_{0,0})}
=
\frac{\theta\bk{\frac12}{0}(\mathfrak{z} \mid 2\tau)}{\theta\bk{\frac12}{0}(0 \mid 2\Pi_{0,0})}
\fe
implicitly defines $\Pi_{0,0}$ as a function of $\mathfrak{z}$ and $\tau$. Let $\Pi(\mathfrak{z}, \tau) \equiv \Pi_{0,0}(\mathfrak{z}, \tau)$, then the rest of $\Pi_{\varepsilon,\D}(\mathfrak{z}, \tau)$ are related via half-integer shifts of $\mathfrak{z}$
\ie
\Pi_{\varepsilon,\D}(\mathfrak{z}, \tau) = \Pi(\mathfrak{z}+\D+\varepsilon\tau, \tau) \, .
\fe
Using the Schottky relation \eqref{Schottky} together with the identities
\ie
\theta\bk{\C}{0}(\mathfrak{z} + \frac12 \mid 2\tau) &= \theta\bk{\C}{\frac12}(\mathfrak{z} \mid 2\tau) \, ,
\\
\theta\bk{\C}{0}(\mathfrak{z} + \frac\tau2 \mid 2\tau) &= \exp(-\frac{i \pi}{16}) \, \theta\bk{\C+\frac14}{0}(\mathfrak{z}\mid 2\tau) \, ,
\\
\theta\bk{\C}{0}(\mathfrak{z} + \frac{1+\tau}2 \mid 2\tau) &= \exp(-\frac{i \pi}{16}) \, \theta\bk{\C+\frac14}{\frac12}(\mathfrak{z} \mid 2\tau) \, ,
\fe
we find that in the limit of the two twist fields colliding $z \to 0$, the period matrix $\Pi$ behaves as
\ie
\label{PiLimit}
\Pi_{0,0}(0, \tau) = \tau \, , \quad \Pi_{0,\frac12}(0, \tau) = i \infty \, , \quad \Pi_{\frac12,0}(0, \tau) = i0^+ \, , \quad \Pi_{\frac12,\frac12}(0, \tau) = -1+i0^+ \, .
\fe

We are now ready to examine the torus Regge limit. To recap, the torus two-point function is a sum of four terms
\ie
\la {\cal E}(z, \bar z) {\cal E}(0) \ra_{\Sigma_{g=1}(\tau, \bar\tau)} = \frac{1}{2 \, |\eta(\tau)|^2} \sum_{\varepsilon, \D \in \frac12 \bZ_2} Z^{\rm cl}(r, \Pi_{\varepsilon, \D}, \overline\Pi_{\varepsilon, \D}) \, \left| c\bk{\varepsilon}{\D}(\tau) \right|^{-2} \, .
\fe
Under $z \to z+1$,
\ie
\label{ZclC}
& Z^{\rm cl}(r, \Pi_{\varepsilon, \D}, \overline\Pi_{\varepsilon, \D}) \to Z^{\rm cl}(r, \Pi_{\varepsilon, \D+\frac12}, \overline\Pi_{\varepsilon, \D}) \, ,
\\
& \left|c\bk{\varepsilon}{\D}(\tau)\right|^{-2} \to e^{\frac{i\pi}{8}} |E(z, 0)|^{-\frac14} \frac{\theta\bk{
\frac{\varepsilon}{2}}{\D+\frac12}(\frac12 \int_0^z \omega \mid 2\tau)}{\theta\bk{0
}{0}(0 \mid 2\Pi_{\varepsilon, \D+\frac12})} 
\frac{\overline{\theta\bk{
\frac{\varepsilon}{2}}{\D}}(\frac12 \int_0^z \omega \mid 2\bar\tau)}{\overline{\theta\bk{0
}{0}}(0 \mid 2\overline\Pi_{\varepsilon, \D})} \, ,
\fe
where we have set $\C = 0$ without loss of generality. The $e^{2i\pi h} = e^{\frac{i\pi}{8}}$ phase will henceforth be stripped off. In the further $z, \, \bar z \to 0$ limit, in each term the limiting $\Pi$ and $\overline\Pi$ each takes one of the four values given in \eqref{PiLimit}, and the combined limits of the four terms are summarized in Table~\ref{Tab:Limits}. It suffices to examine say the first and third limits in Table~\ref{Tab:Limits}, as the remaining two are related by complex conjugation.

For the first limit, $\Pi \to i\infty$ projects the solitonic sum to $p = 0$, where we see a dichotomy between rational and irrational $r^2$. If irrational, then the only term with $p = 0$ is $p = \bar p = 0$; if $r^2 = u/v$ is rational, then $p = 0$ corresponds to
\ie
(m, w) \in \{ n \times (u, v) \mid n \in \bZ \} \, .
\fe
Hence
\ie
\lim_{\Pi \to i\infty} Z^{\rm cl}(r, \Pi, \bar\tau) &= \lim_{\substack{\Pi \to i\infty \\ \overline\Pi \to \bar\tau}}
\sum_{p, \bar p \in \Gamma_r} \exp\left[ \frac{i\pi}{2} (p^2 \Pi - \bar p^2 \bar\tau) \right]
\\
&= \lim_{\Pi \to i\infty} 
\sum_{m, w \in \bZ} \exp\left[ \frac{i\pi}{2} ( \left(\frac{m}{r}+wr)^2 \Pi - (\frac{m}{r}-wr)^2 \bar\tau \right) \right]
\\
&= \begin{cases}
\displaystyle 
\sum_{n \in \bZ} \exp\left[ -2 i \pi u v n^2 \bar\tau \right]
= \overline{\theta\bk{0}{0}}(0|2{uv}\bar\tau) 
& \quad r^2 = u/v \text{ rational} \, ,
\\
\displaystyle 1 & \quad r^2 \text{ irrational} \, .
\end{cases}
\fe

Next consider the third term. The $\Pi, \overline\Pi$-dependent factors are the classical solitonic sum together with the denominator of \eqref{ZclC},
\ie
\label{ZclC2}
\lim_{\substack{\Pi \to i0^+ \\ \overline\Pi \to -1-i0^+}} \frac{Z^{\rm cl}(r, \Pi, \overline\Pi)}{{\theta\bk{0
}{0}(0 \mid 2\Pi)} \, {\overline{\theta\bk{0
}{0}}(0 \mid 2\overline\Pi)}} \, . 
\fe
The limit can be easily taken by first performing a modular transformation. Writing $\Pi' \equiv -1/\Pi$ and $\overline\Pi' \equiv -1/\overline\Pi$, and noting
\ie
Z^{\rm cl}(r, \Pi, \overline\Pi) &= \sqrt{{\Pi' \, \overline\Pi'}} \times Z^{\rm cl}(r, \Pi', \overline\Pi') \, ,
\quad
{\theta\bk{0}{0}(0 \mid 2\Pi)} = \sqrt{\frac{\Pi'}{2}} \times {\theta\bk{0}{0}(0 \mid \frac{\Pi'}{2})} \, ,
\fe
\eqref{ZclC2} becomes
\ie
& \lim_{\substack{\Pi \to i0^+ \\ \overline\Pi \to -1-i0^+}} \frac{Z^{\rm cl}(r, \Pi, \overline\Pi)}{{\theta\bk{0
}{0}(0 \mid 2\Pi)} \, {\overline{\theta\bk{0
}{0}}(0 \mid 2\overline\Pi)}}
= 
\lim_{\substack{\Pi' \to i\infty \\ \overline\Pi' \to 1-i0^+}} \frac{2 \, Z^{\rm cl}(r, \Pi', \overline\Pi')}{{\theta\bk{0
}{0}(0 \mid \frac{\Pi'}{2})} \, {\overline{\theta\bk{0
}{0}}(0 \mid \frac{\overline\Pi'}{2})}}
\\
&= 
\begin{cases}
\displaystyle
\lim_{\overline\Pi' \to 1-i0^+} 
\frac{2 \, \overline{\theta\bk{0}{0}}(0|2{uv}\overline\Pi')}
{{\overline{\theta\bk{0}{0}}(0 \mid \frac{\overline\Pi'}{2})}}
= \sqrt{\frac{2}{uv}} \, \exp(-\frac{i\pi}{4}) & \quad r^2 = u/v \text{ rational} \, ,
\\
\\
\displaystyle
\lim_{\overline\Pi' \to 1-i0^+} 
\frac{2}
{{\overline{\theta\bk{0}{0}}(0 \mid \frac{\overline\Pi'}{2})}}
= 0 & \quad r^2 \text{ irrational} \, .
\end{cases}
\fe

Collecting everything, the final results are summarized as follows. 

\begin{table}[t]
\centering
\begin{tabular}{c|c}
$(\varepsilon, \D)$ & $(\Pi, \overline\Pi)$
\\\hline
$(0,0)$ & $(i\infty \, , \ \bar\tau)$
\\
$(0,\frac12)$ & $(\tau \, , \ -i\infty)$
\\
$(\frac12,0)$ & $(i0^+ \, , \ -1+i0^+)$
\\
$(\frac12,\frac12)$ & $(-1+i0^+ \, , \ -i0^+)$
\end{tabular}
\caption{Limiting values of the moduli $(\Pi, \overline{\Pi})$ in the spatial torus Regge limit for the four terms in the torus two-point function of twist fields.}
\label{Tab:Limits}
\end{table}

\subsubsection{Rational points}

If $r^2 = u/v$ is rational, then the loop-normalized torus partition function twisted by $\cL_{\cal E}$ is
\ie
Z^{\widetilde\cL_{\cal E}}
(\tau, \bar\tau)
=
\frac{Z^{\cL_{\cal E}}(\tau, \bar\tau)}{\vev{\cL_{\cal E}}_{\bR^2}}
&= \frac{1}{2 \, |\eta(\tau)|^2} \Bigg\{ \left( \theta\bk{0}{\frac12}(0 \mid 2\tau) \, \overline{\theta\bk{0}{0}(0 \mid 2{uv}\tau)} + \text{c.c.}\right)
\\
& \quad + \sqrt{\frac{2}{{uv}}} \left( \exp(-\frac{i\pi}{4}) \, \theta\bk{\frac14}{\frac12}(0 \mid 2\tau) \, \overline{\theta\bk{\frac14}{0}(0 \mid 2\tau)} + \text{c.c.}  \right) \Bigg\} \, .
\fe
Using the identities
\ie
{\theta\bk{0}{0}(0 \mid 2{uv}\tau)\over \eta(\tau)}=\sum_{m,w\in\bZ,\,p_{\rm R}=0} \chi_{p_{\rm L}^2\over 4}(\tau) \, ,\quad
{\theta\bk{0}{{1\over 2}}(0 \mid 2\tau)\over \eta(\tau)}=\sum_{n=0}^\infty (-1)^n\chi_{n^2}(\tau) \, ,
\fe
we can decompose the twisted partition function into Virasoro characters
\ie\label{eqn:ZLEdV}
\frac{Z^{\cL_{\cal E}}(\tau, \bar\tau)}{\vev{\cL_{\cal E}}_{\bR^2}}
&= \left( \sum_{\substack{m \in \bZ,\, w\in\bZ_{>0} \\  mv=  wu}} \chi_{p_{\rm L}^2\over 4}(\tau)\sum_{n=0}^\infty (-1)^n\overline{\chi_{n^2}(\tau)} + \text{c.c.} \right) && \text{(one-sided degenerate)}
\\
& +\sum_{n, \bar n \in \bZ_{\ge0} } {(-1)^n+(-1)^{\bar n}
\over 2}\chi_{ n^2}(\tau) \overline{\chi_{ \bar n^2}(\tau)} && \text{(two-sided degenerate)}
\\
& \hspace{-.5in} + \frac{1}{\sqrt{2{uv}}} \sum_{n,\bar n \in \bZ_{\ge0}} (1+\cos\tfrac{(n-\bar n)\pi}{2}-\sin\tfrac{(n+\bar n)\pi}{2}) && \text{(twisted sector)}
\\
&&&
\hspace{-3.5in} \quad \times(-1)^{{1\over 4}(1-\cos{n\pi\over 2}+\sin{n\pi\over 2})(1-\cos{\bar n\pi\over 2}+\sin{\bar n\pi\over 2})}\chi_{{(2n+1)^2\over 16}}(\tau)\overline{\chi_{{(2\bar n+1)^2\over 16}}(\tau)} \, .
\fe
The action of $\cL_{\cal E}$ can be figured out by comparing the decomposition of the twisted partition function \eqref{eqn:ZLEdV} with the decomposition of the partition function with no twist \eqref{RationalDecomp}, which for the ease of reference we reproduce below
\ien
Z(\tau, \bar\tau)
&=\left(\sum_{\substack{m \in \bZ,\,w\in\bZ_{>0} \\  mv\neq\pm wu}}+\sum_{\substack{m \in \bZ_{>0} \\ w=0}}\right) \,\chi_{\frac{p_{\rm L}^2}{4}}(\tau) \overline{\chi_{  \frac{p_{\rm R}^2}{4}}(\tau)} && \text{(non-degenerate)}
\\
& + \left( \sum_{\substack{m \in \bZ,\, w\in\bZ_{>0} \\  mv=  wu}} \chi_{  \frac{p_{\rm L}^2}{4}}(\tau)\sum_{n=0}^\infty \overline{\chi_{n^2}(\tau)}
+ \text{c.c.} \right) && \text{(one-sided degenerate)}
\\
& +
\sum_{n, \bar n \in \bZ_{\ge0}}{1+(-1)^{n+\bar n}\over 2} \chi_{ n^2}(\tau) \overline{\chi_{ \bar n^2}(\tau)} && \text{(two-sided degenerate)}
\\
&\hspace{-.5in} + \sum_{n,\bar n \in \bZ_{\ge0}} (1+\cos\tfrac{(n-\bar n)\pi}{2}-\sin\tfrac{(n+\bar n)\pi}{2}) \, \chi_{{(2n+1)^2\over 16}}(\tau)\overline{\chi_{{(2\bar n+1)^2\over 16}}(\tau)} && \text{(twisted sector)} \, .
\fen
The above decompositions are irreducible if $r$ is irrational (though $r^2$ is rational); otherwise, the characters with $p_{\rm L} \in 2\bZ$ or $p_{\rm R} \in 2\bZ$ are further reducible. For the simplicity of discussion, we assume that $r$ is irrational.

The action on primary states without multiplicity can be directly read off: in the untwisted sector, $\cL_{\cal E}$ annihilates all non-degenerate primaries with no multiplicity, and acts on the (one- and two-sided) degenerate primaries by signs. On primary states with multiplicity, the action of $\cL_{\cal E}$ cannot be unambiguously determined from \eqref{eqn:ZLEdV} alone. Nevertheless, in the following we propose an action that is consistent with the special rational points examined in Appendix~\ref{Sec:RationalOrbifold}, and we believe that this action is correct at all rational points.

The non-degenerate states in the untwisted sector have multiplicity two when $m=m'u$ and $w=w'v$, with $m',\, w'\in\bZ$ and $m'\neq\pm w'$. We propose that in an appropriate basis, $\cL_{\cal E}$ acts on them by 
\ie
\sqrt{{uv}}\begin{pmatrix}
1 && 0
\\
0 && -1
\end{pmatrix} \, .
\fe
In the Ising$^2$ description of the ${{\rm S}^1_{r=\sqrt2}}/{\bZ_2}$ point, $\cL_{\cal E} = \cN_1$ or $\cN_2$ acts on the pair of ${\rm Vir}^2 \times \overline{\rm Vir}^2$ primaries $\varepsilon_1, \, \varepsilon_2$ by $\pm\sqrt{2}$.

The twisted sector primaries all have multiplicity two. The pattern exhibited by the special rational points in Appendix~\ref{Sec:RationalOrbifold} suggests that $\cL_{\cal E}$ is oriented when $u$ and $v$ are both odd, and unoriented otherwise.

When $\cL_{\cal E}$ is unoriented, we propose that in an appropriate basis, its action is
\ie
(-1)^{{1\over 4}(1-\cos{n\pi\over 2}+\sin{n\pi\over 2})(1-\cos{\bar n\pi\over 2}+\sin{\bar n\pi\over 2})}\begin{pmatrix}
\sqrt2 && 0
\\
0 && 0
\end{pmatrix} \, .
\fe
Consider again Ising$^2 = {{\rm S}^1_{r=\sqrt2}}/{\bZ_2}$. The ${\rm Vir}^2 \times \overline{\rm Vir}^2$ primaries in the twisted sector are
$\sigma_1, \, \sigma_2, \, \sigma_1 \varepsilon_2, \, \varepsilon_1 \sigma_2$. In this subspace, the TDL $\cL_{\cal E} = \cN_1$ annihilates $\sigma_1$ and $\sigma_1 \varepsilon_2$, but acts on $\sigma_2$ and $\varepsilon_1 \sigma_2$ by $\pm \sqrt2$.

When $\cL_{\cal E}$ is oriented, we propose that in an appropriate basis, its action is
\ie
\label{Z8Phases}
(-1)^{{1\over 4}(1-\cos{n\pi\over 2}+\sin{n\pi\over 2})(1-\cos{\bar n\pi\over 2}+\sin{\bar n\pi\over 2})}\begin{pmatrix}
e^{{i\pi\over 4}} && 0
\\
0 && e^{-{i\pi\over 4}} 
\end{pmatrix} \, ,
\fe
and the action of its orientation reversal $\ocL_{\cal E}$ is given by the complex conjugate. In the ${\rm S}^1_{r=2}$ description of the ${{\rm S}^1_{r=1}}/{\bZ_2}$ point, the twisted sector ground states correspond to
\ie
p_{\rm L} = \frac14(\frac{m}{2} + 2w)^2 \, , \quad p_{\rm R} = \frac14(\frac{m}{2} - 2w)^2 \, , \quad m = \pm 1 \, , \quad w = 0 \, .
\fe
In this subspace, the $\bZ_8$ line $\cL_{\cal E} = \cL_{(\pm\frac{\pi}{4}, \pi)}^{{\rm S}^1_{r=2}}$ indeed acts by the phases appearing in \eqref{Z8Phases}.

The modular S transform gives the loop-normalized defect partition function\footnote{Note that $\theta\bk{\frac12}{-\frac14}(0 \mid \frac\tau2)$ defined in \eqref{Riemann} as a $q$-series has coefficients that are $\frac{1}{\sqrt2}$ times integers due to the combinations of $e^{\pm\frac{i\pi}{4}}, \, e^{\pm\frac{3i\pi}{4}}$ phases. The $\frac{1}{\sqrt2}$ is compensated by the overall $\sqrt2$ factor to produce integer coefficients. 
}
\ie\label{eqn:rDPF}
Z_{\widetilde\cL_{\cal E}}(\tau, \bar\tau)
&= \frac{1}{4 \sqrt{uv} \, |\eta(\tau)|^2} \Bigg\{ \Bigg( \theta\bk{\frac12}{0}(0 \mid \frac\tau2) \, \overline{\theta\bk{0}{0}(0 \mid \frac{\tau}{2{uv}}) }
+ \text{c.c.}
\Bigg)
\\
& \hspace{1in}
+ \sqrt2 \left( \theta\bk{\frac12}{-\frac14}(0 \mid \frac\tau2) \, \overline{\theta\bk{0}{-\frac14}(0 \mid \frac{\tau}{2})} + \text{c.c.}\right) \Bigg\} \, .
\fe
The planar loop expectation value ${\vev{\cL}_{\bR^2}} = \sqrt{uv}$ is the smallest number such that the Cardy-normalized defect partition function
\ie
Z_{\cL}(\tau, \bar\tau) = {\vev{\cL}_{\bR^2}} \times Z_{\widetilde\cL,\,{\rm S}^1_r/\bZ_2}(\tau, \bar\tau)
\fe
has a character expansion with positive integer coefficients.\footnote{For each of the two pieces in braces in \eqref{eqn:rDPF}, there are terms in the $q, \, \bar q$-expansion with coefficient 2. But when the two pieces are combined, all terms have coefficients that are multiples of 4, canceling the overall $\frac14$.
}

\subsubsection{Irrational points}

If $r^2$ is irrational, and suppose the twist field factorizes through some $\cL_{\cal E}$, then the spatial torus Regge limit gives
\ie
\label{ZS1Z2Irr}
Z^{\widetilde\cL_{\cal E}}(\tau, \bar\tau)
&= \frac{1}{2|\eta(\tau)|^2} \left( \theta\bk{0}{\frac12}(0 \mid 2\tau) + {\overline{\theta\bk{0}{\frac12}}(0 \mid 2\bar\tau)} \right) 
\\
&=\frac{1}{2|\eta(\tau)|^2} \left( 1 + \sum_{n=1}^\infty (-)^n (q^{n^2} + \bar q^{n^2}) \right) \, .
\fe
The partition function and twisted partition function can be decomposed into irreducible Virasoro characters as
\ie
Z(\tau, \bar\tau) &= \frac{1 + \left| \theta\bk{0}{\frac12}(0|2\tau) \right|^2}{2|\eta(\tau)|^2} + \dotsb = \sum_{\substack{n, \bar n = 0 \\ n - \bar n \in 2\bZ}}^\infty \, \chi_{h=n^2} \bar\chi_{\bar h = \bar n^2} + \dotsb \, ,
\\
Z^{\widetilde\cL_{\cal E}}(\tau, \bar\tau) &= \sum_{\substack{n, \bar n = 0 \\ n - \bar n \in 2\bZ}}^\infty \, (-)^n \chi_{h=n^2} \bar\chi_{\bar h = \bar n^2} \, .
\fe
Since all states in the untwisted sector have no multiplicity, it is clear that $\cL$ acts on the (two-sided) degenerate modules by a sign, and annihilates the non-degenerate modules (there are no one-sided degenerate modules when $r^2$ is irrational). The twisted sector has multiplicity two, so the action of $\widetilde\cL_{\cal E}$ on the twisted sector cannot be unambiguously determined from \eqref{ZS1Z2Irr} alone. Nevertheless, this action should be the $u v \to \infty$ limit of the corresponding action of the loop-normalized $\widetilde\cL_{\cal E}$ at rational points: $\widetilde\cL_{\cal E}$ annihilates the twisted sector at irrational points.

The modular S transform of \eqref{ZS1Z2Irr} gives the defect partition function
\ie\label{eqn:irrDH}
Z_{\widetilde\cL_{\cal E}}
(\tau, \bar\tau) &= \frac1{2\sqrt2|\eta(\tau)|^2} \left( \frac{1}{\sqrt{i\bar\tau}} \theta\bk{\frac12}{0}(0 \mid \frac{\tau}{2}) + \frac{1}{\sqrt{-i\tau}}{\overline{\theta\bk{\frac12}{0}}(0 \mid \frac{\bar\tau}{2})} \right)
\\
&= \frac{1}{\sqrt2|\eta(\tau)|^2} \sum_{n=0}^\infty ( \frac{q^{\frac{(2n+1)^2}{16}}}{\sqrt{i\bar\tau}} + \frac{\bar q^{\frac{(2n+1)^2}{16}}}{\sqrt{-i\tau}} )
\\
&= \frac{1}{|\eta(\tau)|^2} \sum_{n=0}^\infty \int_0^\infty dp \left( q^{\frac{(2n+1)^2}{16}} \bar q^{\frac{p^2}{4}} + q^{\frac{p^2}{4}}\bar q^{\frac{(2n+1)^2}{16}} \right) \, ,
\fe
which has a spectrum of primary operators continuous in twist,
\ie
(h,\bar h) = (\frac{(2n+1)^2}{16}, \, \frac{p^2}{4}) \, , \ (\frac{p^2}{4}, \, \frac{(2n+1)^2}{16}) \, ,\quad n\in\bZ, \ p\in\bR \, .
\fe
At irrational points, the defect partition function \eqref{eqn:irrDH} coincides with the $uv \to \infty$ limit of that \eqref{eqn:rDPF} at rational points,
\ie
\lim_{uv \to \infty } Z_{\widetilde\cL_{\cal E}}
(\tau,\bar \tau) &= \frac{1}{4|\eta(\tau)|^2}\lim_{uv\to\infty }\frac{1}{\sqrt{uv}} \left( \theta\bk{\frac12}{0}(0 \mid \frac\tau2) \, \overline{\theta\bk{0}{0}}(0 \mid \frac{\bar\tau}{2{uv}}) 
+ \text{c.c.}
\right)
\\
&= \frac{1}{4|\eta(\tau)|^2}\lim_{uv\to\infty }\frac{1}{\sqrt{uv}} \sum_{m,n\in\bZ}\left( q^{(2n+1)^2\over 16} \bar q^{{m^2\over 4uv}}+q^{{m^2\over 4uv}} \bar q^{(2n+1)^2\over 16}\right)
\\
&= \frac{1}{|\eta(\tau)|^2}\sum^\infty_{n=0}\int^\infty_0 dp\left( q^{\frac{(2n+1)^2}{16}} \bar q^{\frac{p^2}{4}} + q^{\frac{p^2}{4}}\bar q^{\frac{(2n+1)^2}{16}} \right) \, .
\fe
Thus $\widetilde\cL_{\cal E}$ is a non-compact TDL at irrational points.

\subsection{Fusion rules for the non-compact topological defect lines}
\label{LLfusion}

Consider the fusion of the $\cL_{{\cal E}}$ line with its orientation reversal. When $r^2 = u/v$ is rational with $u, v$ coprime, the result can be decomposed into a sum over simple TDLs. One could decode the fusion rule by from the twisted partition function of the loop-normalized $\widetilde\cL_{{\cal E}} \, \widetilde\ocL_{{\cal E}}$,
\ie\label{eqn:ZLELEdV}
Z^{\widetilde\cL_{{\cal E}} \, \widetilde\ocL_{{\cal E}}}(\tau, \bar\tau) &= \frac{
Z^{\cL_{\cal E}\ocL_{\cal E}}(\tau, \bar\tau)
}{\vev{\cL_{\cal E}}_{\bR^2}^2}
\\
&= \sum_{\substack{m \in \bZ,\, w\in\bZ_{>0} \\  mv=wu}}  \chi_{p_{\rm L}^2\over 4}(\tau)\sum_{n=0}^\infty \overline{\chi_{n^2}(\tau)}+\sum_{n=0}^\infty \chi_{n^2}(\tau)  \sum_{\substack{m \in \bZ,\, w\in\bZ_{>0} \\  mv=-wu}}  \overline{\chi_{p_{\rm R}^2\over 4}(\tau)}
\\
&\quad +\left(\sum_{\substack{m \in \bZ,\,w\in\bZ_{>0} \\  m\neq\pm w}}+\sum_{\substack{m \in \bZ_{>0} \\ w=0}}\right)  \chi_{uv(m+w)^2\over 4}(\tau) \overline{\chi_{uv(m-w)^2\over 4}(\tau)}
\\
&\quad +\sum_{n, \bar n \in \bZ_{\ge0} } {1+(-1)^{n+\bar n}
\over 2}\chi_{ n^2}(\tau) \overline{\chi_{ \bar n^2}(\tau)}
\\
& \quad + \frac{1}{{uv}} \sum_{n,\bar n \in \bZ_{\ge0}} (1+\cos\tfrac{(n-\bar n)\pi}{2}-\sin\tfrac{(n+\bar n)\pi}{2})\chi_{{(2n+1)^2\over 16}}(\tau)\overline{\chi_{{(2\bar n+1)^2\over 16}}(\tau)} \, ,
\fe
computed by twice applying the $\cL_{{\cal E}}$ action proposed in Appendix~\ref{Sec:ReggeS1Z2}.

When $u$ is an even integer, the twisted partition function \eqref{eqn:ZLELEdV} can be written as\footnote{As noted before, the pattern exhibited by the special rational points in Appendix~\ref{Sec:RationalOrbifold} suggests that $\cL_{\cal E} = \ocL_{\cal E}$ is unoriented when $u$ or $v$ is even.
}
\ie
Z^{\widetilde\cL_{{\cal E}} \, \widetilde\ocL_{{\cal E}}}(\tau, \bar\tau) &= \frac{
Z^{\cL_{\cal E}\ocL_{\cal E}}(\tau, \bar\tau)
}{\vev{\cL_{\cal E}}_{\bR^2}^2}
\\
&=  {1\over 2|\eta(\tau)|^2}\sum_{m,w\in\bZ}\sum_{\substack{m'=0\\ m'\in2\bZ} }^{2u-2}\sum_{w'=0}^{v-1}{1\over uv}e^{\pi i({ v\over u}m'+w')m+\pi i(m'+{u \over v}w')w}{q^{\frac{p_{\rm L}^2}{4}} \bar q^{\frac{p_{\rm R}^2}{4}}}  
\\
&\quad+ \frac{|\theta_3(\tau) \theta_4(\tau)|}{2 |\eta(\tau)|^2}+ \frac{1}{{uv}}\left[ \frac{|\theta_2(\tau) \theta_3(\tau)|}{2 |\eta(\tau)|^2}+ \frac{|\theta_2(\tau) \theta_4(\tau)|}{2 |\eta(\tau)|^2}\right] 
\\
&=  {1\over uv|\eta(\tau)|^2}\Bigg(\sum_{\substack{m \in \bZ \\ w\in\bZ_{>0}}}+\sum_{\substack{m \in \bZ_{>0} \\ w=0}}\Bigg)\Bigg\{(1+(-1)^m){q^{\frac{p_{\rm L}^2}{4}} \bar q^{\frac{p_{\rm R}^2}{4}}}  
\\
&\quad\quad+\sum_{\substack{m=0\\m\in2\bZ}}^{2u-1} \sum^{v-1}_{\substack{w=2\\m\in2\bZ}}2\cos\left[\pi i({ v\over u}m'+w')m+\pi i(m'+{u \over v}w')w\right]{q^{\frac{p_{\rm L}^2}{4}} \bar q^{\frac{p_{\rm R}^2}{4}}}  
\\
&\quad\quad+\sum_{\substack{m=2\\m\in2\bZ}}^{u-2} 2\cos\left[\pi i({ v\over u}m'+w')m\right]{q^{\frac{p_{\rm L}^2}{4}} \bar q^{\frac{p_{\rm R}^2}{4}}}  \Bigg\}
\\
&\quad+ \frac{|\theta_3(\tau) \theta_4(\tau)|}{2 |\eta(\tau)|^2}+ \frac{1}{{uv}}\left[ \frac{|\theta_2(\tau) \theta_3(\tau)|}{2 |\eta(\tau)|^2}+ \frac{|\theta_2(\tau) \theta_4(\tau)|}{2 |\eta(\tau)|^2}\right] \, ,
\fe
from which we find the decomposition of $\cL_{\cal E}\,\ocL_{\cal E}$ to be
\ie
\cL_{{\cal E}} \, \ocL_{{\cal E}} = \cI+\eta_{\rm m}+\sum_{\substack{m=0\\m\in2\bZ}}^{2u-2} \sum^{v-1}_{\substack{w=2\\m\in2\bZ}} \cL_{-\pi(\frac{m}{r^2}+w, m+w r^2)}+\sum_{\substack{m=2\\m\in2\bZ}}^{u-2} \cL_{-\pi(\frac{m}{r^2}, m)}\, .
\fe
When $u$ is odd and $v$ is even, by an analogous calculation we obtain a similar fusion rule with $u$ and $v$ exchanged and $\eta_{\rm m}$ replaced by $\eta_{\rm w}$, which is expected by T-duality.

When $u$ and $v$ are both odd integers, the twisted partition function can be written as
\ie
Z^{\widetilde\cL_{{\cal E}} \widetilde\ocL_{{\cal E}}}(\tau, \bar\tau) &= \frac{
Z^{\cL_{\cal E}\ocL_{\cal E}}(\tau, \bar\tau)
}{\vev{\cL_{\cal E}}_{\bR^2}^2}
\\
&=  {1\over uv|\eta(\tau)|^2}\Bigg(\sum_{\substack{m \in \bZ \\ w\in\bZ_{>0}}}+\sum_{\substack{m \in \bZ_{>0} \\ w=0}}\Bigg)\Bigg\{{q^{\frac{p_{\rm L}^2}{4}} \bar q^{\frac{p_{\rm R}^2}{4}}}  
\\
&\quad\quad+\sum_{\substack{m=0\\m\in2\bZ}}^{2u-1} \sum^{v-1}_{\substack{w=2\\m\in2\bZ}}2\cos\left[\pi i({ v\over u}m'+w')m+\pi i(m'+{u \over v}w')w\right]{q^{\frac{p_{\rm L}^2}{4}} \bar q^{\frac{p_{\rm R}^2}{4}}}  
\\
&\quad\quad+\sum_{\substack{m=2\\m\in2\bZ}}^{u-1} 2\cos\left[\pi i({ v\over u}m'+w')m\right]{q^{\frac{p_{\rm L}^2}{4}} \bar q^{\frac{p_{\rm R}^2}{4}}}  \Bigg\}
\\
&\quad+ \frac{|\theta_3(\tau) \theta_4(\tau)|}{2 |\eta(\tau)|^2}+ \frac{1}{{uv}}\left[ \frac{|\theta_2(\tau) \theta_3(\tau)|}{2 |\eta(\tau)|^2}+ \frac{|\theta_2(\tau) \theta_4(\tau)|}{2 |\eta(\tau)|^2}\right] \, ,
\fe
from which we find the decomposition of $\cL_{\cal E}\,\ocL_{\cal E}$ to be
\ie
\cL_{{\cal E}} \, \ocL_{{\cal E}} = \cI+\sum_{\substack{m=0\\m\in2\bZ}}^{2u-2} \sum^{v-1}_{\substack{w=2\\m\in2\bZ}} \cL_{-\pi(\frac{m}{r^2}+w, m+w r^2)}+\sum_{\substack{m=2\\m\in2\bZ}}^{u-1} \cL_{-\pi(\frac{m}{r^2}, m)}\, .
\fe

For arbitrary $u$ and $v$, the twisted partition function can also be written as
\ie
& Z^{\widetilde\cL_{{\cal E}} \widetilde\ocL_{{\cal E}}}(\tau, \bar\tau) = \frac{
Z^{\cL_{\cal E}\ocL_{\cal E}}(\tau, \bar\tau)
}{\vev{\cL_{\cal E}}_{\bR^2}^2}
\\
&=  {1\over 2|\eta(\tau)|^2}\sum_{\substack{m\in u\bZ\\w\in v\bZ}}{q^{\frac{p_{\rm L}^2}{4}} \bar q^{\frac{p_{\rm R}^2}{4}}}  + \frac{|\theta_3(\tau) \theta_4(\tau)|}{2 |\eta(\tau)|^2}+ \frac{1}{{uv}}\left[ \frac{|\theta_2(\tau) \theta_3(\tau)|}{2 |\eta(\tau)|^2}+ \frac{|\theta_2(\tau) \theta_4(\tau)|}{2 |\eta(\tau)|^2}\right] 
\\
&=  {|\theta_3({uv\tau\over 2})|^2+|\theta_4({uv\tau\over 2})|^2\over 4|\eta(\tau)|^2}+ \frac{|\theta_3(\tau) \theta_4(\tau)|}{2 |\eta(\tau)|^2}+ \frac{1}{{uv}}\left[ \frac{|\theta_2(\tau) \theta_3(\tau)|}{2 |\eta(\tau)|^2}+ \frac{|\theta_2(\tau) \theta_4(\tau)|}{2 |\eta(\tau)|^2}\right] \, .
\fe
The defect partition function of the loop-normalized  $\widetilde\cL_{\cal E} \, \widetilde\ocL_{\cal E}$ is obtained by a modular S transform to be
\ie
\hspace{-.2in}
&Z_{\widetilde\cL_{\cal E} \widetilde\ocL_{\cal E}}
(\tau, \bar\tau)
\\
&=  {|\theta_3({2\tau\over uv})|^2+|\theta_2({2\tau\over uv})|^2\over 2uv|\eta(\tau)|^2}+ \frac{|\theta_2(\tau) \theta_3(\tau)|}{2 |\eta(\tau)|^2}+ \frac{1}{{uv}}\left[ \frac{|\theta_4(\tau) \theta_3(\tau)|}{2 |\eta(\tau)|^2}+ \frac{|\theta_2(\tau) \theta_4(\tau)|}{2 |\eta(\tau)|^2}\right] \, .
\fe
The Cardy-normalized defect partition function is the above multiplied by $\vev{\cL_{\cal E}}_{\bR^2}^2 = uv$, and has a discrete spectrum with integer multiplicities.

The defect partition function at irrational $r^2$ is obtained by taking the $uv\to\infty$ limit
\ie
& \lim_{uv\to \infty}Z_{\widetilde\cL_{\cal E} \,  \widetilde\ocL_{\cal E}}
(\tau, \bar\tau)
\\
&= \frac{1}{2|\eta(\tau)|^2}\lim_{uv\to\infty }\frac{1}{uv} \sum_{m,n\in\bZ}\left( q^{m^2\over uv} \bar q^{{n^2\over uv}}+q^{{(m-{1\over2})^2\over uv}} \bar q^{{(n-{1\over2})^2\over uv}}\right)+ \frac{|\theta_2(\tau) \theta_3(\tau)|}{2 |\eta(\tau)|^2}
\\
&=\frac{1}{|\eta(\tau)|^2}\int^\infty_{-\infty}dpd\bar p\, q^{p^2}\bar q^{p^2}+ \frac{|\theta_2(\tau) \theta_3(\tau)|}{2 |\eta(\tau)|^2}
\\
&={1\over 2}\int^{2\pi}_0{d\theta_{\rm w}d\theta_{\rm m}\over (2\pi)^2}Z_{\cL_{(\theta_{\rm m}, \theta_{\rm w})}}(\tau, \bar\tau) \, ,
\fe
from which we deduce the fusion rule
\ie
\widetilde\cL_{\cal E} \, \widetilde\ocL_{\cal E} = {1\over 2}\int^{2\pi}_0{d\theta_{\rm w}d\theta_{\rm m}\over (2\pi)^2}\cL_{(\theta_{\rm m}, \theta_{\rm w})} \, .
\fe

\bibliography{refs} 
\bibliographystyle{JHEP}

\end{document}